\documentclass[]{aa}  

\usepackage{graphicx}
\usepackage{epstopdf}
\usepackage{subcaption}
\usepackage{tablefootnote}
\usepackage{esvect}

\usepackage{txfonts}
\usepackage{color}
\begin{document} 

   \title{Simulating the cloudy atmospheres of HD 209458 b \\and HD 189733 b with the 3D Met Office Unified Model}

   \author{S. Lines
          \inst{1}
          \and
          N. J. Mayne\inst{1}\fnmsep
          \and
          I. A. Boutle\inst{1,3}\fnmsep
          \and
          J. Manners\inst{1,3}\fnmsep
          \and
          G. K. H. Lee\inst{4,5,6}\fnmsep
          \and
          Ch. Helling\inst{4,5}\fnmsep
          \and
          B. Drummond\inst{1}\fnmsep
          \and
          D. S. Amundsen\inst{1,7,8}\fnmsep
          \and
          J. Goyal\inst{1}\fnmsep
          \and
          D. M. Acreman\inst{1,2}\fnmsep
          \and
          P. Tremblin\inst{9}\fnmsep
           \and
          M. Kerslake\inst{1}\fnmsep        
          }

   \institute{Physics and Astronomy, College of Engineering, Mathematics and Physical Sciences, University of Exeter, EX4 4QL\\
              \email{s.lines@exeter.ac.uk}
         \and
              Computer Science, College of Engineering, Mathematics and Physical Sciences, University of Exeter, EX4 4QF
         \and
             Met Office, FitzRoy Road, Exeter, Devon EX1 3PB, UK
          \and
             Centre for Exoplanet Science, University of St Andrews, North Haugh, St Andrews, Fife, KY16 9SS, UK
          \and
             School of Physics and Astronomy, University of St Andrews, North Haugh, St Andrews, Fife, KY16 9SS, UK
          \and
             Atmospheric, Oceanic \& Planetary Physics, Department of Physics, University of Oxford, Oxford OX1 3PU, UK 
          \and
             NASA Goddard Institute for Space Studies, 2880 Broadway, New York, NY 10025
          \and
             Department of Applied Physics and Applied Mathematics, Columbia University, New York, NY 10027
	 \and
             Maison de la Simulation, CEA, CNRS, Univ. Paris-Sud, UVSQ, Universit\'e Paris-Saclay, 91191 Gif-Sur-Yvette, France
             }

   \date{Accepted February 28th 2018}

 
 \abstract
   {}
   {To understand and compare the 3D atmospheric structure of HD 209458 b and HD 189733 b, focusing on the formation and distribution of cloud particles, as well as their feedback on the dynamics and thermal profile.}
   {We couple the 3D Met Office Unified Model (UM), including detailed treatments of atmospheric radiative transfer and dynamics, to a kinetic cloud formation scheme. The resulting model self--consistently solves for the formation of condensation seeds, surface growth and evaporation, gravitational settling and advection, cloud radiative feedback via absorption and, crucially, scattering. We use fluxes directly obtained from the UM to produce synthetic spectral energy distributions and phase curves.}
   {Our simulations show extensive cloud formation in both HD 209458 b and HD 189733 b. However, cooler temperatures in the latter result in higher cloud particle number densities. Large particles, reaching 1 $\mu$m in diameter, can form due to high particle growth velocities, and sub-$\mu$m particles are suspended by vertical flows leading to extensive upper-atmosphere cloud cover. A combination of meridional advection and efficient cloud formation in cooler high latitude regions, result in enhanced cloud coverage for latitudes $>$ 30$^{\textrm{o}}$ and leads to a zonally banded structure for all our simulations. The cloud bands extend around the entire planet, for HD 209458 b and HD 189733 b, as the temperatures, even on the day side, remain below the condensation temperature of silicates and oxides. Therefore, the simulated optical phase curve for HD 209458 b shows no `offset', in contrast to observations. Efficient scattering of stellar irradiation by cloud particles results in a local maximum cooling of up to 250 K in the upper atmosphere, and an advection-driven fluctuating cloud opacity causes temporal variability in the thermal emission. The inclusion of this fundamental cloud-atmosphere radiative feedback leads to significant differences with approaches neglecting these physical elements, which have been employed to interpret observations and determine thermal profiles for these planets. This suggests both a note of caution of interpretations neglecting such cloud feedback and scattering, and merits further study.}
   {}


   \maketitle
%

\section{Introduction}\label{sec:intro}

Two decades of exoplanet research has yielded over 3600 confirmed detections\footnote{www.exoplanet.eu} and has instilled the knowledge that with great numbers comes great diversity. Planets have been found with a wide range of orbital parameters, inferred bulk compositions and, most importantly for the question of habitability, atmospheric properties. With technological constraints limiting our ability to probe small terrestrial planets like our own, some of the best observational data regarding atmospheric composition and dynamics comes from the well studied planetary class of hot-Jupiters. Their large planetary radii coupled with their short-period, and often transiting nature leads to the retrieval of high-resolution spectral datasets. Such data has revealed the presence of a number of chemical species including water, sodium and potassium \citep[e.g.][]{charbonneau02,sing11,birkby17}. One of the most important results is that recent observations show, even within this limited sub-set of planets, that there exists a large range in atmosphere types which encompass clear, hazy and cloudy skies \citep{sing16}.

Clouds can be challenging because they obscure our view on the underlying atmospheric composition. Their wavelength dependent opacity can act as a blanket that mutes water features \citep{deming13,helling16} and other key species in the infra-red (IR) region of the transmission spectrum \citep{pont13,iyer16}. \cite{helling16} show that obscuration by high altitude cloud in their simulations of HD 209458 b (as opposed to deeper clouds in HD 189733 b) can explain the reduced H$_2$O amplitude in atmosphere observations of HD 209458 b, and underscores the impact of the vertical cloud structure on direct observables. Additionally, photochemically produced complex hydrocarbons \citep{zahnle16} or small cloud particles lofted into the upper atmosphere both contribute to an enhanced Rayleigh scattering slope in the visual that can, via an increased continuum opacity, mask, sometimes completely, the alkali metal lines \citep{kirk17}. The masking of absorption lines in transmission spectra makes placing constraints on absolute and even relative atmospheric chemical abundances difficult. Spectral lines are also key in measuring the doppler shift caused by the fast global winds detected on some of these planets \citep{snellen10,birkby13,louden15,birkby17}.

\cite{cushing06} obtain spectra of the L-type brown dwarf DENIS 0255 and infer a silicate feature from a dip in the emission around 10 $\mu$m which is not seen in clear skies atmosphere models. No such detection has been made in an exoplanet atmosphere; \cite{richardson07} claimed to have detected a broad emission silicate feature at 9.65$\mu$m in HD 209458 b, but it was not found in subsequent studies \citep{swain08}. The James Webb Space Telescope (JWST) however will enhance available constraints on models, with the 10 $\mu$m wavelength coverage potentially allowing for the first direct detections of clouds via their vibrational mode absorption features (Wakeford et al., 2016). \citep{sing16} suggest that silicate detection requires significant vertical mixing to sustain condensates at observable pressures. However, the primary difficulty in obtaining direct signatures of clouds in the spectra, currently, is due to the quality of the data obtained from exoplanet observations.

Aside from the aforementioned modifications to both the visual and near-IR transmission spectra from cloud scattering and absorption, clouds have also been inferred by other means. One of the more compelling observational indications of inhomogeneous cloud coverage is the westwards offset in the peak optical phase curve, first detected for Kepler-7b by \cite{demory13}. A westwards shift in the visual curve suggests more efficient backscattering of optical photons near the night-day terminator. \cite{oreshenko16} investigates the sensitivity of this offset in Kepler-7b to aerosol and condensate composition. Such a phase curve offset has also been observed in thermal emission (although the presence of clouds is not required to produce this effect); an eastwards offset in the IR phase curve which shows the hot-spot shift from the sub-stellar point \citep{knutson07,crossfield10}. \cite{evans13} find a decreasing albedo towards longer wavelengths of their $HST$ / STIS observations; evidence for day-side optically thick, reflective clouds. A large horizontal temperature gradient, expected for hot-Jupiters, can lead to a cyclic mechanism of night-side condensation to day-side evaporation \citep{lee16}. Clouds advected onto the day-side could possibly introduce a higher visual albedo compared to the day-night transition region for a clear skies atmosphere. Further studies have shown that offsets in the visual for Kepler planets are common \citep{angerhausen15,esteves15,shporer15}. \cite{armstrong16} observed the first temporal variability of the phase curve from HAT-P-7b which is indicative of variable cloud coverage potentially driven and moderated by global winds. Additionally, \emph{Spitzer} data from the emission of WASP-43b shows a distinct lack of night-side flux, consistent with increased cloud opacity in this region \citep{stevenson17}. For the hottest exoplanets that are expected to exhibit such a hemispherical dichotomy in cloud coverage, ionising processes, for example by the stellar irradiation and cosmic ray impact \citep{helling16c} may open new pathways to form cloud particles even higher in the atmosphere. We would consequently expect the formation of an ionosphere also on exoplanets \citep{helling17}. This may be the case for the hot super-earth CoRoT-7b \citep{helling16b,mahapatra17}.

\cite{louden15,brogi16} inferred wind speeds of up to 5.3 kms$^{-1}$ on HD 189733 b, and models have shown that high-velocity, super-rotating jets should be common for tidally locked hot-Jupiters \citep{showman02,mayne14}. Zonally asymmetric radiative forcing is responsible for the generation of strong eastwards equatorial flow on these hot-Jupiter exoplanets; a combination of eddy and mean flow momentum transport specifically involved in the jet pumping \citep{showman11,tsai14,mayne17}. For the formation of clouds this is particularly relevant as the advection of cloud particles from the night-side to the day-side could result in their subsequent evaporation. Likewise, the advection of metal enriched gas can provide condensible material to the night-side. Moreover, the vertical transport of cloud from its coupling to the vertical advection of the gas (from mean flows) and the gravitational settling of cloud particles due to precipitation, can play a large role in the distribution of cloud and gases. As cloud particles drift into the deeper, hotter regions of the atmosphere they can evaporate and enrich specific element abundances \citep{woitke04,witte09}. 


We choose to focus on two well studied hot-Jupiters; HD 209458 b and HD 189733 b. Their similar orbital motions, physical size/mass, equilibrium temperatures and the discovery of water vapour, neutral oxygen, carbon monoxide and methane, suggest that models indicating they share similar cloud properties are unsurprising \citep{benneke15,helling16}. Given these similarities, the observational differences such as the weak H$_2$O absorption depth in HD 209458 b compared to HD 189733 b is surprising \citep{sing16}. \cite{helling16} suggest that the vertical distribution of cloud, and hence modification of the atmosphere optical depth, could be responsible for these contrasting observables.

The formation, structure, composition and evolution of clouds are all sensitive to the local thermal, chemical and dynamical conditions of the atmosphere which in turn respond to the radiative influence of such clouds on their host atmospheres. This complex interplay can be best understood by consistently treating cloud formation as part of a 3D atmosphere or global circulation model (GCM). Much progress has been made recently in modelling cloud-free planetary atmospheres, particularly hot-Jupiters. Simulation results \cite[e.g.][]{dobbs07,dobbs13,carone14,mayne14,heng15,amundsen16,boutle17} show a variety of extremely complex atmospheres that can be dominated by radiative flux, momentum-coupled horizontal and vertical advection and driving from the deep atmospheric thermal flux (in turn driven by overshoot from the convective interior).

Recently, \cite{lee16} combined the kinetic cloud formation model developed by \cite{woitke03,woitke04,helling06,helling08} \citep[for a summary see][]{helling13} with a 3D global circulation model \citep{dobbs13}. The cloud model describes the micro-physics of seed formation (nucleation) and surface growth/evaporation consistently coupled with element conservation, gas phase chemistry, hydrodynamical mixing and gravitational settling. We utilise the time-dependent representation of the cloud formation model as derived in \cite{helling06}, and couple it to the Met Office GCM, the `Unified Model' \citep{wood14,mayne14}, otherwise known as the UM. We present our model results for the atmospheres of the hot Jupiters HD 209458 b and HD 189733 b. We use this advanced, cloud-coupled UM version in order to present a comparative study between the two similar giant gas planets which only differ significantly in their gravitational strength. \cite{helling16} concentrated on the cloud structures for 1D atmospheres, \cite{lee16} presented the first consistent 3D atmosphere solution with cloud formation, and our paper presents a continuation of these modelling efforts, including full radiative treatment (absorption and scattering) of the cloud.

In section \ref{sec:method} we discuss the hydrodynamics, thermodynamics, radiative transfer and cloud formation formalism, including initial and boundary conditions. Our results are presented in section \ref{sec:results} and discussed in the context of previous and complementary studies in section \ref{sec:dis}. We summarise our results and discuss the model's limitation and potential improvements in section \ref{sec:summary}.

\section{Numerical methods}\label{sec:method}

We couple our state-of-the-art numerical 3D GCM with a state-of-the-art microphysical cloud model in order to simulate the dynamics of a cloudy hot-Jupiter atmosphere. To solve for the thermodynamical properties and hydrodynamical motions of the atmosphere, we use the Met Office UM which has been well tested, not only for our own atmosphere \citep{brown12}, but also for hot-Jupiter atmospheres in \cite{mayne14,amundsen14,amundsen16} and \cite{mayne17}. 

\begin{figure}[t]
\centering
\includegraphics[scale=0.16]{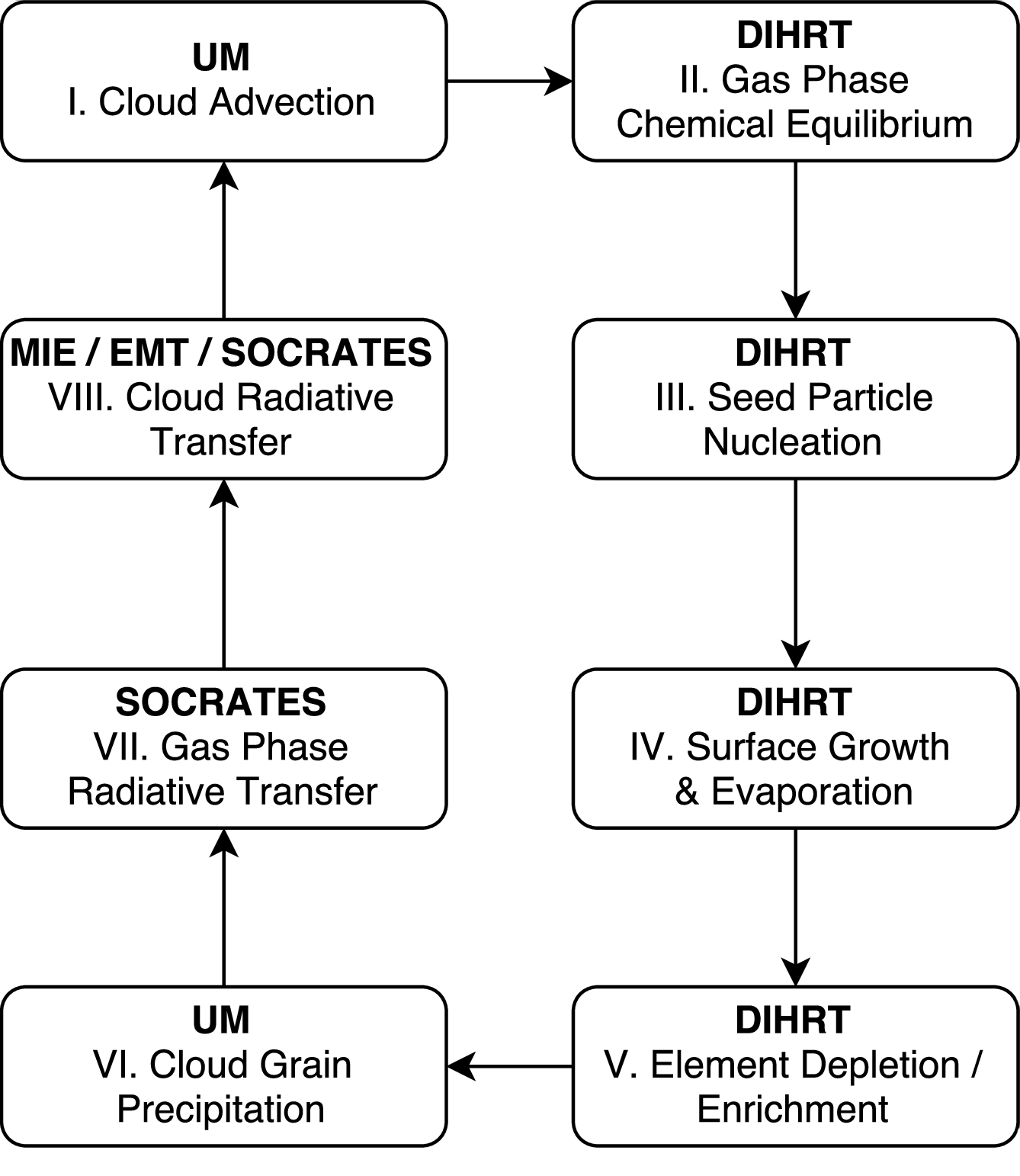}
\caption{The eight-stage DIHRT-UM cloud formation and evolution model flow. Heating rate solutions to the gas and cloud radiative transfer are applied to the UM and feedback onto the dynamics. Note: Absorption and scattering coefficients are found independently for the cloud and gas phase but the radiative transfer equations are only solved once for the combined coefficients.}
\label{fig:vec}
\end{figure}

The microphysical cloud model that we utilise here is summerised in \cite{helling13}, and has already been tested in its stationary form for brown dwarfs and giant gas planets by \cite{helling08-b1,helling08-b2} and \cite{witte09,witte11}, and by \cite{juncher17}, in addition to the time-dependent version as part of 2D turbulence modelling in \cite{helling01,helling04} and the already mentioned 3D GCM for HD 189733 b in \cite{lee16}. Figure \ref{fig:vec} is a flow schematic of the physical processes of our coupled model which are described in the following sections.

\subsection{Hydro- and thermodynamical modelling}

The UM uses the dynamical core and finite-difference code {\emph{ENDGame}}, \citep[Even Newer Dynamics for General Atmospheric Modelling of the Environment,][]{wood14}. For each vertical layer the atmosphere is discretised onto a latitude-longitude grid by using a staggered Arakawa-C grid \citep{arakawa77}. The vertical structure uses a geometric height (as opposed to pressure height) vertical grid using a vertically staggered Charney-Phillips grid \citep{charney53}. ENDGame is semi-Lagrangian and time stepping uses a semi-implicit scheme \citep{crank96}.

The UM's dynamical core solves five equations that correspond to the solution for a full, deep and non-hydrostatic atmosphere \citep[see][for more information]{mayne14}:

\begin{equation}
\frac{Du}{Dt} = \frac{uv\, \textrm{tan}\phi}{r} - \frac{uw}{r} + fv - f'w - \frac{c_{p}\theta}{r\, \textrm{cos}\phi}\frac{\partial \Pi}{\partial \lambda} + {\bf{D}}(u),
\end{equation}

\begin{equation}
\frac{Dv}{Dt} = \frac{u^{2}\, \textrm{tan}\phi}{r} - \frac{vw}{r} - uf - \frac{c_{p}\theta}{r}\frac{\partial \Pi}{\partial \phi} + {\bf{D}}(v),
\end{equation}

\begin{equation}
\frac{Dw}{Dt} = \frac{u^{2} + v^{2}}{r} + uf - g(r) - c_{p}\theta \frac{\partial \Pi}{\partial r},
\end{equation}

\begin{equation}
\delta \frac{D\rho_{\textrm{gas}}}{Dt} = -\rho_{\textrm{gas}} \left [ \frac{1}{r\, \textrm{cos}\phi} \frac{\partial u}{\partial \lambda} + \frac{1}{r\, \textrm{cos} \phi} \frac{\partial(v\, \textrm{cos}\phi)}{\partial \phi} + \frac{1}{r^{2}} \frac{\partial (r^{2}w)}{\partial r} \right ],
\end{equation}

\begin{equation}
\frac{D\theta}{Dt}=\frac{Q}{\Pi}+{\bf{D}}(\theta),
\end{equation}

\begin{equation}
\Pi ^{\frac{1-\kappa}{\kappa}} = \frac{R\rho_{\textrm{gas}} \theta}{p_0}.
\end{equation}
We define the coordinates $\lambda$ as the longitude, $\phi$ as the latitude, $r$ as radial distance from the planetary centre, and $t$ as time. The equations are solved for the zonal ($u$), meridional ($v$) and vertical ($w$) wind velocity components as well as the fluid density, $\rho_{\textrm{gas}}$ and the potential temperature (temperature at a fixed reference pressure), $\theta$. $c_{p}$ is the fluid specific heat capacity, R is gas constant, Q is the heating rate, {\bf{D}} is the diffusion operator, $\kappa$ is the ratio $R / c_{p}$, $\Pi$ is the Exner pressure function and the Coriolis parameters $f = 2 \Omega \textrm{sin}\phi$ and $f' = 2 \Omega \textrm{cos}\phi$, with $\Omega$ the rotation frequency of the planet. g(r) is the height-varying gravitational acceleration at radial position $r$ calculated via,

\begin{equation}
g(r) = g_{p}\left( \frac{R_{p}}{r} \right)^{2},
\end{equation}
where $g_p$ is the gravitational acceleration, $R_p$ is the radial value for the planetary inner boundary and g(r) is always directed towards the lower boundary in each cell. 

To smooth out grid-scale noise and to account for unresolved sub-grid processes such as eddies and turbulence we use the diffusion scheme\footnote{The diffusion scheme is incorrectly described in \cite{amundsen16} and prior studies.} in \cite{mayne17}. Additionally, at the upper boundary we implement a sponge layer as described in \cite{mayne14}.


\subsection{The cloud formation model}
\label{sec:cm}

Cloud formation is a phase transition process from a gas to a solid (or liquid) phase and progresses through two major steps: seed formation (nucleation) and surface growth/evaporation. These processes are complemented by element conservation, gravitational settling and advection. \cite{helling13} present the kinetic cloud formation approach derived by \cite{woitke03,woitke04,helling06,helling08}. We therefore only summarise the main equation body, applicable to this work, that we solve in combination with chemical equilibrium in our 3D radiation-hydrodynamics code. In the following work, `cloud' is used to refer to an ensemble of particles that have a local and a global size distribution, that are made of a mix of materials which can change in time and in space, and which has a certain geometrical extension.

The cloud particle properties in each cell are described in terms of a set of dust moments $L_{j}$($\bf{r,t}$) [cm$^j$g$^{-1}$] ($j$ = 0, 1, 2, 3). Dust moments are integrated volume-weighted particle size distributions \citep{woitke03,woitke04,helling06,helling08}:

\begin{equation}
\rho_{\textrm{gas}}({\bf{r}},t) L_{j}({\bf{r}},t) = \int_{V_{l}}^{\infty}f(V,{\bf{r}},t)V^{j/3}dV,
\end{equation}
where $j$ is the moment index, $\rho_{gas}$({\bf{r}},t) is the gas density, $V_l$ is the seed particle volume [cm$^3$] and $f(V,{\bf{r}})$ is the cloud particle size distribution function [cm$^{-6}$].

The dust moments L$_{j}({\bf{r}},t)$ are calculated from the dust moment equation \citep{woitke03}:

\begin{equation}
\frac{D (\rho_{\textrm{gas}}L_j)}{Dt} = V_l^{j/3}J_* + \frac{j}{3}\chi^{\textrm{net}}\rho_{\textrm{gas}}L_{j-1}.
\label{eq:dm}
\end{equation}
The evolution of the dust moments in time (left hand side) is dependent on the nucleation of seed particles (first term right hand side) and the net surface growth velocity (second term right hand side). $J_*$ is the nucleation rate [cm$^{-3}$s$^{-1}$] and $\chi^{\textrm{net}}$ is the net surface growth velocity [cms$^{-1}$] which has a negative value for evaporation ($\leq$ 0 cms$^{-1}$) and positive for growth ($\geq$ 0 cms$^{-1}$). Applying the operator splitting method, equation \ref{eq:dm} is integrated in time during each hydrodynamic timestep. For this, we apply the implicit ordinary differential equation solver LIMEX \citep{deuflhard87}.

The cloud model describes the nucleation process by applying the modified classical nucleation theory to calculate the nucleation rate \citep[e.g.][]{helling13,gail14,lee15b}. The formation of seed particles during the nucleation process provides a surface for further chemical reactions. The cloud particle growth or evaporation velocity, $\chi^{\textrm{net}}(\bf{r},t)$ [cm s$^{-1}$], is calculated considering the summation of the growth of each dust species, $s$ from each contributing reaction $r$ \citep{gail86,helling06,helling13}.

The local number density of cloud particles, $n_{\textrm{d}}$, [cm$^{-3}$] is calculated by

\begin{equation}
n_{\textrm{d}}(\textbf{r}) = \rho_{\textrm{gas}} L_{0}(\textbf{r}),
\label{eq:nd}
\end{equation}
and the mean volume of a cloud particle (which we refer to in the rest of the manuscript as `mean cloud particle volume') by

\begin{equation}
\langle V \rangle =  \frac{L_{3}}{L_{0}}.
\label{eq:volume}
\end{equation}


\begin{figure}
\includegraphics[scale=0.6,angle=0]{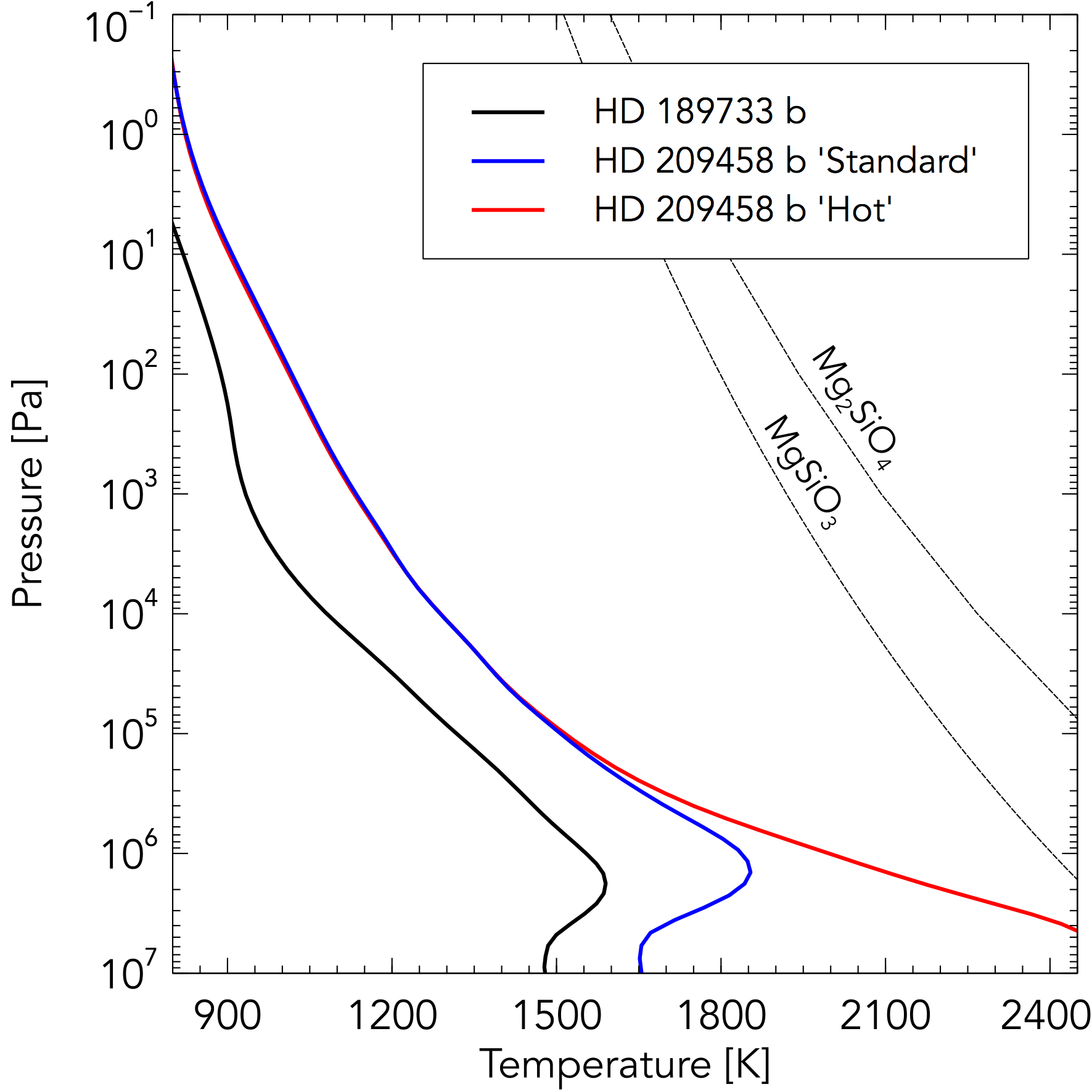}
\caption{Horizontally averaged pressure-temperature profiles of our three atmosphere models, prior to modification by the cloud radiative feedback, at t$_{\textrm{cloud}}$ = 0 days (see Table \ref{tab:params} and Section \ref{sec:na}). Dashed lines show condensation curve data for enstatite (MgSiO$_3$) and forsterite (Mg$_2$SiO$_4$). }
\label{fig:ptinit}
\end{figure}


Elements are depleted or enriched as a response to the formation or evaporation of cloud. The change in elemental abundance is accounted for via the following equation,

\begin{equation}
\begin{split}
\frac{D (n_{\langle H \rangle}\epsilon_i)}{Dt} = \sqrt[3]{36\pi} \rho_{\textrm{gas}} L_{2} \sum_{r=1}^{R} \frac{\Delta V_r^s n_r^{\textrm{key}}v_r^{\textrm{rel}}\alpha_r}{v_r^{\textrm{key}}} \left( 1 - \frac{1}{S_r} \frac{1}{b_{\textrm{surf}}^s} \right),
\end{split}
\label{eq:elemcon}
\end{equation}
where $n_{\langle H \rangle}$ is the total number density of hydrogen molecules and $\epsilon_i$ is an element's relative abundance to hydrogen. For elements where the abundance can be enriched/depleted, the value of $\epsilon_i$ is also advected with atmospheric flow, as solved in the UM by Eqs. 1-6.


\begin{figure*}[t!]

\begin{subfigure}{0.48\textwidth}
\includegraphics[scale = 0.085, angle = 0]{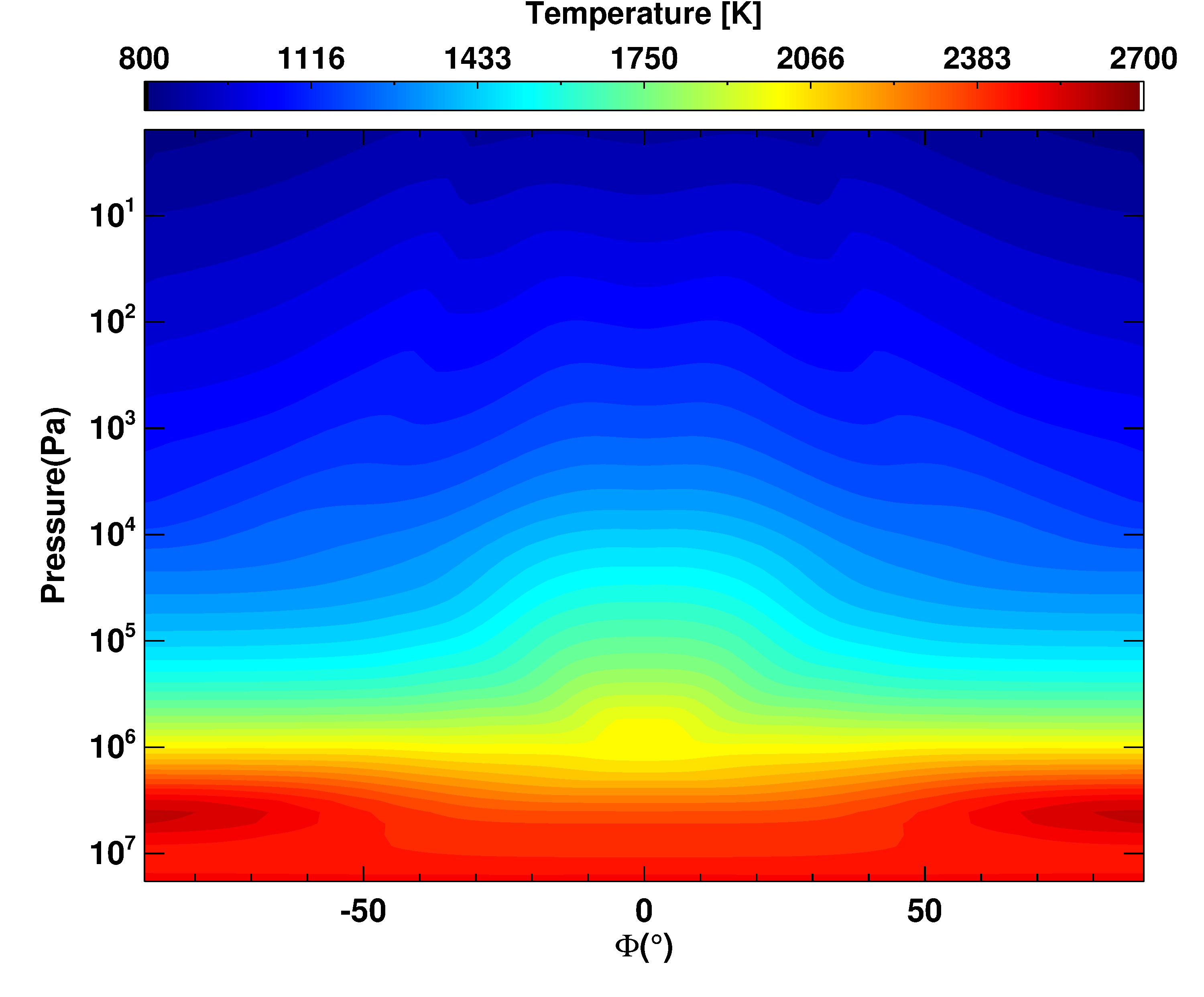}
\end{subfigure}\hspace*{\fill}
\begin{subfigure}{0.48\textwidth}
\includegraphics[scale = 0.085, angle = 0]{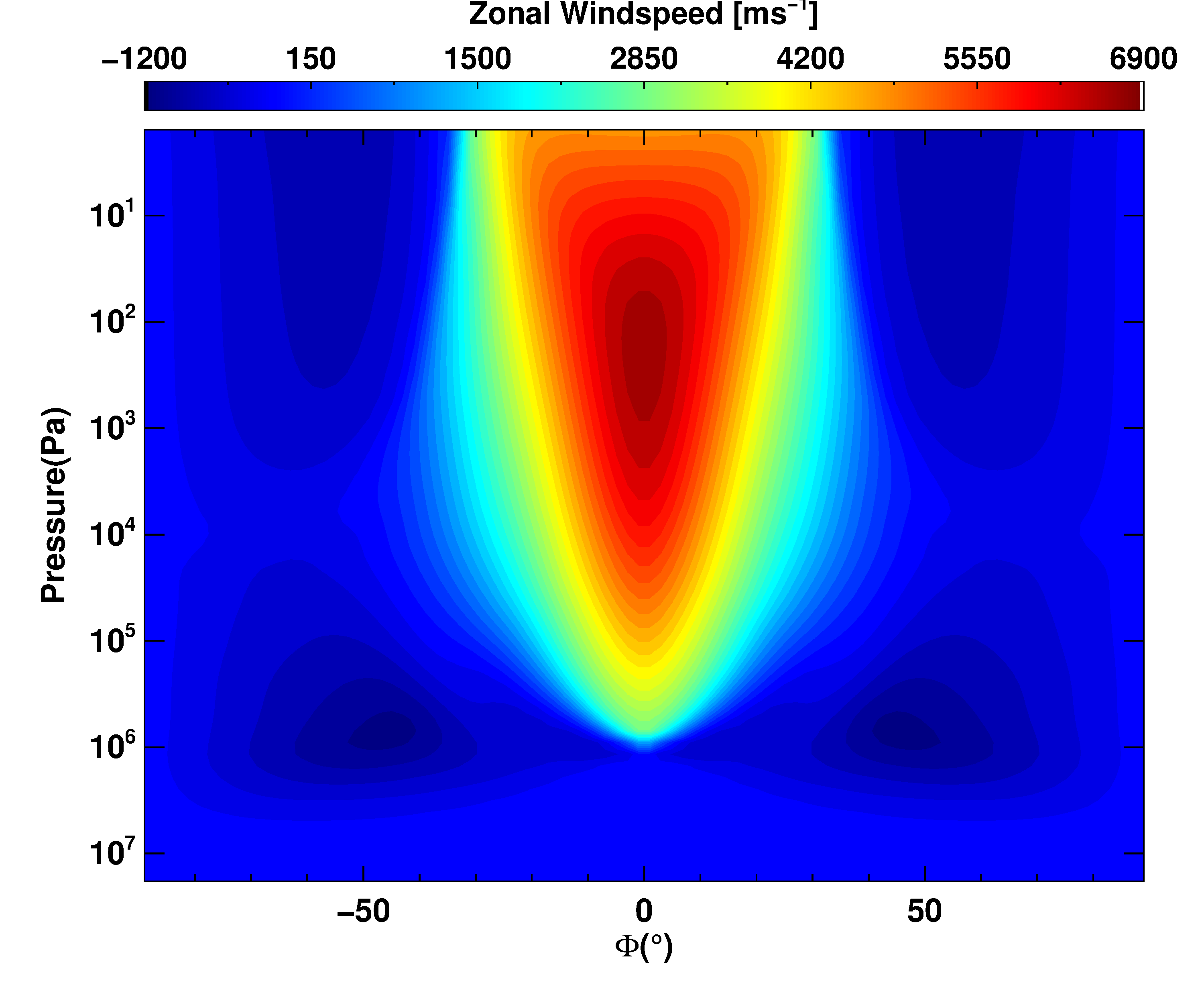}
\end{subfigure}

\medskip
\begin{subfigure}{0.48\textwidth}
\includegraphics[scale = 0.085, angle = 0]{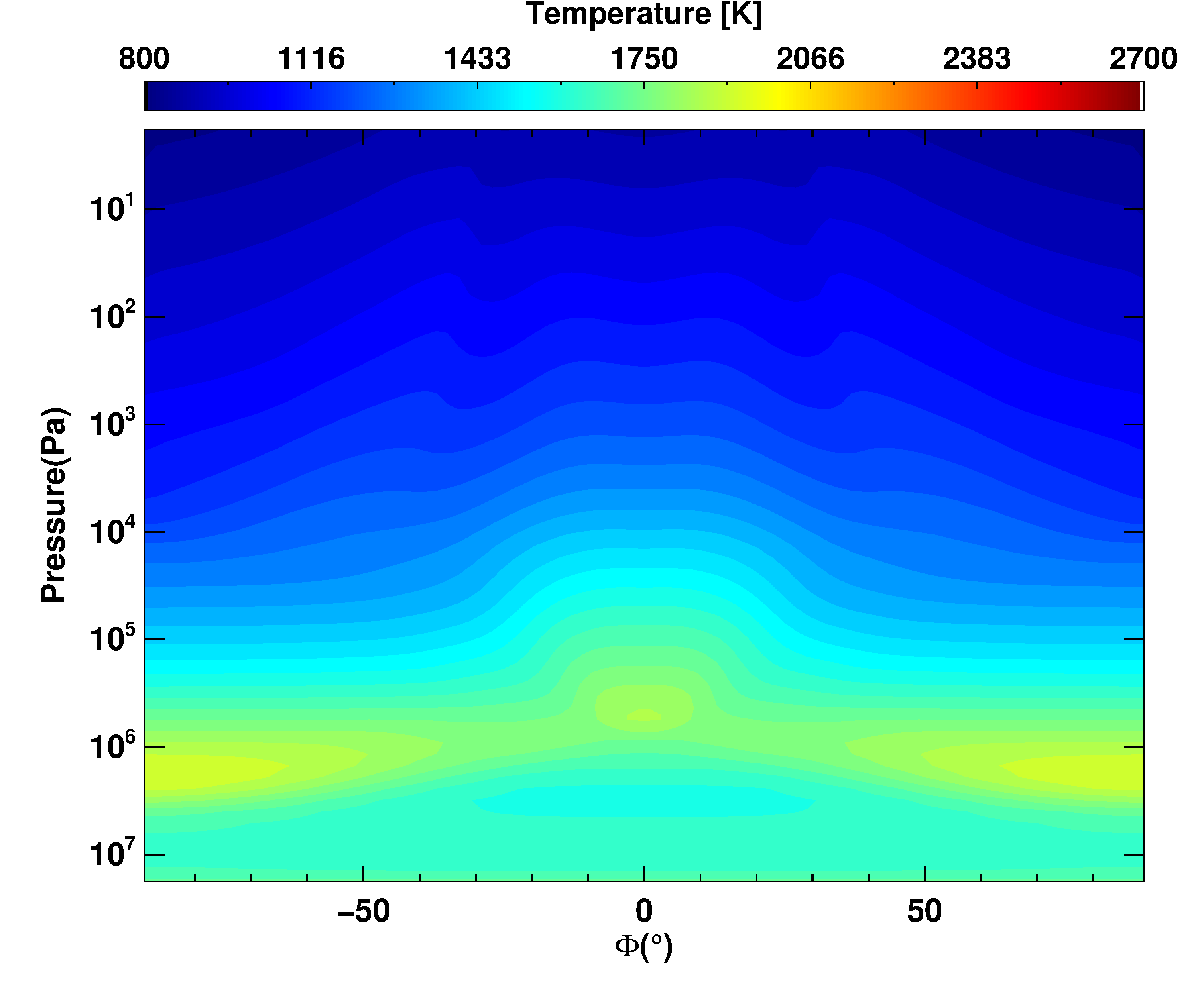}
\end{subfigure}\hspace*{\fill}
\begin{subfigure}{0.48\textwidth}
\includegraphics[scale = 0.085, angle = 0]{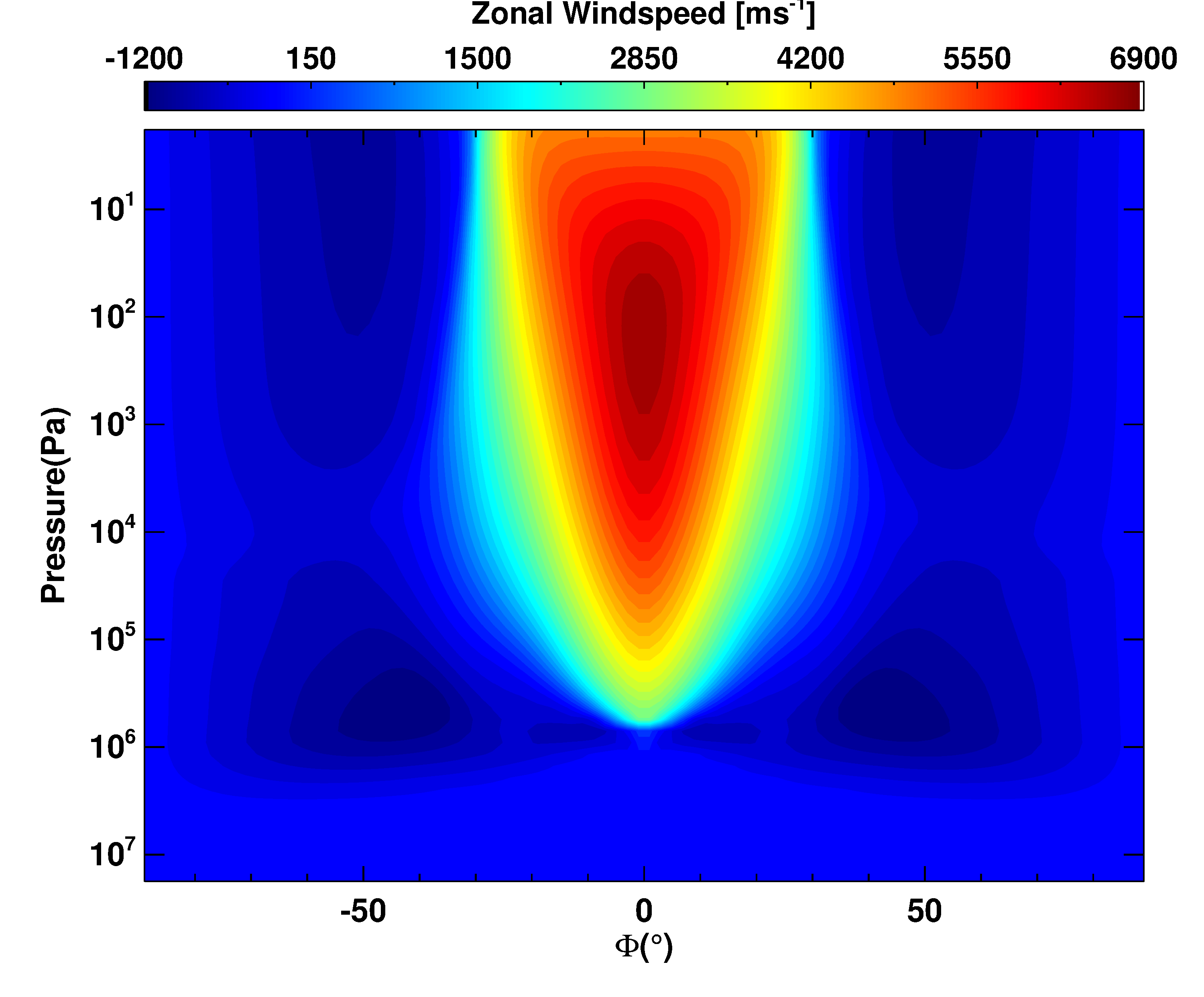}
\end{subfigure}

\caption{Zonally averaged temperature [K] (left) and zonal wind speed [ms$^{-1}$] (right) of both the hot HD 209458 b (upper) and standard HD 209458 b (lower) atmospheres at t$_{\textrm{cloud}}$ = 0 days (see Table \ref{tab:params} and Section \ref{sec:na}).} 

\label{fig:hd209_start}

\end{figure*}


Cloud particles may settle or `precipitate' through the atmosphere with a velocity dependent on the local gas density and cloud particle size. To account for this precipitation, a drift velocity is calculated and used to advect the cloud particles downwards in a column. For the case of free molecular flow (Knudsen number $\ge$ 1) \citep{woitke03}:

\begin{equation}
v_{\textrm{dr}} = - \frac{\sqrt{\pi}}{2} \frac{g\rho_d \langle a \rangle}{\rho_{\textrm{gas}}c_s}.
\end{equation}
Here, $g$ is again the gravitational acceleration, $\rho_d$ is the cloud particle mass density, $c_s$ is the speed of sound and $\langle$a$\rangle$ is the mean particle size,

\begin{equation}
\langle a \rangle = \sqrt[3]{\frac{3}{4\pi}}\frac{L_1}{L_0}.
\end{equation}

The sedimentation routine uses a first order Eulerian advection scheme, working downwards in each column from the model top. To maintain stability, an exponential limiter \citep{rotstayn97} is used to prevent violation of the Courant-Friedrichs-Lewy (CFL) condition \citep{courant28}.

\subsection{Cloud Specifics}

In our model, we utilise the nucleation of TiO$_2$ seed formation. Cloud particle growth from five dust species (TiO$_2$[s], SiO[s], SiO$_2$[s], MgSiO$_3$[s] and Mg$_2$SiO$_4$) is modelled to occur by 22 surface reactions which involve 13 gas phase species forming the solid materials. The elements involved in the cloud formation processes (nucleation, growth and evaporation) are Si, Mg, Ti and O and these elements (with the exception of oxygen) will be depleted or enriched according to Eq. \ref{eq:elemcon} as cloud formation progresses in time and in space. The relevant material data (surface tension, vapour pressures) are the same as in \cite{lee16}. The initial, unaltered element abundances are chosen to be solar and we adopt the values from \cite{asplund09} in order to remain comparable to the work by \cite{lee16}. The element abundances are input material properties for our gas phase chemistry calculation from which we derive the gas phase composition in local thermodynamic equilibrium \citep{helling01}, hence, the number densities of the 13 gas species that are directly involved in the cloud formation processes.

\subsection{Gas Phase Radiative Transfer}

\begin {table*}
\begin{center}
\begin{tabular}{ l|c|c|c }
{\bf{Quantity}} & {\bf{Hot HD 209458 b}} & {\bf{HD 209458 b}} & {\bf{HD 189733 b}} \\
\hline
P-T Profile Name & Hot & Standard & Standard \\
Horizontal resolution (Grid Cells) & $\lambda$ = 144, $\phi$ = 90 & $\lambda$ = 144, $\phi$ = 90 & $\lambda$ = 144, $\phi$ = 90 \\ 
Vertical resolution (levels) & 66 & 66 & 66 \\ 
Hydrodynamical timestep (s) & 30 & 30 & 30 \\ 
Radiative timestep (s) & 150 & 150 & 150 \\ 
Intrinsic temperature (K) & 100 & 100 & 100 \\
Initial inner boundary pressure (Pa) & 2.0 x 10$^7$ & 2.0 x 10$^7$ & 2.0 x 10$^7$ \\
Upper boundary height (m) & 1.0 x 10$^7$ & 9.0 x 10$^6$ & 3.2 x 10$^6$ \\
Diffusion coefficient $K$ & 0.158 & 0.158 & 0.158 \\
Damping coefficient & 0.15 & 0.15 & 0.15 \\
Damping geometry & Linear$^{(1)}$ & Linear & Linear \\
\hline
Transient RT damping coefficient & 0.17 & 0.20 & 0.17 \\
Transient RT damping geometry & Parametric$^{(2)}$ & Parametric & Parametric \\
Transient RT transparent layers & 10 & 10 & 14 \\
Transient Heating Rate Limit (Ks$^{-1}$) & 1.3 x 10$^{-3}$ & 1.3 x 10$^{-3}$ & 1.3 x 10$^{-3}$ \\
\hline
Ideal gas constant, $R$ (Jkg$^{-1}$K$^{-1}$) & 3556.8 & 3556.8 & 4593\\
Specific heat capacity, $c_{\textrm{p}}$ (Jkg$^{-1}$K$^{-1}$) & 1.3 x 10$^4$ & 1.3 x 10$^4$ & 1.3 x 10$^4$  \\
Radius, $R_{\textrm{p}}$ (m) & 9.00 x 10$^7$ & 9.00 x 10$^7$ & 8.05 x 10$^7$ \\
Rotation rate, $\Omega$ (s$^{-1}$) & 2.06 x 10$^{-5}$ & 2.06 x 10$^{-5}$ & 3.28 x 10$^{-5}$ \\
Surface gravity, $g_{\textrm{p}}$ (ms$^{-2}$) & 10.79 & 10.79 & 22.49 \\
Semi-major axis, $a_{\textrm{p}}$ (au) & 4.75 x 10$^{-2}$ & 4.75 x 10$^{-2}$ & 3.14 x 10$^{-2}$ \\
\hline
{\bf{Run Lengths (Earth days)}} &  &  &  \\
\hline
Cloud Free & 800 & 1200 & 800 \\
Transparent Cloud & 50 & 50 & 50 \\ 
Transient Radiative Cloud & 14 & 14 & 20 \\ 
Full Radiative Cloud & 36 & 36 & 30 \\
Total Cloud & 100 & 100 & 100 \\ 
Total, t$_{\textrm{end}}$ & 900 & 1300 & 900 \\
\hline
\end{tabular}
\vspace{+10pt}
\caption{Model parameters of our three simulated planetary atmospheres, covering grid setup, run-lengths and planet constants. See (1) \cite{mayne14} and (2) Drummond et al., (submitted) for more information.}
\label{tab:params}
\end{center}
\end{table*}

A chemical equilibrium model solves for the abundances of key species involved in cloud formation. For computational efficiency however, we use the fast analytical formulation of \cite{burrows99} \citep[used in][]{amundsen16} that calculates abundances of key absorbing species involved in our radiative hydrodynamics. This parameterised gas phase chemistry assumes solar metallicity. Analytical formulae for exoplanet chemistry has also been published by \cite{heng16}. In addition, the abundances of the alkali species Na, K, Li, Cs and Rb are approximated using the same parameterisations described in \cite{amundsen16}. Since the chemistry scheme used in the radiative transfer stage does not use the elemental abundances depleted by cloud formation, the abundances of key absorbers may well be overestimated.

We use the Suite of Community Radiative Transfer codes based on Edwards and Slingo (SOCRATES) code which is coupled to the UM \citep{edwards96}. SOCRATES implements the plane-parallel two-stream approximation and uses the sophisticated correlated-$k$ method to avoid computationally demanding line-by-line calculations \citep{lacis91,amundsen14,amundsen16}. By solving the radiative transfer equations, we improve upon the Newtonian Cooling approximation used in many previous models \citep{cooper05,menou09,heng11,mayne14}. 

Rayleigh scattering is included for H$_2$ and He with their numerical factors described in \cite{amundsen16}. Opacities are determined, where possible, from the latest ExoMol line lists \citep{tennyson12,tennyson16} and the model includes line absorption from H$_2$O, CO, CH$_4$, NH$_3$, Li, Na, K, Rb, Cs and H$_2$-H$_2$ and H$_2$-He collision induced absorption (CIA). Further details regarding the opacity implementation can be found in \cite{amundsen14,amundsen16}. Additionally, the treatment of overlapping absorption in the gas phase is described in \cite{amundsen17}.

A ghost layer above the numerically modelled domain is used to account for the radiative processes taking place in the atmosphere extending from the lowest pressures of our simulated domain to P $\approx$ 0 Pa. Thereby the flux at the top boundary is not the exact stellar flux. For more details on this layer, see \cite{amundsen15}.

\subsection{Cloud Radiative Transfer}
\label{sec:crt}

Clouds have been shown to have a dramatic effect on atmospheric opacities; their presence can alter the local thermal conditions by way of back-warming \citep{lee16}. Mie Theory \citep{mie08} and Effective Medium Theory (EMT) are used to obtain the cloud absorption and scattering coefficients, over the 32 bands described in \cite{amundsen14}, which are appended to the coefficients for the gas phase and used in the calculation for the full radiative transfer performed by SOCRATES. All cloud properties relevant for the opacity calculations (cloud particle size, cloud particle number density, material composition) are assumed to be homogeneous across a computational grid cell, hence, they are represented by one value. This represents the effect of the grid resolution on our radiative transfer calculation.

The simultaneous condensation of multiple minerals onto the nucleated seed particles leads to a bulk composition that is the sum of the different volume fractions from the individually condensed materials. Since a single particle is composed of multiple materials it contributes mixed optical properties. A quantification of the bulk optical property of the mixed particle is needed, and EMT is used to account for this. We take our material real and imaginary refractive indices and extinction coefficients from \cite{lee15a} and \cite{lee16}, using the same extrapolation methods to cover the necessary wavelength range that our radiative transfer scheme extends over. Following the Bruggeman method \citep{bruggeman35}, we solve for the average dielectric function of the particle, $\epsilon_{\textrm{av}}$, using a Newton-Raphson minimisation scheme. For cases of non-convergence, the solution is obtained via the analytic Landau-Lifshitz-Looyenga method \citep{looyenga65}. This methodology is applied inboth \cite{lee16} and \cite{juncher17}.

The dielectric function is a necessary input to the Mie Theory solution. \cite{bohren83} {\emph{BHMie}} routines are used to solve for the absorption, Q$_{\textrm{abs}}$($\lambda$,$\langle$a$\rangle$), and scattering, Q$_{\textrm{sca}}$($\lambda$,$\langle$a$\rangle$), efficiencies. Here, $\lambda$ is our considered wavelength. Using the expression of size parameter,

\begin{equation}
x = \frac{2\pi \langle a \rangle}{\lambda},
\end{equation}
for particles of similar size to the photon wavelength (10$^{-6}$ $\le$ x $\le$ 10$^3$) where the interaction between the light and particle is non-trivial, the full Mie theory calculation is required. In \cite{lee16}, analytical expressions are used for the limiting cases of Mie theory; the large particle, hard sphere scattering case (where x $\ge$ 1000) and the small metallic sphere case (where x $\le$ 10$^{-6}$). Improved stability in our Mie code means we can instead approximate these regimes using the full Mie code solution but fixing size parameters that exceed these limiting values (for x $\ge$ 1000, x = 1000 and x $\le$ 10$^{-6}$, x = 10$^{-6}$.)

The resulting absorption and scattering coefficients are scaled by the cloud cross section which is defined as,

\begin{equation}
\frac{n_{\textrm{d}}}{\rho_{\textrm{gas}}}\pi \langle a \rangle^2.
\end{equation}
These cloud scattering and absorption coefficients are then appended to the same coefficients from the gas phase for a total (cloud and gas) opacity.

Using SOCRATES, radiative fluxes and heating rates are calculated using a two-stream solver with a full treatment of scattering for both the stellar and thermal fluxes using the Practical Improved Flux Method \citep{zdunkowski80}. Delta-rescaling is performed to reduce errors due to strong forward scattering by cloud particles. This configuration of SOCRATES is also used by the Met Office climate model \citep{walters17} and has undergone extensive validation for clouds on Earth.

\begin{figure} 

\begin{subfigure}{\linewidth}
\includegraphics[scale = 0.085, angle = 0]{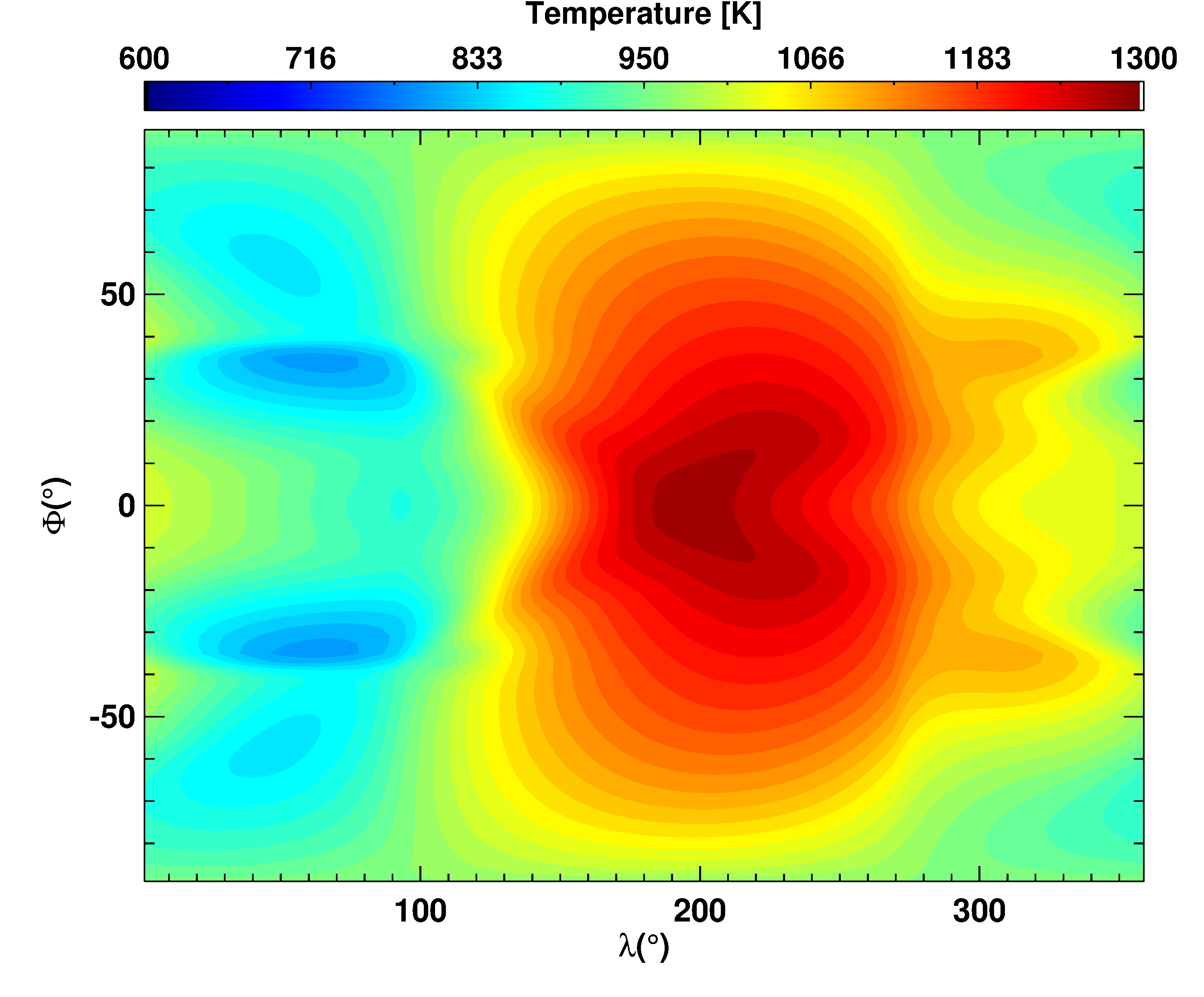}
\end{subfigure}

\medskip

\begin{subfigure}{\linewidth}
\includegraphics[scale = 0.085, angle = 0]{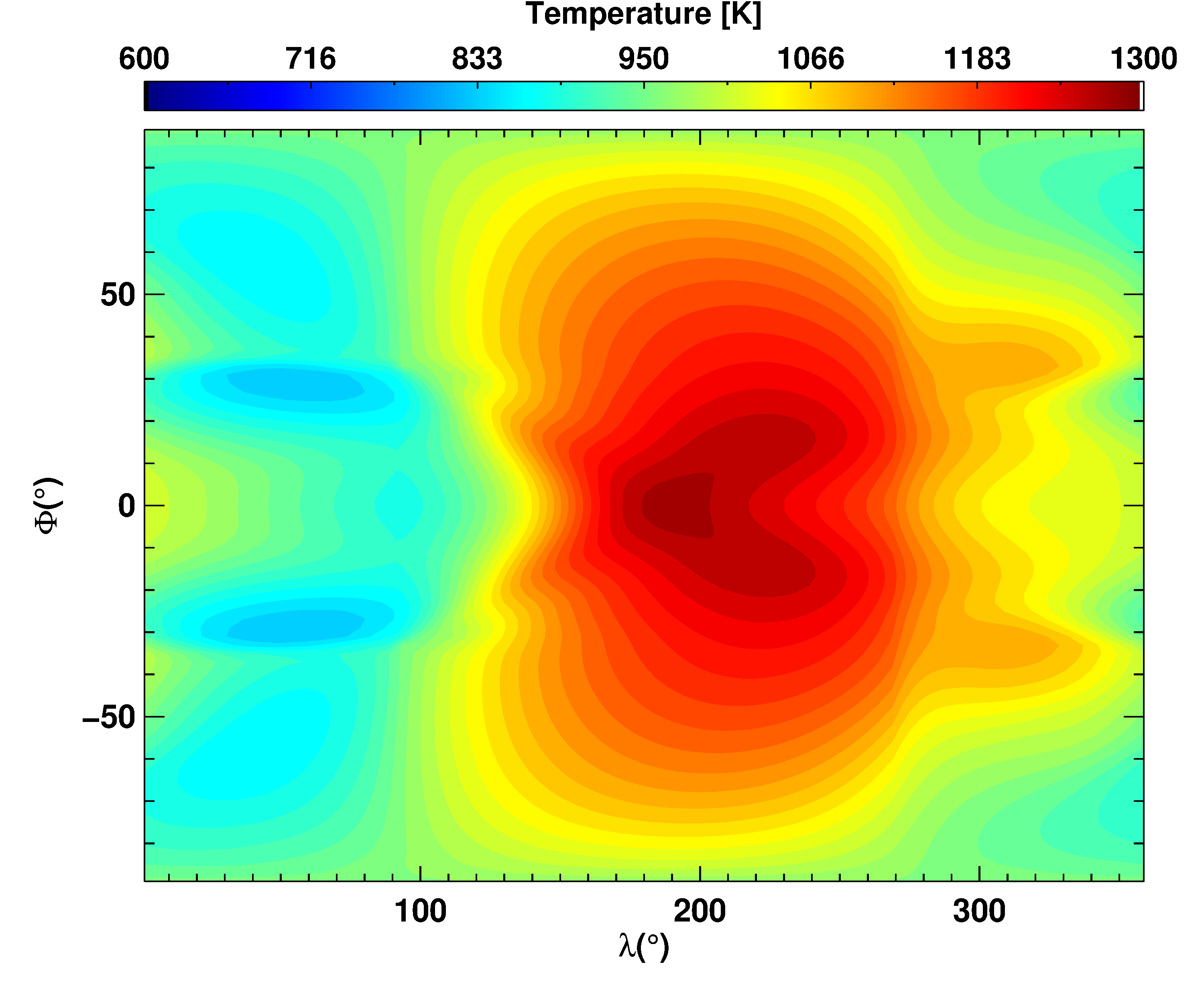}
\end{subfigure}

\medskip

\centering
\begin{subfigure}{\linewidth}
\includegraphics[scale = 0.085, angle = 0]{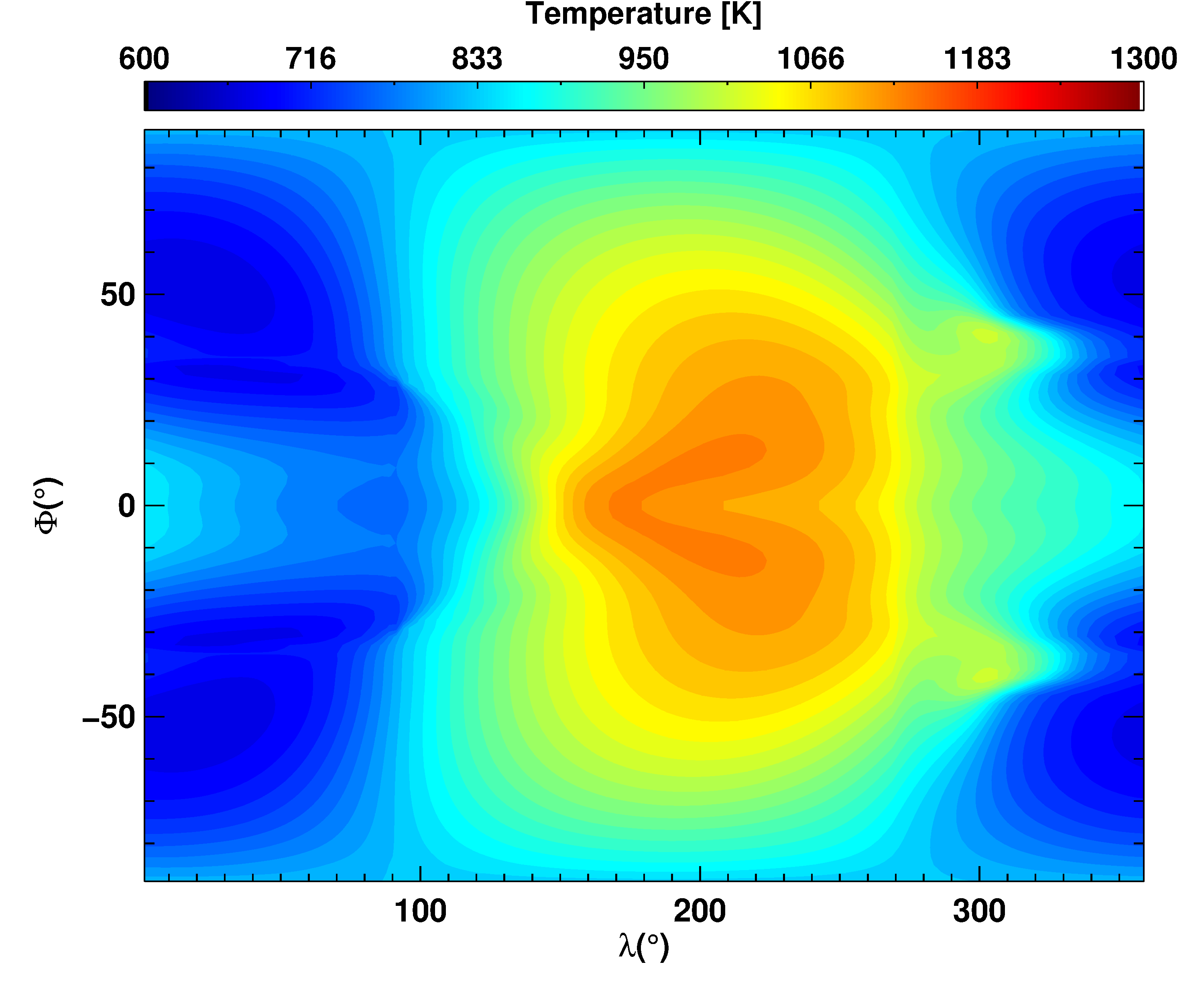}
\end{subfigure}

\caption{Horizontal temperature [K] distribution of both the hot HD 209458 b (upper), standard HD 209458 b (middle), and HD 189733 b (lower) atmospheres at t$_{\textrm{cloud}}$ = 0 days (see Table \ref{tab:params} and Section \ref{sec:na}). Plots at constant pressures of 10$^2$ Pa and 2 x 10$^2$ Pa for HD 209458 b and HD 189733 b respectively.} \label{fig:hoz_t_start}

\end{figure}


\subsection{Numerical Approach}
\label{sec:na}

We perform three simulations that encompass two well-known transiting hot-Jupiters, HD 209458 b and HD 189733 b. Each of our simulations is composed of three numerical phases or `stages', primarily to isolate the cloud formation dependence on the atmosphere's thermal profile, and the radiative effect of cloud back onto the atmosphere. These stages are described in detail in subsections \ref{sec:cf}, \ref{sec:transparent} and \ref{sec:rt}. The first considers a clear-skies, cloudless atmosphere which we term the `cloud free' stage. A further period of time follows in which clouds can form and evolve in response to the atmospheric thermal, chemical and dynamical conditions, but do not radiatively interact with (feedback onto) the atmosphere; this stage we call `transparent cloud'. The final and third stage allows the clouds to both scatter and absorb radiation and hence modify the atmospheric conditions. This stage we term `radiatively active cloud'. To evolve through transient instabilities arising from large heating rates\footnote{The heating rate can be negative in the event of cooling.}, caused by steep gradients in the cloud opacity, we introduce a sub-period termed `transient radiative cloud', which includes a number of numerical aids to maintain model stability during this period. The total combined simulation length extends from time t = 0 to t = t$_{\textrm{end}}$ days (note that throughout this paper, `day' refers to an Earth day unless stated otherwise), with t$_{\textrm{end}}$, alongside full model parameters, defined in Table \ref{tab:params}. Since the length of the cloud free stage differs between our three atmosphere models, for ease of comparison we define a new time, t$_{\textrm{cloud}}$, which defines the length of time since clouds are included.

\subsubsection{Initial Conditions}

Our models start from rest with an initial, spherically symmetric, temperature-pressure (TP) profile. For HD 209458 b, these profiles represent both non-adjusted and hotter interior atmospheres which we subsequently refer to as the `standard' and `hot' cases respectively \cite[see][for details]{amundsen16}. The TP profile of HD 189733 b is calculated using a similar method to that of \cite{amundsen16}, but only a standard profile is considered. For P $\leq$ 10$^5$ Pa, and by t = 400 days (under cloud-free conditions), non-evolving wind speed maxima and well resolved flows indicate the atmosphere has reached a quasi-steady-state. For higher pressures the thermal timescale increases significantly \cite{mayne14}. The physical motivation for a hotter `deep' interior is raised by the works of \cite{cooper05,mayne14,amundsen16} and \cite{mayne17} which have shown the deep atmosphere continues to evolve beyond our longest current simulation time of 1200 days, indicating that the thermal structure of a non-adjusted standard atmosphere is likely to be too cool. Further studies have shown that dynamical processes can be responsible for the slow heating of the radiatively inactive regions of the atmosphere, including the downwards flux of kinetic energy \citep{showman02} and transportation of potential temperature \citep{tremblin17}. Simulations in 1D are unable to capture these physical processes due to the omission of horizontal advection. In 3D, computational feasibility limits the ability to capture such processes acting over long timescales. Therefore, we use a TP profile with + 800 K to mimic the increased deep atmosphere temperature effect of \cite{tremblin17}; this is the method used in \cite{amundsen16}.

In all models, the zeroth vertical level (z = 0) is set to an initial pressure of P = 2 x 10$^7$ Pa. Since the surface gravity of HD 189733 b is larger than that of HD 209458 b, the atmosphere is more compressed and hence the vertical geometric extent of the atmosphere is reduced from $\sim$ 10,000 km (HD 209458 b) to 3,200 km. Correspondingly, the outer boundary pressures vary between our simulations; in HD 189733 b, we simulate up to P $\sim$ 10 Pa, whereas much lower pressures of P $\sim$ 10$^{-1}$ Pa are probed for HD 209458 b. 

The hydrodynamical time step is set to t = 30 s, shorter than that of \cite{mayne14}, but consistent with \cite{amundsen16}. In our case, a shorter time step is required to account for cloud thermal adjustment from advection in high velocity (e.g. jet) regions. Taking a maximum wind speed of 5400 ms$^{-1}$ which is obtained numerically for HD 209458 b in \cite{mayne14} and observationally for HD 189733 b in \cite{louden15}, the dynamical timescale of the jet is $\sim$ 1 day. Hence, the dynamics of the atmosphere are temporally highly-resolved with almost 10$^5$ steps per the advective timescale in the most dynamically active regions. Our radiative time step is consistent with the optimised value of t = 150 s in \cite{amundsen16}, such that radiative transfer is solved every five hydrodynamical steps. 

Vertical velocities are damped using the standard `sponge' technique \citep[see][]{mayne14,amundsen16}. The damping magnitude is set via the damping coefficient, initially set to 0.15. A linear profile is used whereby the damping coefficient is constant, for a given height, with latitude and is implemented from 0.75 x z$_{\textrm{top}}$ (where z$_{\textrm{top}}$ is the upper boundary height) increasing steeply to z$_{\textrm{top}}$ . We term this damping geometry, defined for each simulation in Table \ref{tab:params}, `Linear' \citep[see][for more information on damping.]{mayne13,mayne14}

\subsubsection{Cloud Free}
\label{sec:cf}
Our cloud free simulations represent the prior state-of-the-art, atmospheric circulation modelling, and include full radiative transfer for gaseous absorbers and scatterers. In this cloud free stage we evolve our initial profiles from rest ($\vec{\bf{v}}$ = 0), without clouds ($J_{*}$ = 0 and $\chi^{\textrm{net}}$ = 0), until the atmosphere has reached a quasi-steady-state, defined by the formation of planetary scale flows (super-rotating equatorial jet) with no significant changes in the maximum zonal wind speed over time. While such conditions are typically met by 400 days, we choose to evolve the cloud-free stage for longer (see Table \ref{tab:params} for details). 

\subsubsection{Transparent Cloud}
\label{sec:transparent}

Subsequent to the cloud free stage, each model runs for a further period of time totalling 50 days (t$_{\textrm{cloud}}$ = 0 - 50 days) in which clouds can form and evolve in response to the local atmospheric properties, but are transparent to radiation; no scattering or absorption from cloud particles is considered. This method allows us to determine the dynamical evolution of the cloud in response to the cloud free temperature profile and to more clearly identify the effects of cloud absorption and scattering on the final atmospheric profile.


\begin{figure*}[]

\begin{subfigure}{0.48\textwidth}
\includegraphics[scale = 0.085, angle = 0]{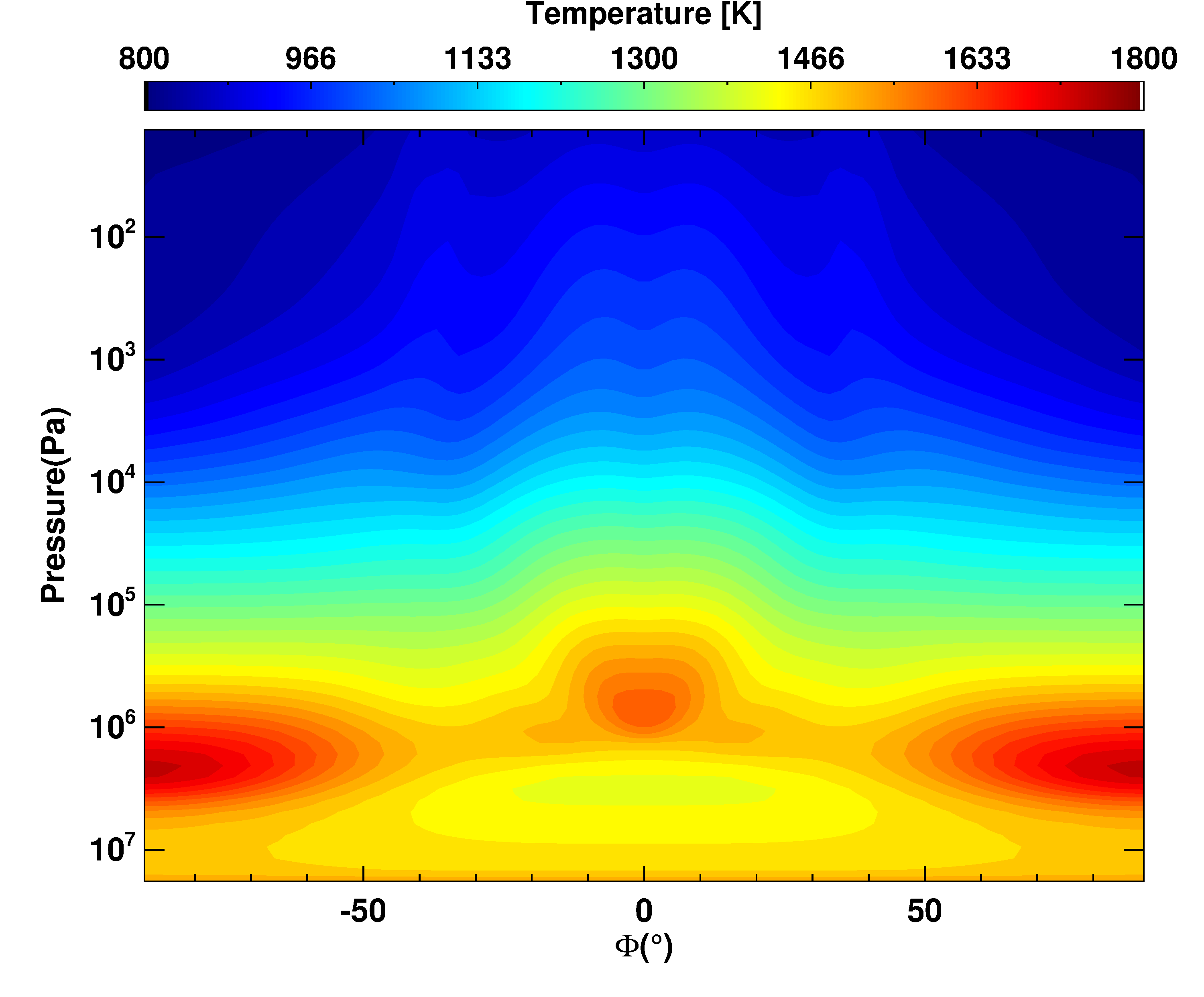}
\end{subfigure}\hspace*{\fill}
\begin{subfigure}{0.48\textwidth}
\includegraphics[scale = 0.085, angle = 0]{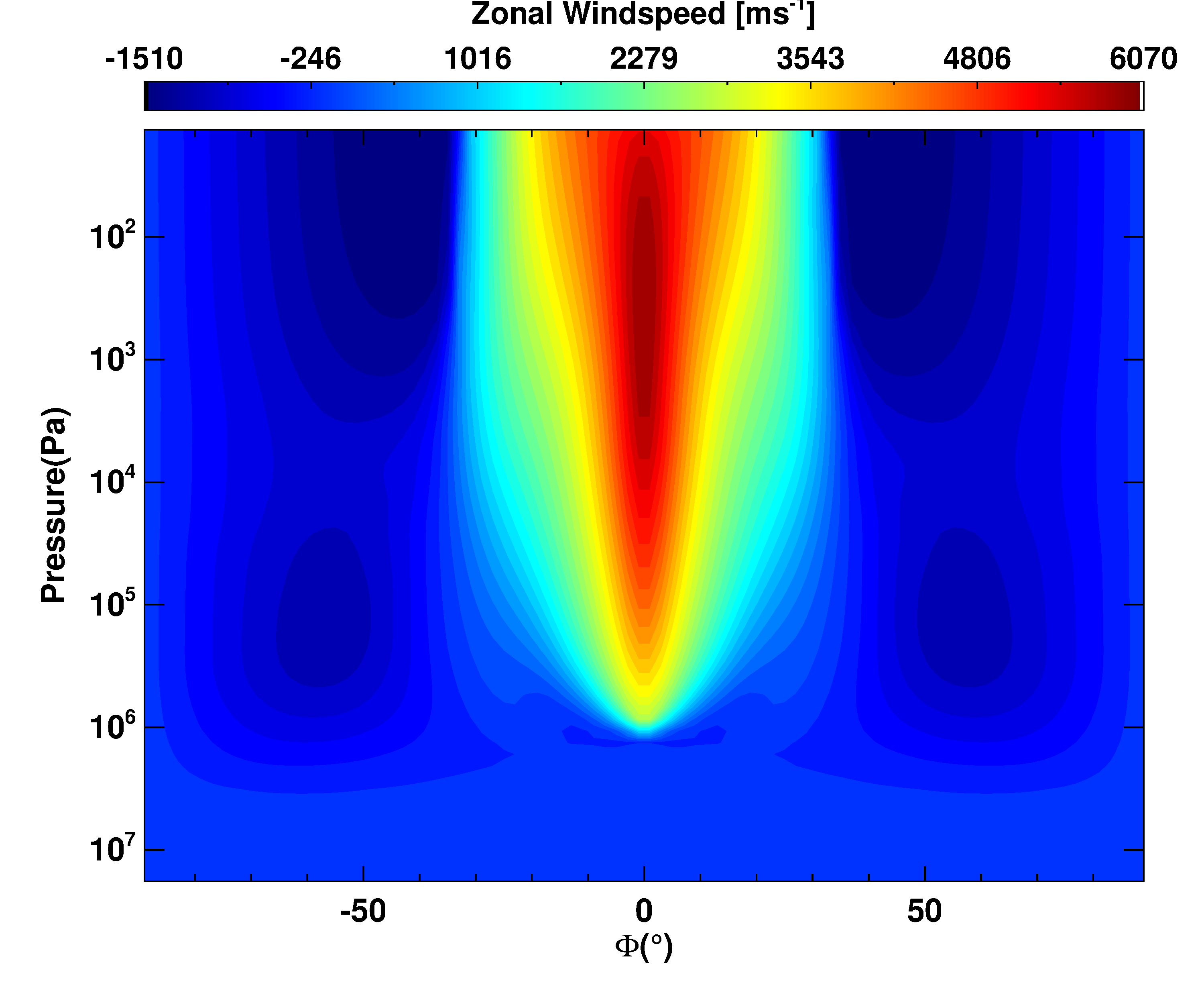}
\end{subfigure}

\caption{Zonally averaged atmospheric profiles of HD 189733 b at t$_{\textrm{cloud}}$ = 0 days (see Table \ref{tab:params} and Section \ref{sec:na}).} 
\label{fig:c189_start}

\end{figure*}


One of the difficulties of this self-consistent coupled-cloud model, identified in \cite{lee16} is the integration time of the cloud chemistry step. For the first 100 hydrodynamical time steps, we solve for the cloud chemistry at each step. After this period, we then solve the cloud chemistry equations every 10 hydrodynamical time steps. This may slightly increase the number of cloud particles that are transported across the night-day terminator, in response to the increased day-side temperature (i.e. cloud that may have evaporated on the next step due to the temperature contrast will remain in the condensed phase for up to nine further hydrodynamical time steps until it is able to respond to the new thermal conditions). This decreased computation frequency means the whole cloud formation complex, including the drift velocity and cloud opacity (both of which are updated after the cloud formation chemistry integration), are also updated less frequently. While this advection of cloud onto the day-side may lead us to slightly overestimate the day-side cloud coverage, we expect this to only apply to the most dynamically active regions (super-rotating equatorial jet) where wind velocities can carry the cloud particles further over the response window of 5 minutes (10 x 30s time step). This interval updating is required to obtain the necessary accuracy to resolve the cloud dynamics while maintaining computational feasibility.

Where evaporation occurs quickly (i.e. in response to a sudden thermal change), the solution to the evaporation velocity can become numerically stiff and hence we employ the same technique used in \cite{lee16} to avoid this issue. The method involves removing a large amount of the evaporating, volatile material at the start of the step, if the time step required is small. The return of elements to the gas phase, from each contribution dust species $s$ is \citep{woitke06}:

\begin{equation}
\epsilon_i = \epsilon_i^b + \frac{\nu_{i,s} 1.427 m_u}{V_{0,s}}\Delta L_{3,s}.
\end{equation}
where $\epsilon_i^b$ is the element abundance prior to the instantaneous evaporation step, $\nu_{i,s}$ is the stoichiometric coefficient of element $i$ in species $s$ and 1.427$m_u$ (where $m_u$ is the atomic mass unit) is the factor for converting between gaseous mass density $\rho_{\textrm{gas}}$ and Hydrogen nuclei density $n_{\langle H \rangle}$.

In cells which report a negative growth velocity ($\chi$ $<$ 0 [cm s$^{-1}$]), but for which the solution is not numerically stiff, standard evaporation, described in Section \ref{sec:cm}, applies. Material is removed from the particle down to the evaporation limit which is set to the seed particle size. When the evaporation limit is reached, the seed particle still remains.  At this point a separate test of the thermal stability of the seed particle ($\chi$ of TiO$_2$) is used. If $\chi_{\textrm{TiO}_2}$ $<$ 0 [cm s$^{-1}$] then seed particles are removed from the reservoir. 

Cloud chemistry is not integrated for cells which contain little to no cloud providing the nucleation rates are small (n$_{\textrm{d}}$ $<$ 10$^{-10}$  and $J_{\textrm{*}}$ $<$ 10$^{-10}$ [cm$^{-3}$s$^{-1}$]). This is computationally efficient and removes problems where if $\chi^{\textrm{net}}$ is large but $J_{\textrm{*}}$ is small, very large particles occur which have drift velocities that become difficult to handle due to CFL condition concerns. Similarly, if $\chi^{\textrm{net}}$ $<$ 10$^{-20}$ [cm s$^{-1}$] then there is no particle growth; only significant departures from equilibrium are solved in time. This value has been tested and optimised in \cite{lee16}.


\begin{figure*}[]
\includegraphics[scale=0.8,angle=0]{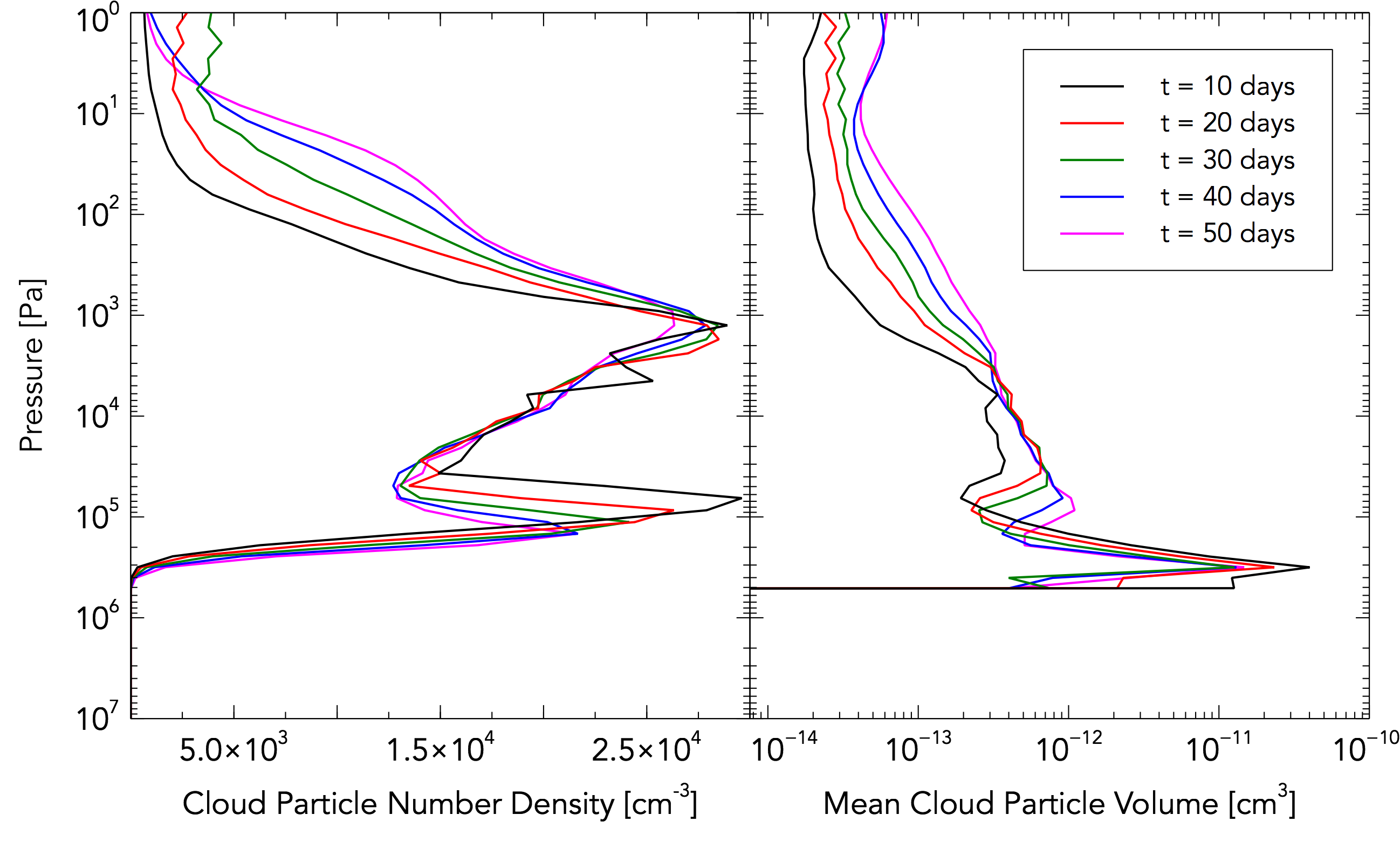}
\caption{Time evolution of the horizontally averaged cloud particle number density, n$_{\textrm{d}}$ (left) and mean cloud particle volume, $\langle V \rangle$ (right) during the transparent cloud stage of the hot HD 209458 b model at between t$_{\textrm{cloud}}$ = 0 and t$_{\textrm{cloud}}$ = 50 days (see Table \ref{tab:params} and Section \ref{sec:na}).}

\label{fig:h209evo}
\end{figure*}


\subsubsection{Radiatively Active Cloud}
\label{sec:rt}
In the radiatively active cloud stage, which immediately follows the transparent cloud stage, we take into account the influence of radiative energy transfer by absorption and scattering from the cloud. Hence, cloud particles can effect the local temperature (hence pressure, density and velocity) only during this stage. For each of our three gas giant atmosphere models, we simulate this stage for a further period of time totalling 50 days (t$_{\textrm{cloud}}$ = 50 - 100 days) which is split into two stages: transient radiative cloud and full radiative cloud.

{\emph{Transient Radiative Cloud:}} during the transient radiative cloud stage, cloud particles are now able to scatter and absorb radiation and thus change the thermal properties of the atmosphere by way of heating rates calculated by the radiative transfer scheme. Due to the large cloud opacity gradients, particularly at the cloud `top', the heating rates can in turn become large. Given our time step, limited by computational feasibility and selected for accuracy and stability, in some circumstances (e.g. regions of low pressure), heating rates in cloud regions can modify the thermal structure of the atmosphere so quickly that transient instabilities appear. 

To deal with these instabilities, the following conditions are necessarily applied to HD 209458 b and HD 189733 b for 14 days and 20 days respectively (the time required before the model can be run without the following measures). We introduce a heating rate cap or limit ($Q_{\textrm{max}}$ = 1.3 x 10$^{-3}$ Ks$^{-1}$) which slows the thermal evolution of the atmosphere. An additional stability enhancing measure includes prescribing a number of upper atmosphere layers in which clouds that extend into these regions are transparent to radiation. These transparent layers are set to have a zero cloud opacity (no contribution from scattering and absorption). This stops the low pressure upper atmosphere from evolving too quickly. To assist with stability we also increase the damping coefficient. Additionally, the sponge layer geometry is switched from a `Linear' profile, to one where the damping effect at polar regions is enhanced; we term this damping geometry `Parametric' (see Drummond et al., submitted, and Table \ref{tab:params} for more details). {\emph{These conditions, listed in Table \ref{tab:params}, are removed at the end of this transient radiative cloud stage to allow for a further period of time with fully active clouds where no restrictions apply; the heating rate limit and transparent layers are switched off, and the damping magnitude and geometry and returned to their original values.}}

Conditions are also applied to the cloud model during the radiatively active cloud stage to aid numerical stability and enhance computational efficiency. To avoid expensive Mie calculations for cloud depleted cells we enforce a limit of n$_{\textrm{d}}$ $<$ 10$^{-2}$ cm$^{-3}$ for the opacity calculations. For values less than this the cloud opacity is set to zero. Should the cloud opacity be calculated, a minimum value of $\kappa_{\textrm{cloud}}$ = 10$^{-7}$ cm$^2$g$^{-1}$ applies. Drift velocities are calculated only for regions with $n_d$ $>$ 10$^{-10}$ cm$^{-3}$. Rarely, the cloud chemistry integration can partially fail causing the generation of erroneous cells that contain tiny but non-zero values for situations where no cloud should form. Floating point precision in these almost-cloud-free cells can therefore produce fictitious particles with extremely large sizes. For both these reason, we limit the radiation transfer coupling to cells containing particles with $\langle$a$\rangle$ $<$ 50 micron (since this value vastly exceeds the sizes of particles expected to form) otherwise the opacity is set to zero. {\emph{These conditions are not removed at the end of the transient radiative cloud stage and continue for the rest of the simulation.}}

{\emph{Full Radiative Cloud:}} clouds are fully coupled to the radiation scheme during this stage, which continues for a further period of time after the transient radiative stage, until t$_{\textrm{end}}$. For all simulations, this stage ends at t$_{\textrm{cloud}}$ = 100 days, which is the summation of all the cloud-enabled stages; transparent, transient radiative and full radiative. It is worth noting that while the UM does not inherently conserve axial angular momentum (AAM), we track its value which is conserved to better than 0.01$\%$ in all simulations, by the end of the full radiative cloud stage.

\subsubsection{Extended Period}

At t$_{\textrm{cloud}}$ = 100 days, for the hot HD 209458 b model only, we continue the run for a further period of time (six planetary periods) but use higher resolution (500 band) spectral files for a second diagnostic call to the radiation scheme. This allows us to calculate the emission spectra and phase curves to a much higher accuracy than the standard 32 bands used for the heating rates \cite[see][]{boutle17}. This process is additionally carried out for a cloud free atmosphere so that comparisons can be made between the fluxes from the clear and cloudy atmospheres. 


\begin{figure*}[]

\begin{subfigure}{0.48\textwidth}
\includegraphics[scale = 0.085, angle = 0]{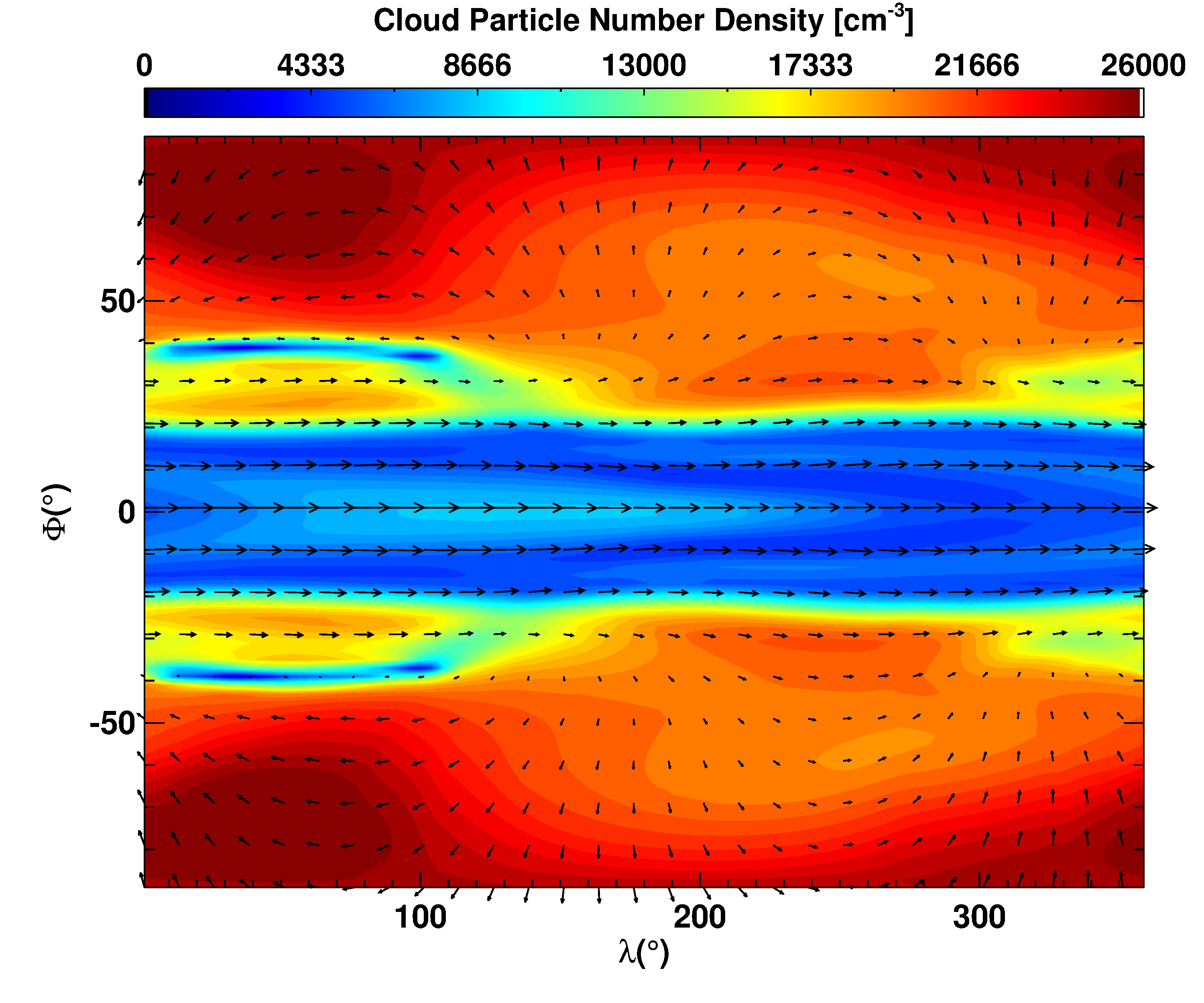}
\end{subfigure}\hspace*{\fill}
\begin{subfigure}{0.48\textwidth}
\includegraphics[scale = 0.085, angle = 0]{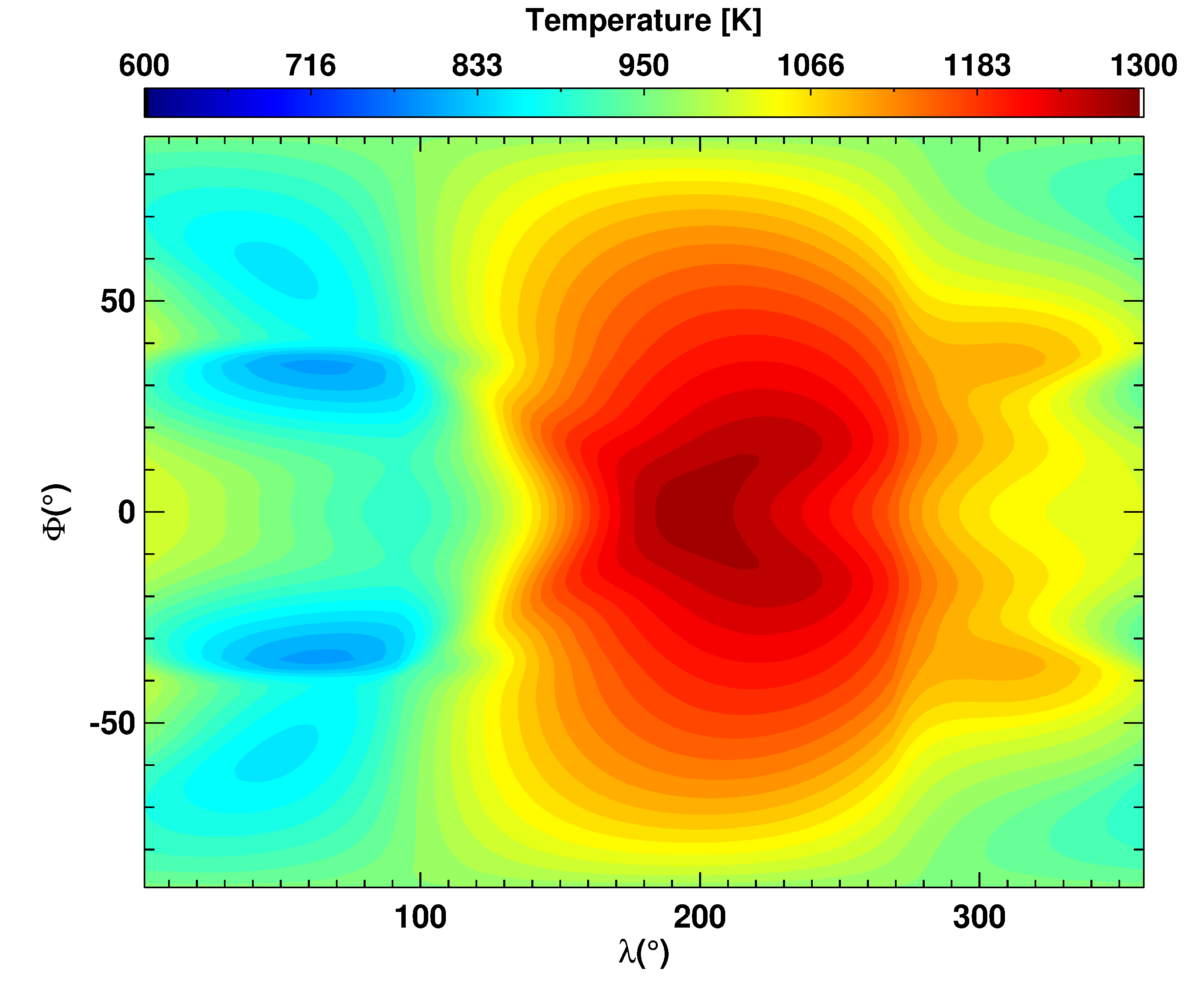}
\end{subfigure}

\medskip
\begin{subfigure}{0.48\textwidth}
\includegraphics[scale = 0.085, angle = 0]{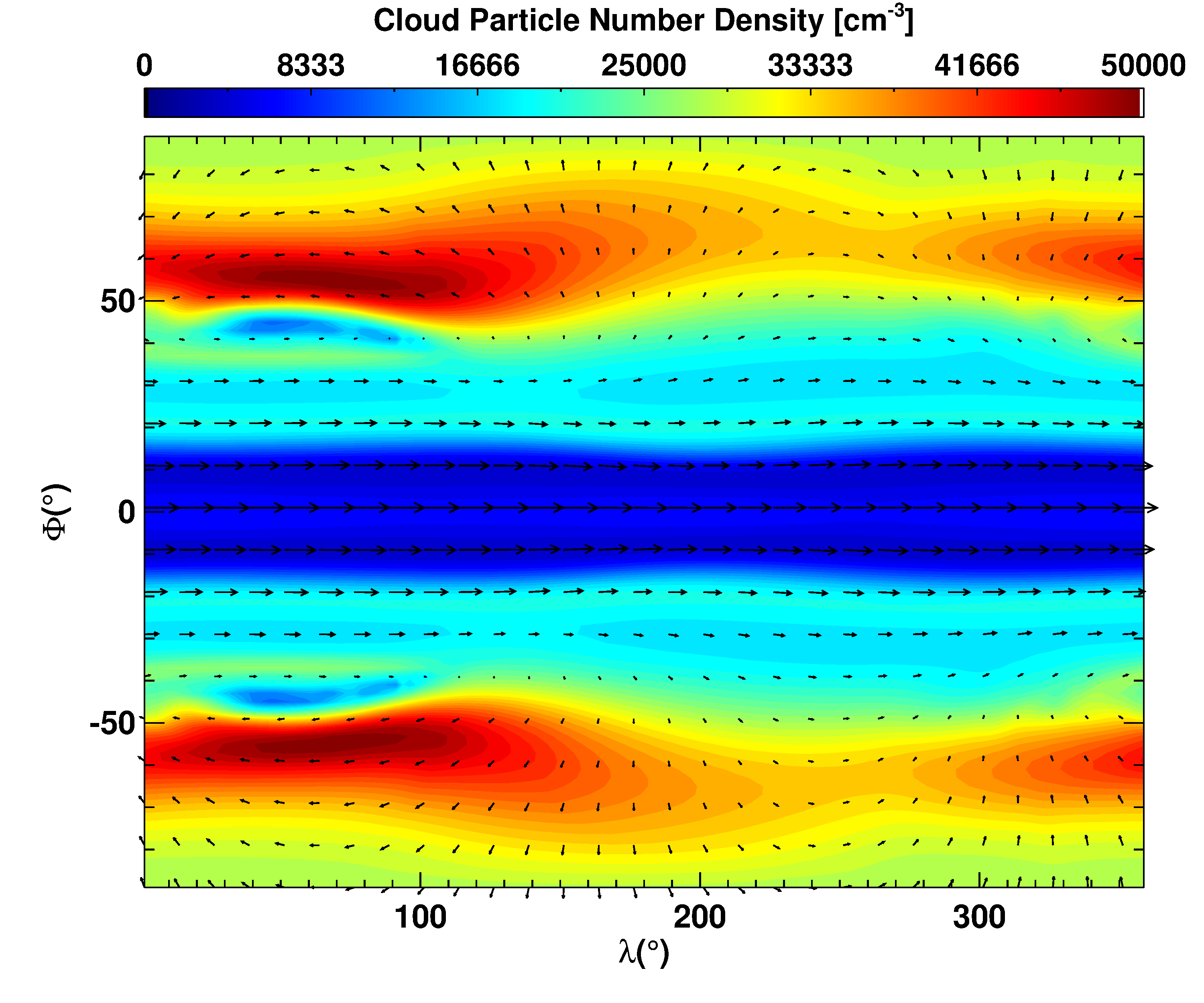}
\end{subfigure}\hspace*{\fill}
\begin{subfigure}{0.48\textwidth}
\includegraphics[scale = 0.085, angle = 0]{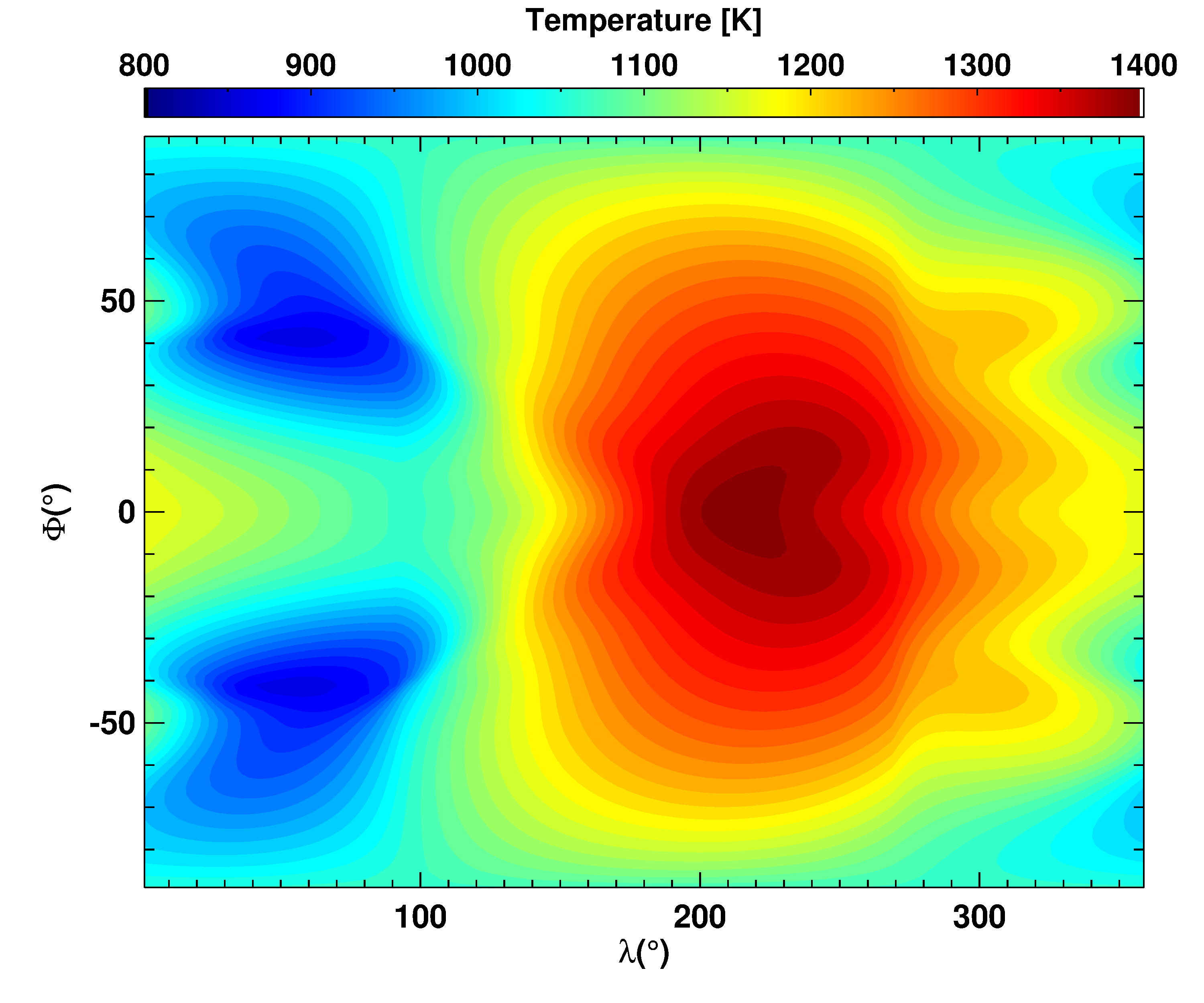}
\end{subfigure}

\medskip
\begin{subfigure}{0.48\textwidth}
\includegraphics[scale = 0.085, angle = 0]{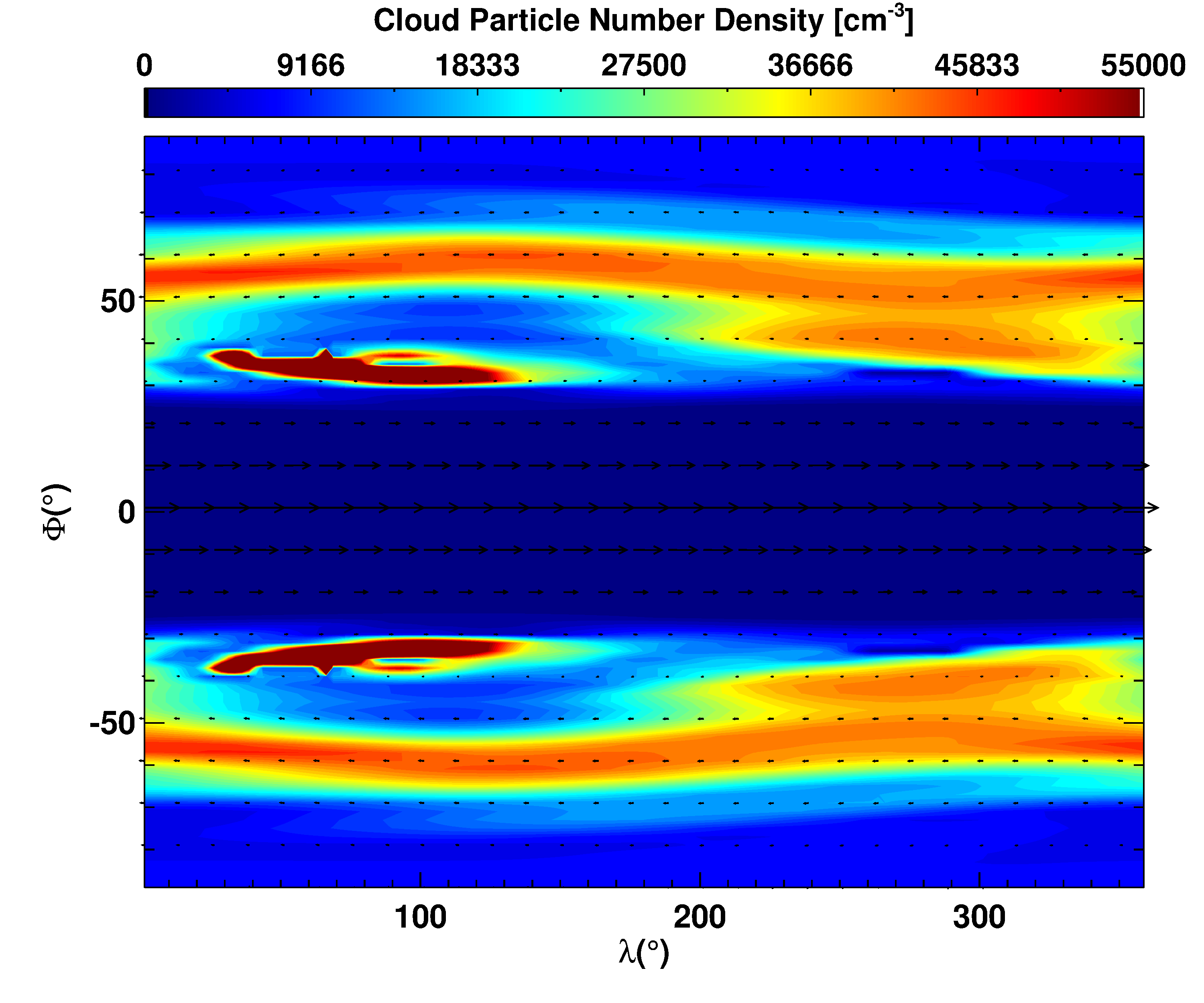}
\end{subfigure}\hspace*{\fill}
\begin{subfigure}{0.48\textwidth}
\includegraphics[scale = 0.085, angle = 0]{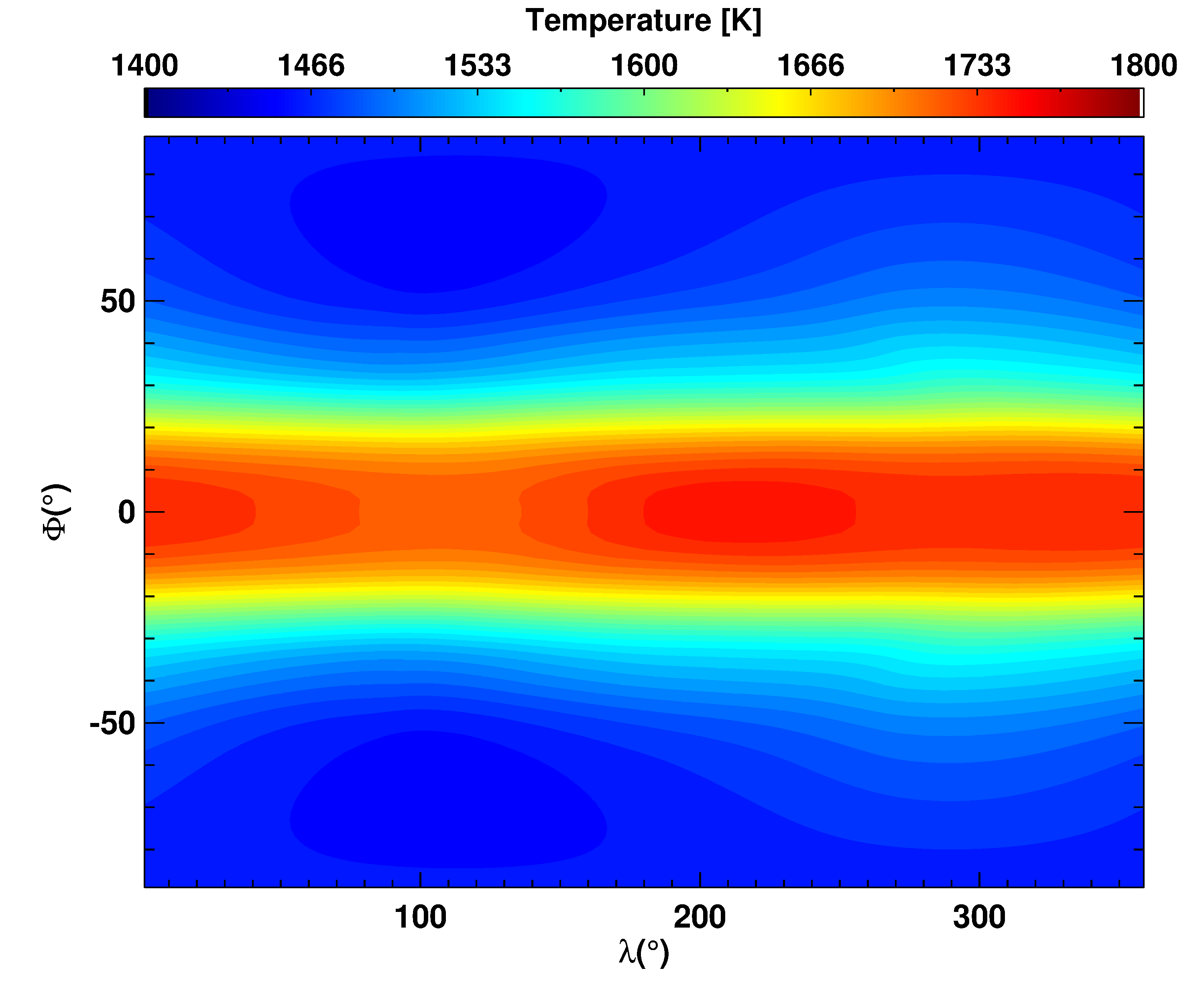}
\end{subfigure}

\caption{Hot HD 209458 b slices at P = 10$^2$ Pa (upper), P = 10$^3$ Pa (middle) and P = 10$^5$ Pa (lower) for the cloud particle number density, $n_d$ [cm$^{-3}$] and horizontal wind velocity [ms$^{-1}$], (left) and temperature [K] (right). Sampled at t$_{\textrm{cloud}}$ = 50 days (see Table \ref{tab:params} and Section \ref{sec:na}), prior to cloud radiation coupling.}

\label{fig:passive_h209}

\end{figure*}



\begin{figure}[]

\begin{subfigure}{\linewidth}
\includegraphics[scale=0.085,angle=0]{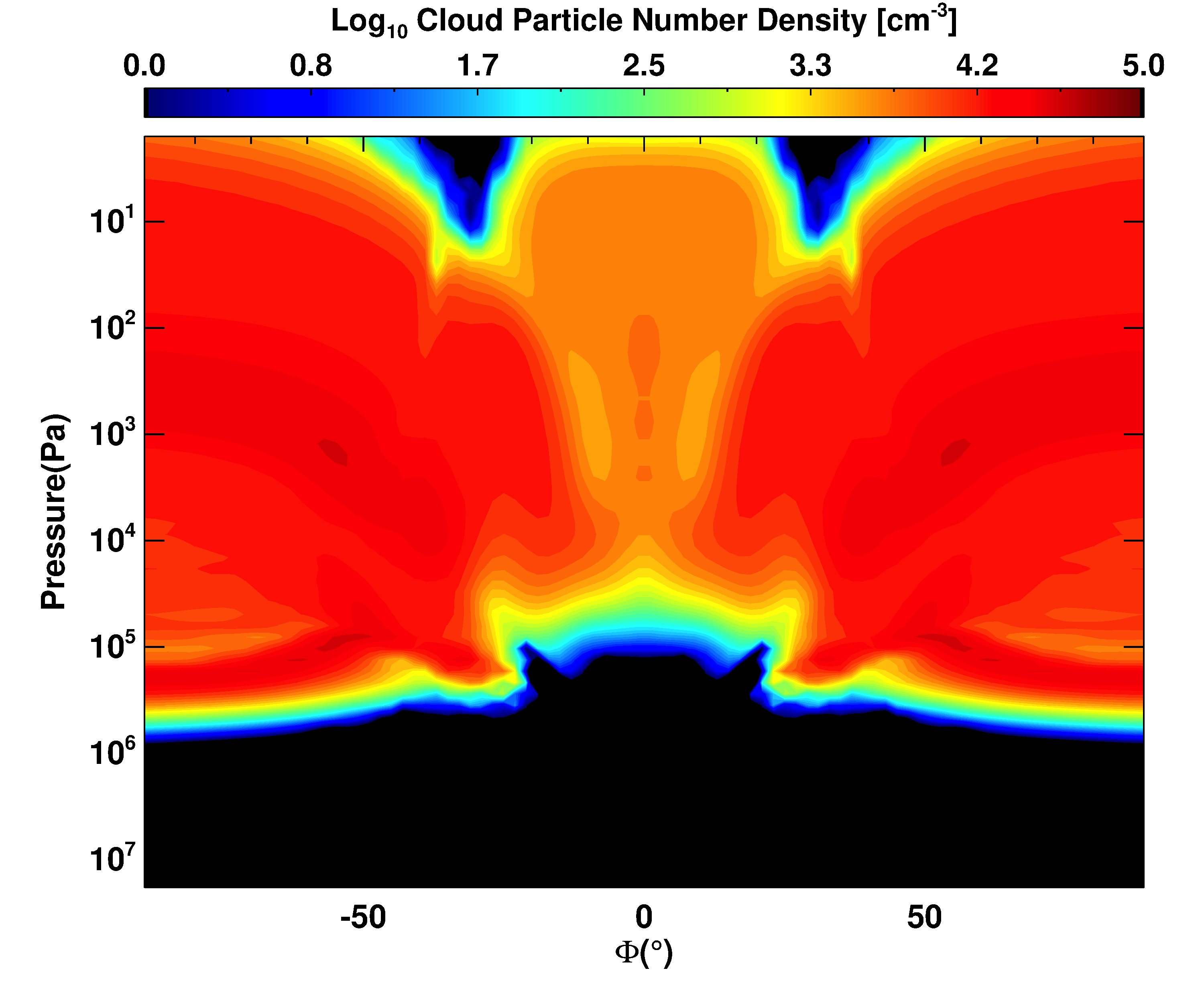}
\end{subfigure}

\vspace{-5pt}

\begin{subfigure}{\linewidth}
\includegraphics[scale=0.085,angle=0]{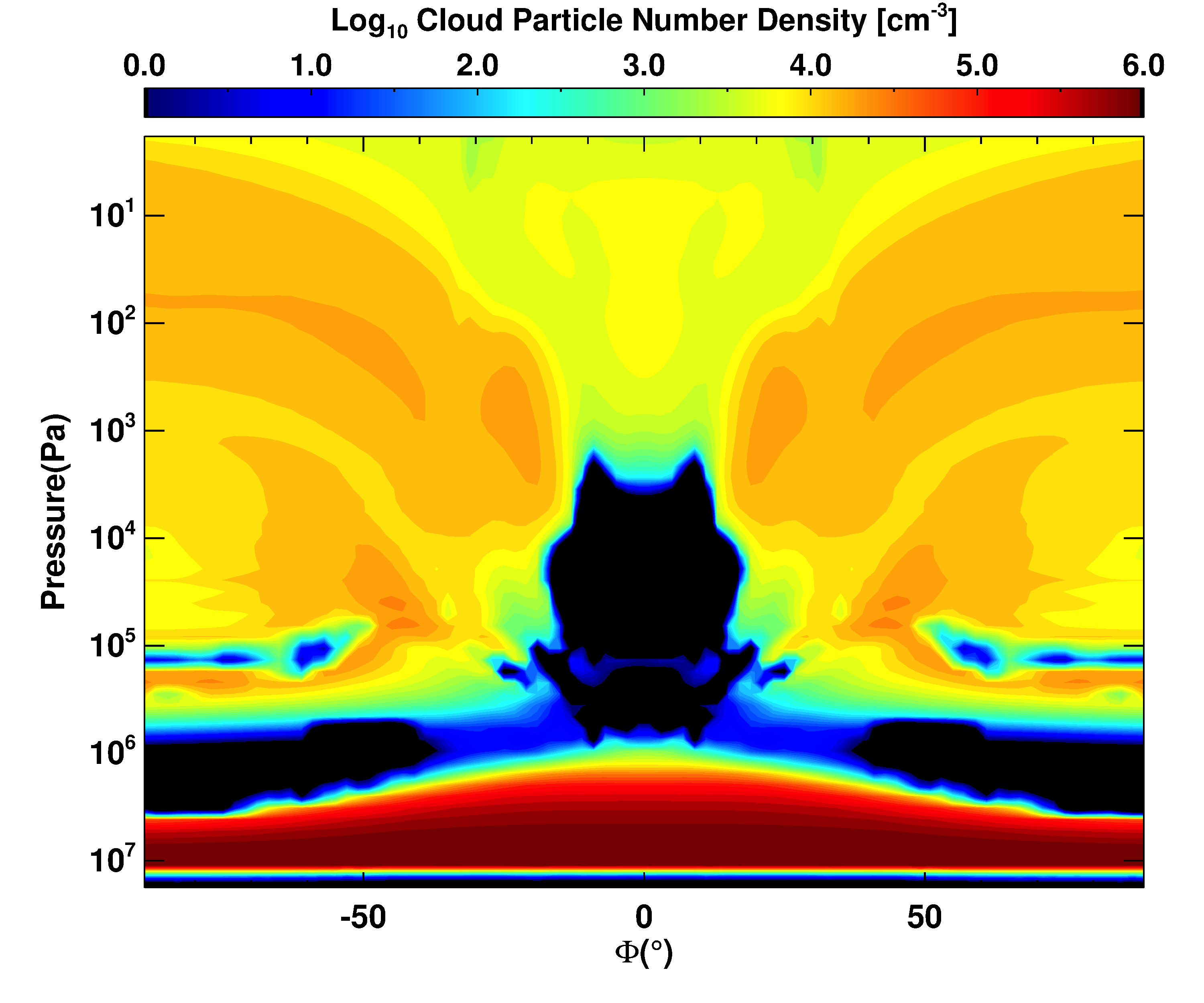}
\end{subfigure}

\vspace{-5pt}

\begin{subfigure}{\linewidth}
\includegraphics[scale=0.085,angle=0]{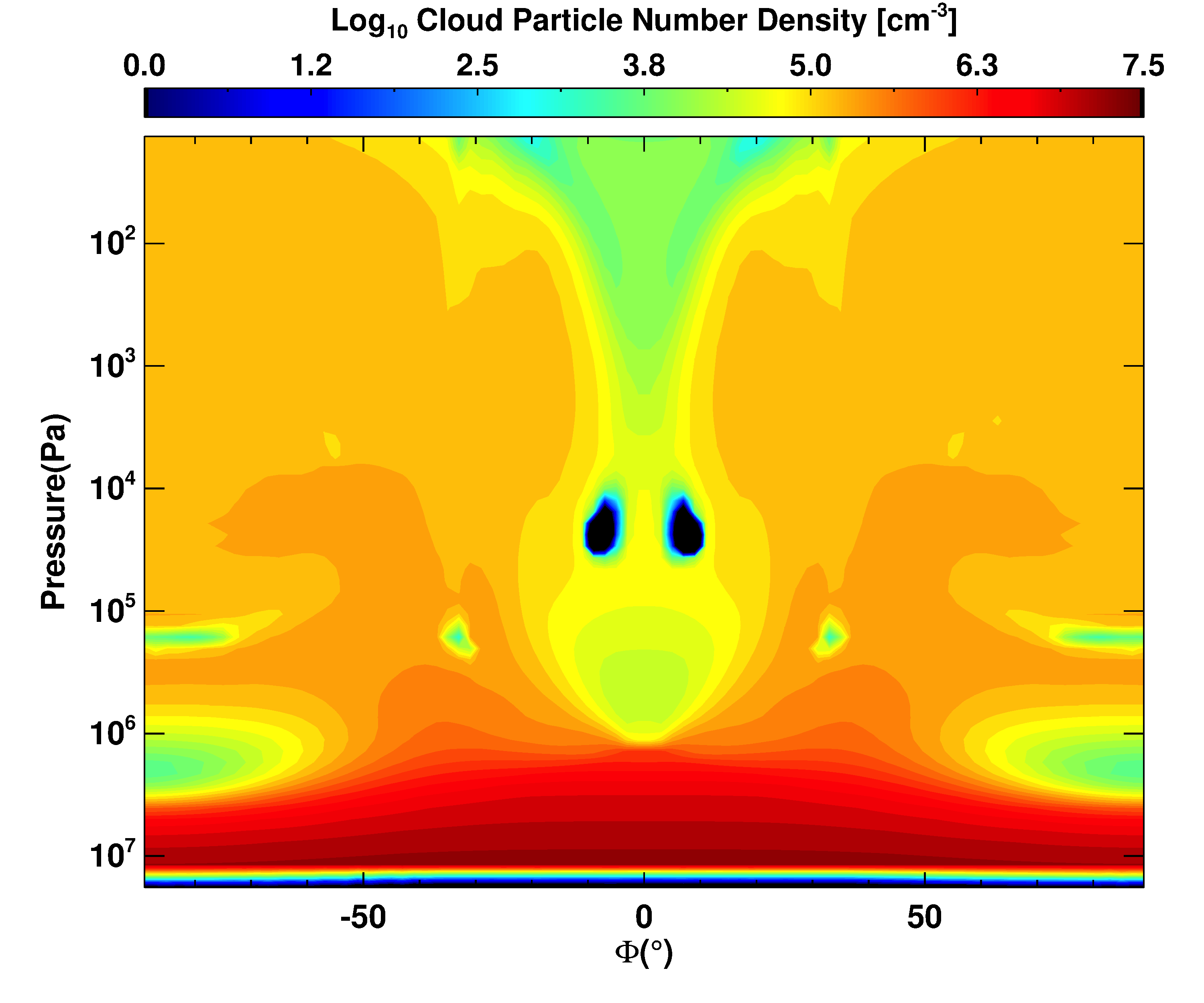}
\end{subfigure}

\vspace{-15pt}
\caption{Zonal mean of the cloud particle number density, log(n$_{\textrm{d}}$) [cm$^{-3}$] for hot HD 209458 b (upper), standard HD 209458 b (middle) and HD 189733 b (lower) atmospheres at t$_{\textrm{cloud}}$ = 50 days (see Table \ref{tab:params} and Section \ref{sec:na}).}

\label{fig:passive_vert}

\end{figure}


\section{Results}\label{sec:results}

We begin in Section \ref{sec:cloudfree} by examining the first stage of our simulations; a clear-skies, cloud-free atmosphere. By doing so we can identify the key thermal and dynamical features, as well as the differences between our chosen profiles, which can then be used to explain the transparent cloud results in Section \ref{sec:transparentres}. Finally in Section \ref{sec:radiative}, we focus on the results of our radiatively active, cloudy atmospheres. As described in Section \ref{sec:na}, we simulate two planetary atmospheres; HD 209458 b and HD 189733 b. We remind the reader that for HD 209458 b we use two temperature profiles, a `standard' atmosphere (solved for radiative-convective balance in 1D models) and a `hot' atmosphere with a hotter deep atmosphere to account for predicted dynamical processes, unable to be captured by 1D solutions and unable to be retrieved in practical timeframes in 3D simulations, that drive the deep atmosphere to much higher temperatures (+ 800 K at P = 10 bar) than the standard solution.

\subsection{Cloud Free}
\label{sec:cloudfree}

As discussed in Section \ref{sec:na}, we evolve our three atmospheric profiles (HD 209458 b (Standard + Hot) and HD 189733 b) without clouds with the UM for at least 800 days. Figure \ref{fig:ptinit}, shows the horizontally averaged TP profiles for each atmosphere at the end of the cloud free stage. 

\subsubsection{HD 209458 b}


\begin{figure*}[]
\begin{subfigure}{0.48\textwidth}
\vspace{+20pt}
\includegraphics[scale=0.55,angle=0]{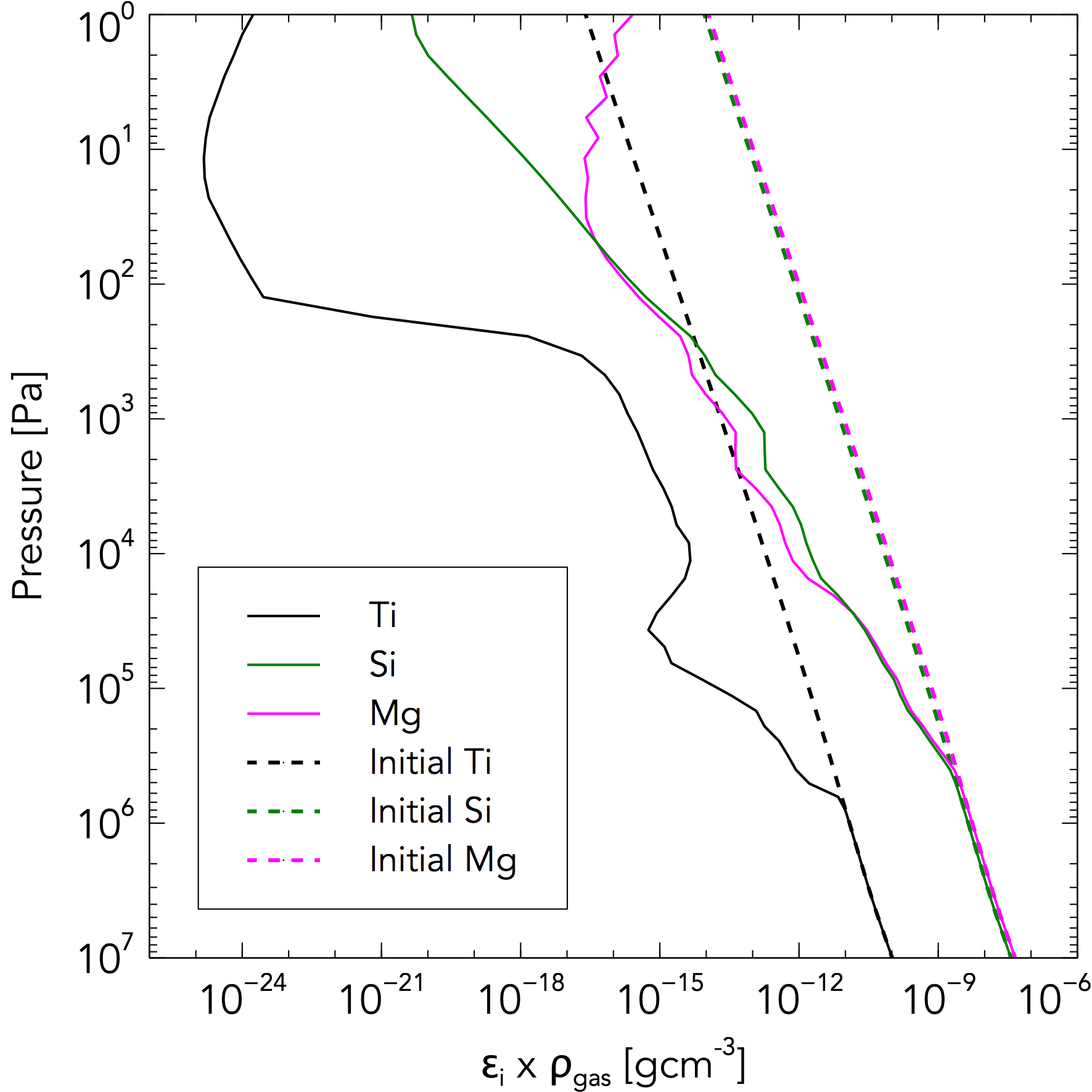}
\caption{Zonal and meridional mean, density scaled, elemental abundances of Ti, Si and Mg. The initial, undepleted values are the solar element abundances.}
\label{fig:h209_elem_passive}
\end{subfigure}\hspace*{\fill}
\begin{subfigure}{0.48\textwidth}
\includegraphics[scale=0.55,angle=0]{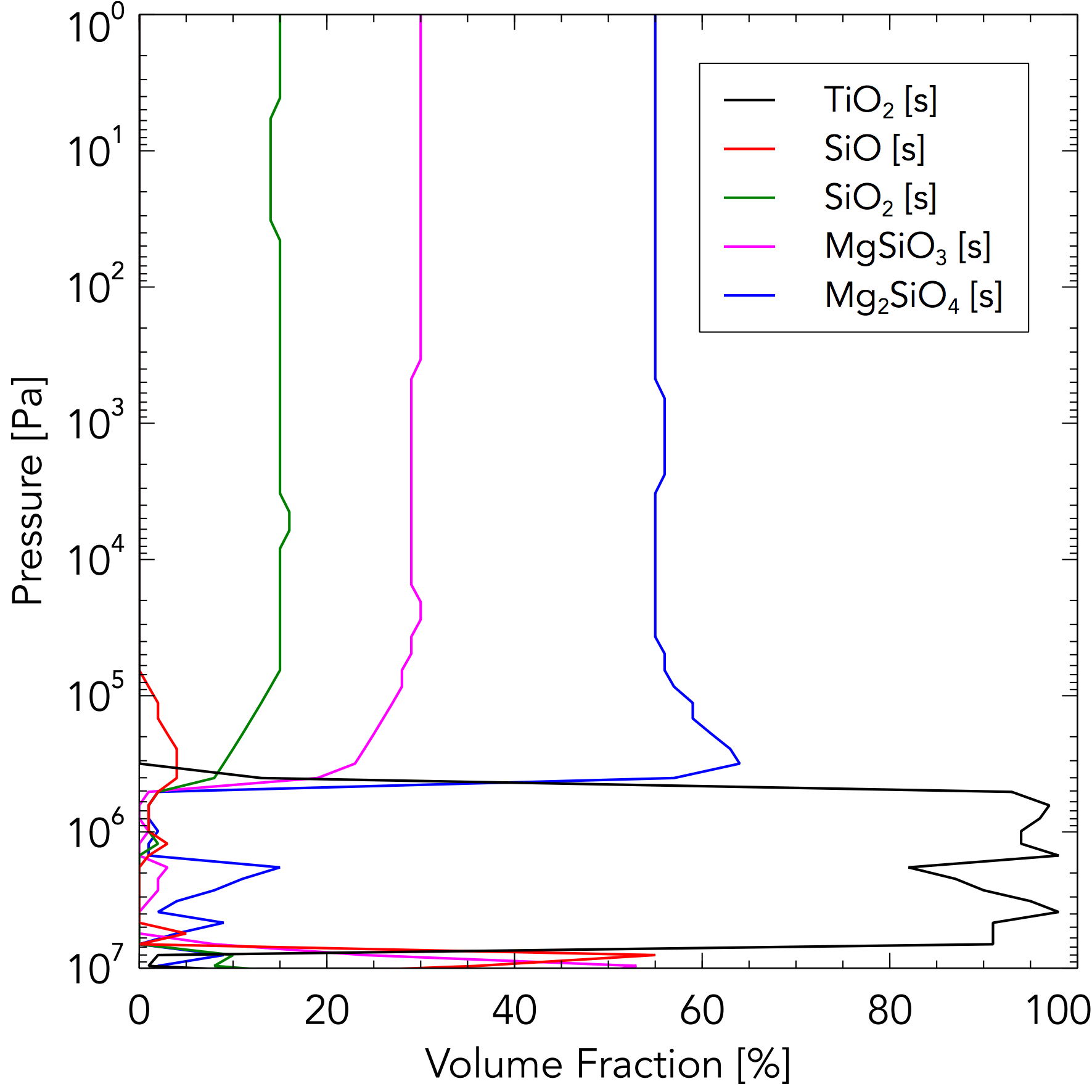}
\caption{Zonal and meridional mean dust volume fractions.}
\label{fig:h209_dust_passive}
\end{subfigure}

\caption{Chemistry in the hot HD 209458 b atmosphere: depletion of elemental abundances and volume contribution of dust species at t$_{\textrm{cloud}}$ = 50 days (see Table \ref{tab:params} and Section \ref{sec:na}).}

\end{figure*}


Alongside Figure \ref{fig:hd209_start} which shows the temperature and zonal windspeed of HD 209458 b, we can see that for P < 10$^6$ Pa, our hotter HD 209458 b is thermally and dynamically similar to that of the standard model which shows only minor differences compared with the results of \cite{mayne14} and \cite{amundsen16}. Since the irradiation is the dominant heating source for these tidally locked planets, the energy deposition is largest at the equator, with a corresponding increase in temperature near the equator with respect to higher latitudes. This trend reverses for P $>$ 10$^6$ Pa where the polar regions become warmer than the equator \citep[see discussion in][]{mayne14,mayne17}. The key identifying feature of the {\emph{cloud free}} standard model is the temperature inversion, where similar temperatures are predicted at both P = 10$^5$ Pa and P = 10$^7$ Pa. This local temperature inversion does not exist for the hot model. At the millibar pressure level there is no change between simulations in the horizontal temperature contrast, shown in Figure \ref{fig:hoz_t_start}. The fast super-rotating equatorial jet is clearly resolved in Figure \ref{fig:hd209_start} (right) with minimal differences between the two thermal profiles. Both jets possess an almost identical shape and velocity ($\approx$ 6.8 kms$^{-1}$), as well as the depth, approximately P = 10$^6$ Pa, at which the jet core extends into the atmosphere (the penetration depth). Cooler temperatures are found at high latitudes away from the maximum stellar energy deposition at $\phi$ = 0$^{\textrm{o}}$, $\lambda$ = 0$^{\textrm{o}}$.

\subsubsection{HD 189733 b}

From the horizontally averaged pressure-temperature profiles in Figure \ref{fig:ptinit}, the HD 189733 b profile is shown to be cooler than the standard HD 209458 b atmosphere, with a maximum temperature difference of $\Delta$T$_{max}$ $\approx$ -200 K. However, similar to the standard HD 209458 b simulation, for a {\emph{cloud free}} atmosphere, a temperature inversion occurs at P = 10$^6$ Pa. In Figure \ref{fig:hoz_t_start} minor differences, compared to HD 209458 b, in the horizontal temperature contrast are observed, with decreased temperatures on the night-side. The vertical distributions of temperature and zonal wind velocity for the clear skies stage are shown in Figure \ref{fig:c189_start}. A comparison between the HD 189733 b and the standard HD 209458 b show very few differences; the most noticeable change is that for HD 189733 b, the jet appears to advect heat more efficiently than HD 209458 b due to the higher equatorial night-side temperatures.

\subsection{Transparent Cloud}
\label{sec:transparentres}

In this section we consider the cloud evolution and structure prior to including cloud scattering and absorption. Cloud formation (seed nucleation and surface growth) and evolution processes (advection, precipitation and compositional changes) are reacting to the dynamical, thermal and chemical properties of our clear skies, cloud-free atmospheres. At this stage, clouds cannot modify the atmospheric thermal and dynamical properties, as this situation applies in the following radiatively active clouds stage. We stress that although transparent to radiation, clouds are not post-processed but form and evolve with the atmosphere's local thermochemical properties.

\subsubsection{Hot HD 209458 b}

The convergence tests of \cite{lee16} indicate that cloud is in a relatively steady-state throughout their simulation (from 0 to 65 days) with no significant change in the cloud particle number density from 20 days onwards. In Figure \ref{fig:h209evo}, analysis of the cloud particle number density over the first 50 days of our simulation indicate that the cloud is in a quasi-steady-state, with only small changes in particle number density in the region of P $\sim$ 10$^2$ Pa. Cloud particles form in dense decks ($n_{\textrm{d}}$ > 5000 cm$^{-3}$) between P = 8 x 10$^5$ and P = 10 Pa, with cloud particle number density tapering off significantly to the upper simulation boundary. The slow evolution of the cloud particles in regions accessible to observations is most likely attributed to low nucleation rates from the low pressure environment. The evolution of the cloud particle volume, $\langle V \rangle$, which reveals the surface growth efficiency after nucleation, is seen in Figure \ref{fig:h209evo} to increase from P = 10$^3$ Pa to P = 10 Pa. Both evolutionary plots show a swift decline of cloud particles and volume, at the base around 5 x 10$^5$ Pa. This is due to the settling of cloud particles into the deeper atmosphere which is too hot for seed nucleation and particle surface growth. It is important to note that while we may obtain a quasi-steady-state over the fast cloud formation timescale, the slow processes of vertical mixing and settling mean we are unable to probe the long term evolutionary effects of the cloud. Tackling this problem in 3D GCMs for a fully consistent, active cloud model is beyond the scope of this work.

The temperature and cloud particle number density at various pressures within the atmosphere are shown in Figure \ref{fig:passive_h209}. In all regions where nucleation and surface growth is possible, bands of cloud particles form outside of the equatorial jet. The distribution of particles in the jet at P = 10$^2$ Pa (see Figure \ref{fig:passive_h209}a) is time dependent, with the less thermally stable components of the mixed-composition particles evaporating in response to advection over the high temperature, `hot-spot' region (centred at $\lambda$ = 200$^{\textrm{o}}$) and re-condensing on the night-side. There are small longitudinal asymmetries in the particle distribution with enhanced numbers on the cooler night side. This is particularly clear in Figure \ref{fig:passive_h209}c where at the deeper P = 10$^3$ Pa level the hot-spot offset \citep[advected heat from the maximum irradiation at the sub-stellar point, e.g.][]{showman02,mayne14,mayne17} is more clearly defined, leading to a more pronounced eastwards shift in cloud particle depletion. At P = 10$^5$ Pa vortices that form across the night-day terminator lead to the trapping of cloud that is seen as enhanced particle number densities in these regions.

The zonally averaged cloud particle number density (Figure \ref{fig:passive_vert}) reveals a more complete picture of the cloud structure. The cloud bands are part of two out-of-equator decks that extend from the top of the simulation domain (P $\approx$ 10 Pa) at the poles, down into the atmosphere to P $\approx$ 10$^6$ Pa for all latitudes. Diverging winds from the jet cause seed particles that form there to advect to higher latitudes, although a decreased number remain trapped at the equator. These winds also constantly advect condensible gas to these higher latitudes where the cooler temperatures on the night side increase the saturation and lead to more efficient cloud formation. This can be seen clearly in Figure \ref{fig:passive_h209} where for all pressures plotted, there are higher cloud particle abundances found on the night-side. This is particularly true for P = 10$^3$ Pa where the hot-spot etches out a contrasting area of low cloud particle number density.

The cloud top has a wing-like shape where the cloud particle number density drops with decreasing latitude due to the increasing temperatures towards the hotter jet region. The cloud base for the hot HD 209458 b is well defined since the temperature increase is linear at pressures of around P = 10$^6$ Pa with no strong latitudinal temperature gradient (see Figure \ref{fig:hd209_start}). This means the transition at which nucleation is no longer possible ($J_{\textrm{*}}$ = 0) due to an inhibiting temperature, occurs at the same pressure level. There are two regions of enhanced particle number density; one extending from P = 10$^3$ Pa to P = 5 x 10$^4$ Pa and another for a narrow pressure range at the cloud base. The former is likely due to the balance between settling cloud particles from the upper atmosphere and preferential thermal conditions for nucleation. At the cloud base, while the temperature is low enough for seed particle nucleation, it is too hot for the surface growth of silicates. Therefore, all available TiO$_2$ material either goes into nucleating seed particles (and increasing seed number densities) or is the only species available for surface growth on these seeds. The latter effect is shown to be true in \cite{lee15b}. 


\begin{figure*}[]

\begin{subfigure}{0.48\textwidth}
\includegraphics[scale=0.085,angle=0]{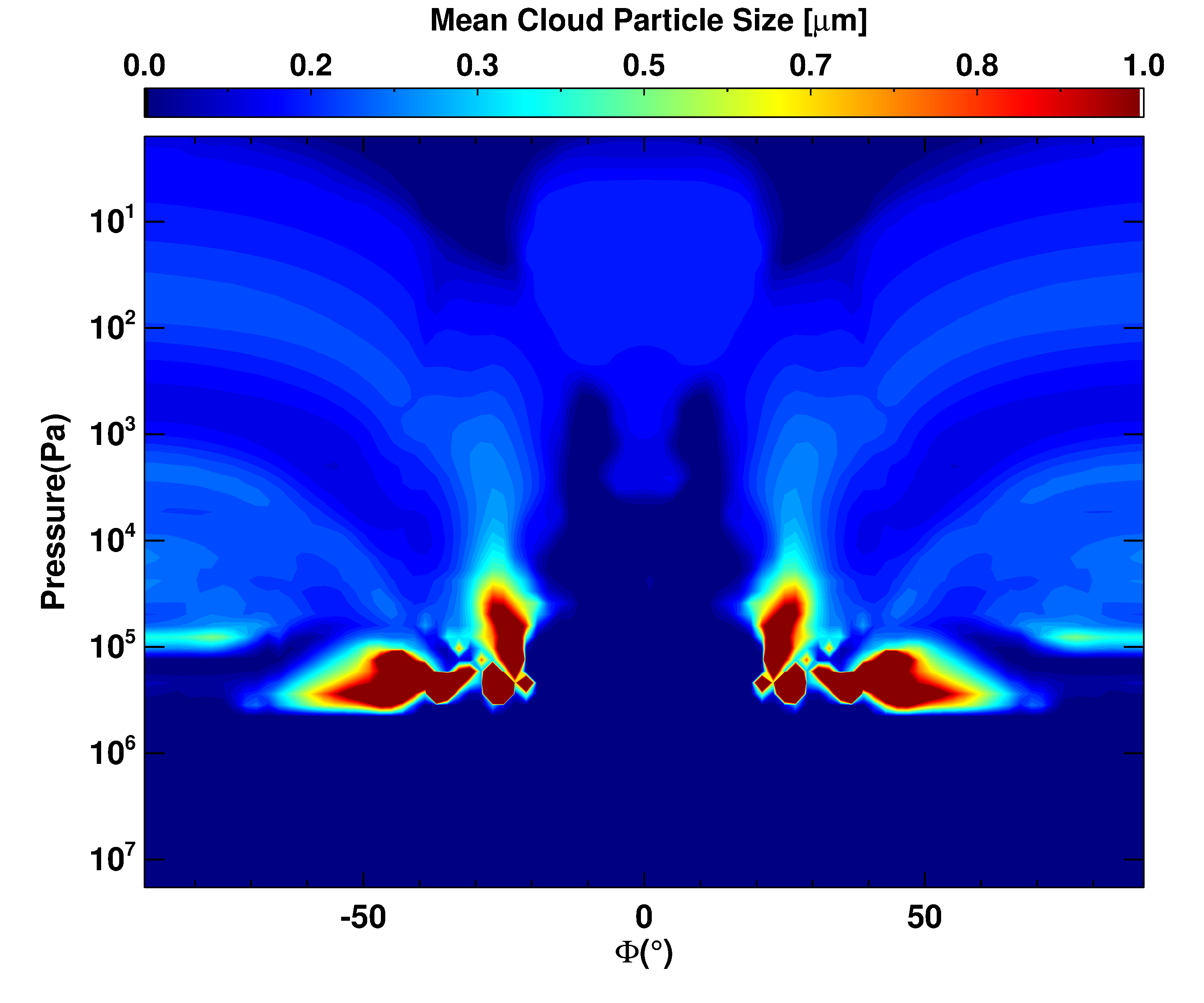}
\end{subfigure}\hspace*{\fill}
\begin{subfigure}{0.48\textwidth}
\includegraphics[scale=0.085,angle=0]{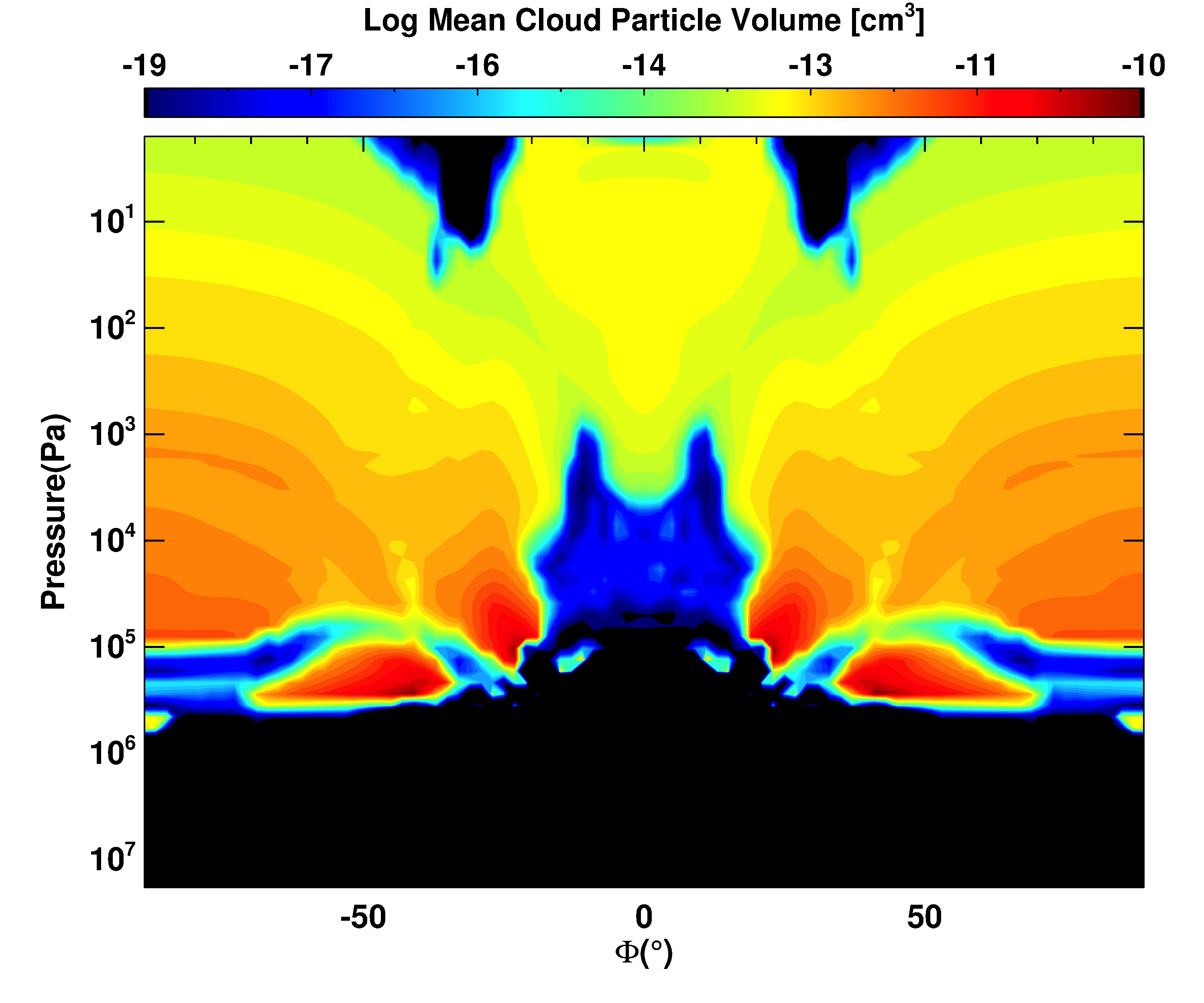}
\end{subfigure}

\caption{Zonal average of the (left) mean cloud particle size, $\langle a \rangle$ [$\mu$m] and (right) mean cloud particle volume, $\langle V \rangle$ [cm$^{-3}$], for the hot HD 209458 b atmosphere at the end of the transparent cloud stage at t$_{\textrm{cloud}}$ = 50 days (see Table \ref{tab:params} and Section \ref{sec:na}).}
\label{fig:h209_passive_a_vol_mean}

\end{figure*}


The depletion and enrichment of elemental abundances is a consequence of the cloud formation and evaporation processes. In Figure \ref{fig:h209_elem_passive} the density scaled abundances of our three tracked elements are shown as a function of atmospheric pressure. Titanium is depleted down towards the cloud base at P = 8 x 10$^5$ Pa. At higher pressures the cloud particles are unable to form and therefore the elemental abundances remain at their initial solar metallicity values. Very small enhancements are made to the metallicity below the cloud base due to the evaporation of particles settling into the hot-zone. Silicon and magnesium are depleted for P $\leq$ 4 x 10$^5$ Pa leading to the narrow TiO$_2$[s] only layer at the very base of the cloud between P = 4 x 10$^5$ Pa and P = 8 x 10$^6$ Pa. The growth velocities between these pressures for our magnesium and silicon based condensates is zero since the temperature in this region is too high for thermal stability, whereas TiO$_2$[s] is able to exist at the higher temperatures. Figure \ref{fig:h209_dust_passive} shows the contribution, by volume, of each surface growth species. This TiO$_2$ layer is clearly identified, with the volume fraction increasing for pressures lower than P = 4 x 10$^5$. A preference for the formation of SiO$_2$[s] means that SiO[s] composes only a small fraction of the cloud particles.

The depletion of cloud particles at the equator observed from the horizontal slices of the atmosphere shown in Figure \ref{fig:passive_h209} is seen clearly in Figure \ref{fig:passive_vert}. Number densities five times smaller than those seen in the areas of highest concentration, exist in the jet between P = 10$^{2}$ Pa and P = 10$^5$ Pa, with these numbers decreasing with time. While a smaller number of particles persist in the jet, many of the particles have advected out of the equatorial region and into the cloud decks. Since the jet region is hotter than at higher latitudes, it does not allow for large surface growth velocities. This means that the particles near the equator are typically smaller in size than those in the higher latitude bands.

The mean particle size (Figure \ref{fig:h209_passive_a_vol_mean} [left]) shows a clear correlation with the cloud particle number density; where cloud particles exist in large number densities the cloud particle size is small and where the particle number density is low, the cloud particle sizes are large. In areas with small numbers of particles, more condensible material grows on a single particle, increasing the size. This effect is similarly seen in the mean cloud particle volume (Figure \ref{fig:h209_passive_a_vol_mean} [right]) where the volumes are largest in regions with the lowest particle number densities. With the exception of the vortex regions near the cloud base, cloud particles are consistently sub-micron sized. In HD 209458 b, since clouds scatter strongly as a function of the number and size of their particles, a combination of the cloud particle number density and the mean particle size is a more useful metric than the cloud particle volume. In further analysis, we will consider the cloud distribution to be defined primarily by the number of cloud particles and not total volume.

\subsubsection{Standard HD 209458 b}

Since the standard HD 209458 b temperature profile is almost identical to the hotter one for P $\leq$ 5 x 10$^5$ Pa it is not surprising that differences in cloud structure between these simulations are minimal. The zonally averaged particle number density (Figure \ref{fig:passive_vert}) shows a cloud deck that extends from the top of the simulation domain to P $\sim$ 10$^6$ Pa. The cloud base matches the hottest part of the atmosphere, with the temperature decreasing again at higher pressures. The temperature inversion seen in Figure \ref{fig:hd209_start} (lower, left) is enough to allow for nucleation and surface growth, with high pressures supporting cloud formation. This leads to the formation of two cloud decks, the deep atmosphere one with vastly higher cloud particle number densities.

There is a similar increase in cloud particles at the cloud base, and areas of depletion between P = 10$^4$ Pa and the cloud base at polar regions. Unlike for the hot profile, enhanced cloud wings just below P = 10$^3$ Pa do not exist, but we instead see an increase in cloud particles from P = 10$^3$ Pa and upwards.  Again, particles persist in the jet and more so at high latitudes in the upper atmosphere, suggesting a less efficient settling process.


\begin{figure*}[]

\begin{subfigure}{0.48\textwidth}
\includegraphics[scale=0.085,angle=0]{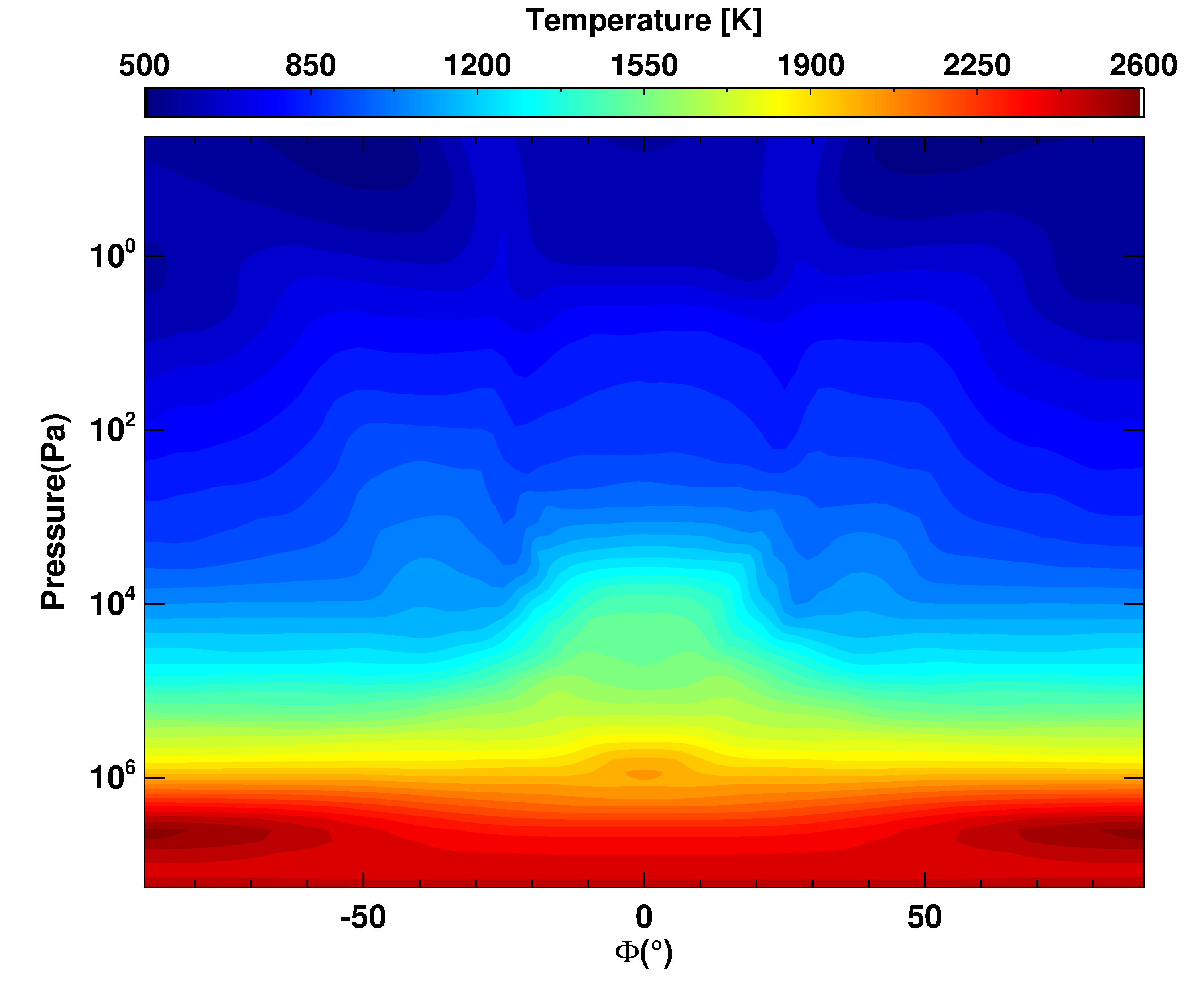}
\end{subfigure}\hspace*{\fill}
\begin{subfigure}{0.48\textwidth}
\includegraphics[scale=0.085,angle=0]{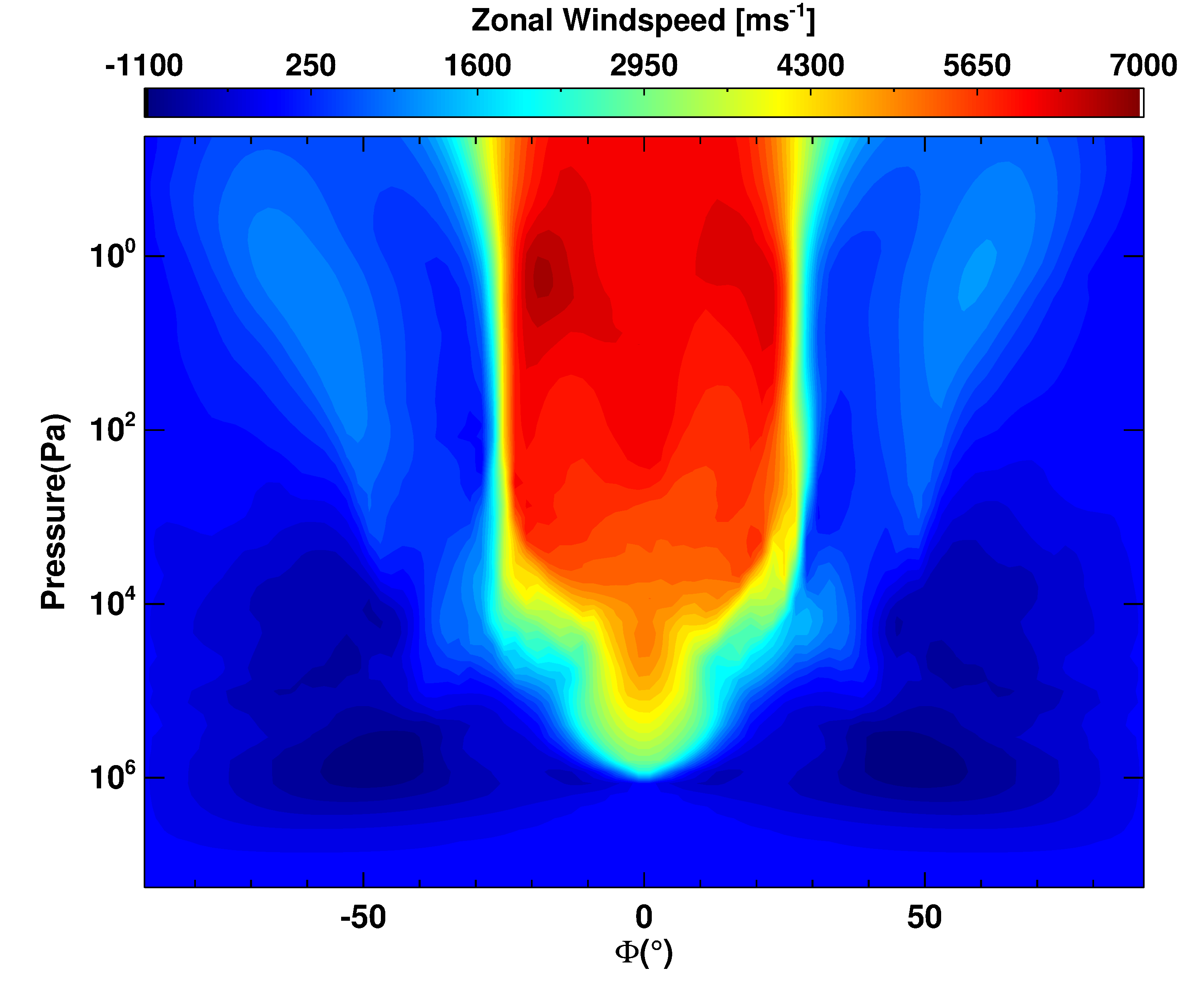}
\end{subfigure}

\begin{subfigure}{0.48\textwidth}
\includegraphics[scale = 0.085, angle = 0]{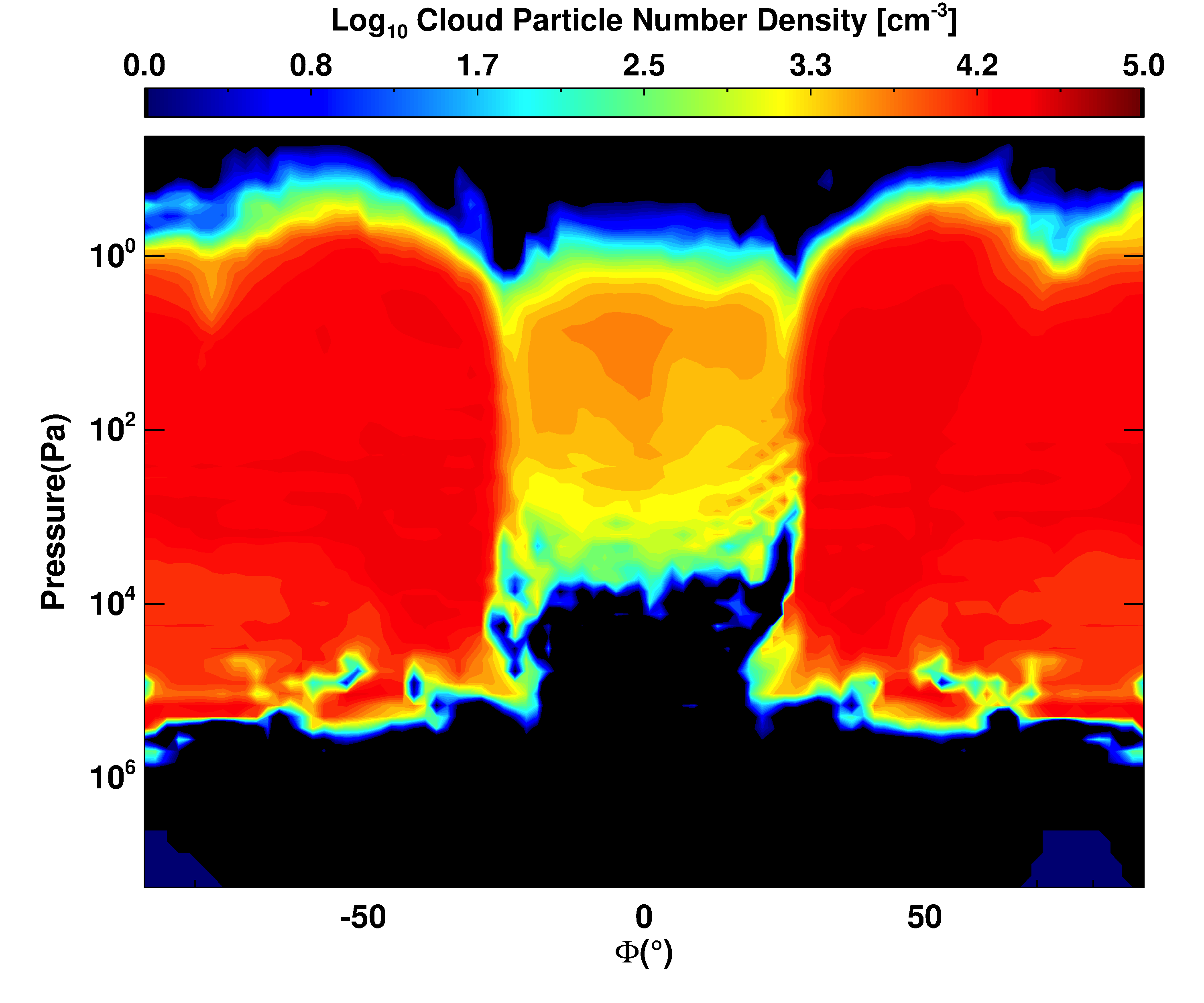}
\end{subfigure}\hspace*{\fill}
\begin{subfigure}{0.48\textwidth}
\includegraphics[scale=0.085,angle=0]{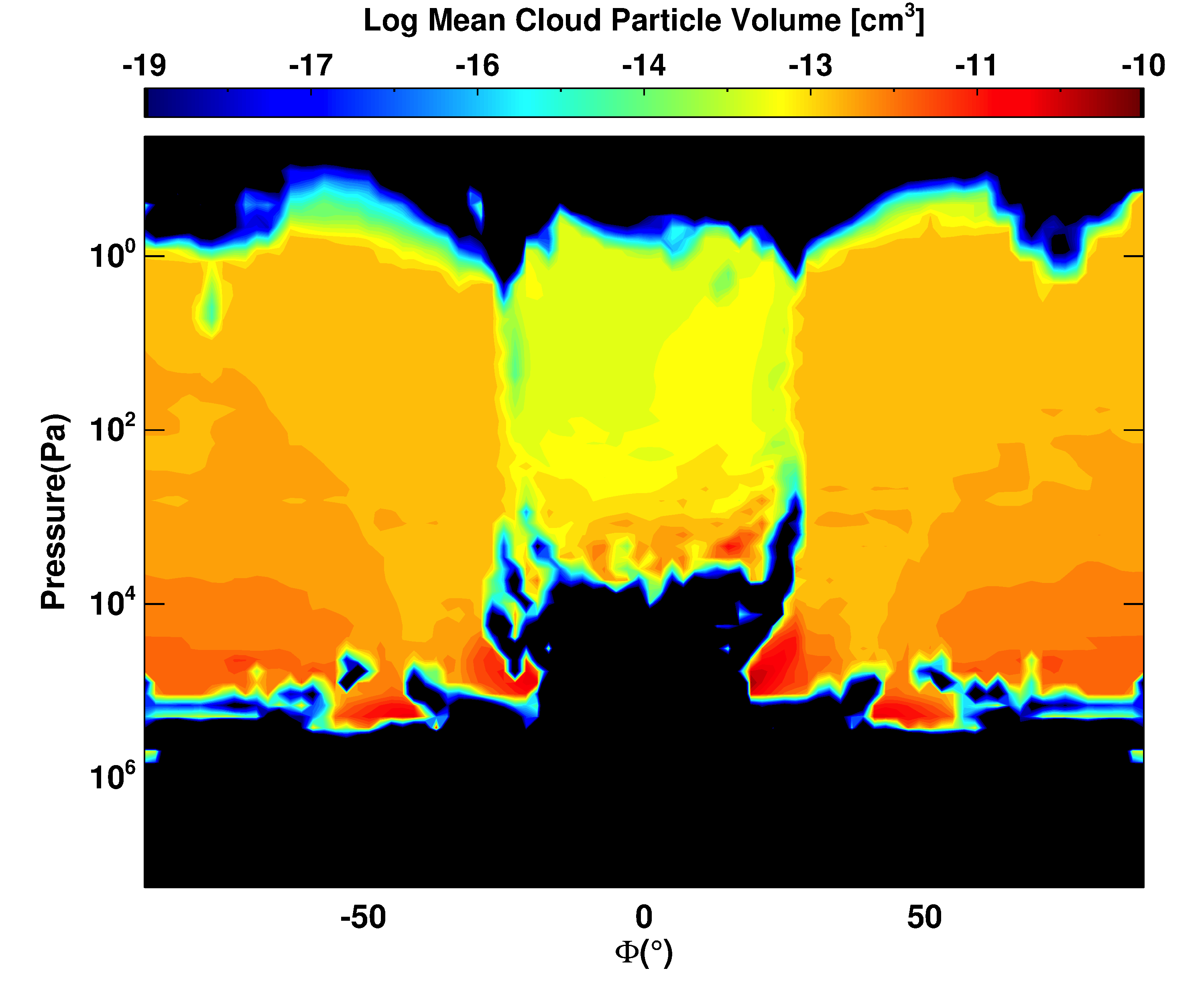}
\end{subfigure}

\begin{subfigure}{0.48\textwidth}
\includegraphics[scale=0.085,angle=0]{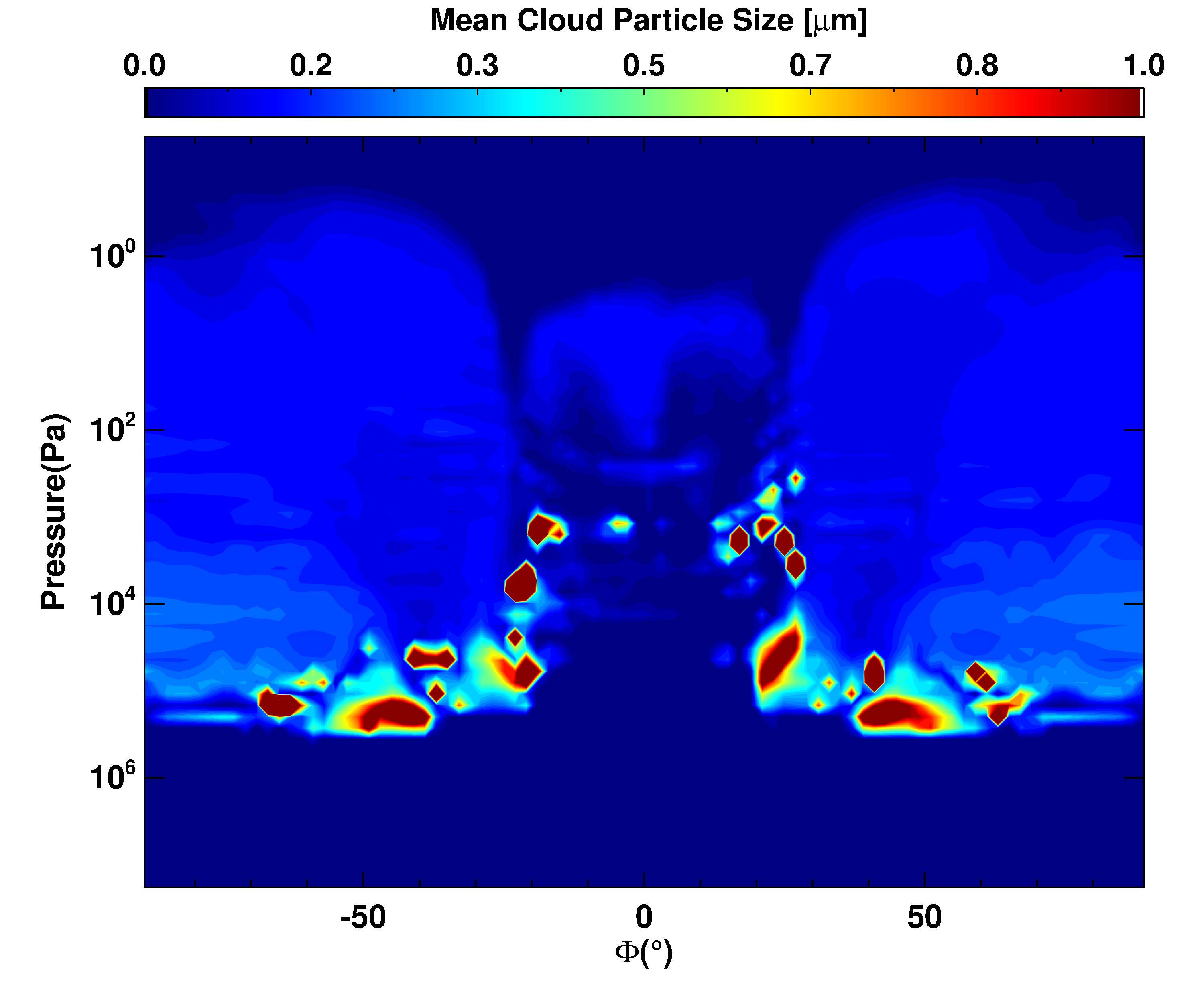}
\end{subfigure}\hspace*{\fill}
\begin{subfigure}{0.48\textwidth}
\includegraphics[scale=0.085,angle=0]{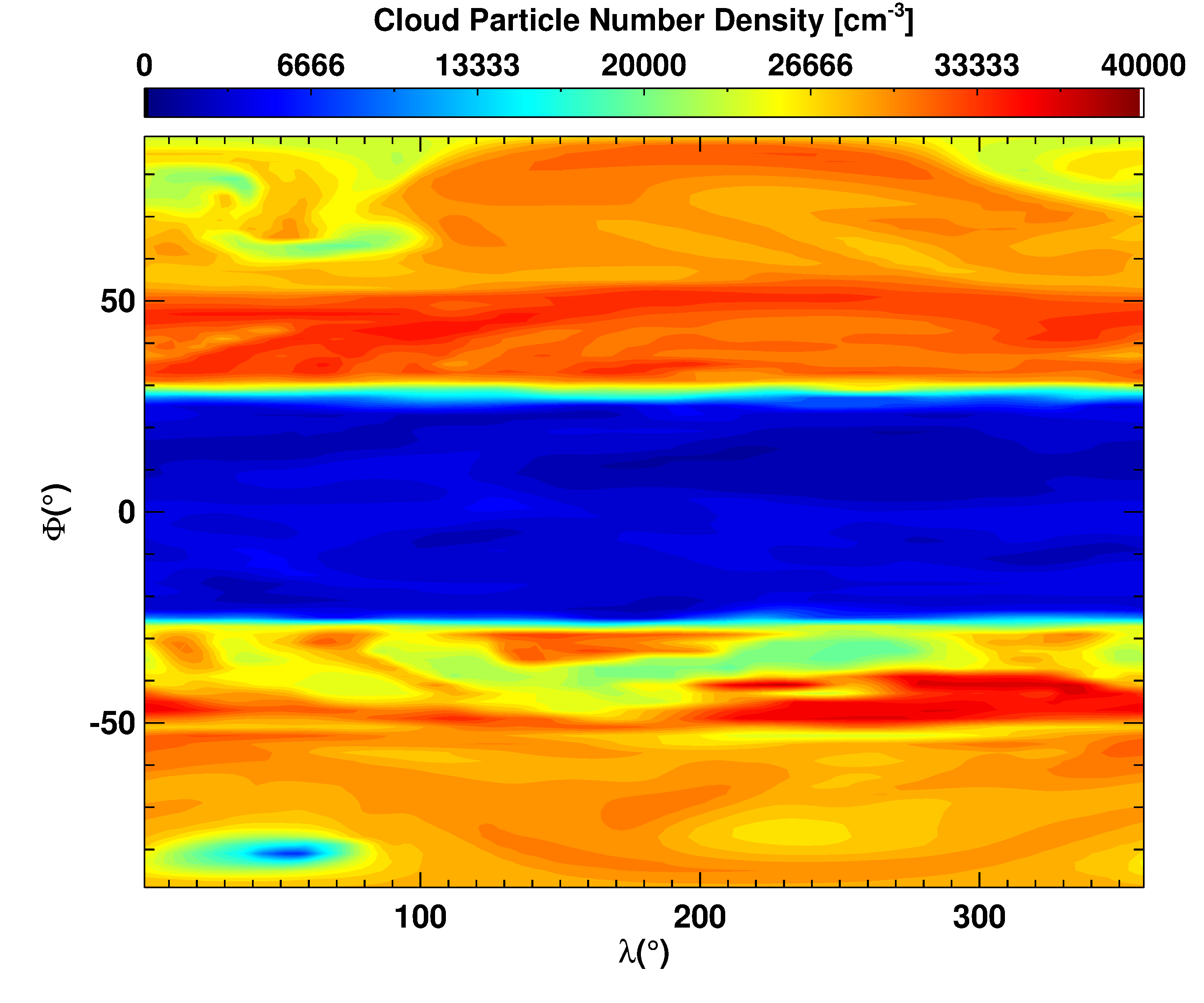}
\end{subfigure}

\caption{Thermal, dynamical and cloud properties of the hot HD 209458 b atmosphere after t$_{\textrm{cloud}}$ = 100 days (see Table \ref{tab:params} and Section \ref{sec:na}) with 50 days of radiatively active clouds.}
\label{fig:h209}

\end{figure*}



\begin{figure*}
\includegraphics[scale=0.75,angle=0]{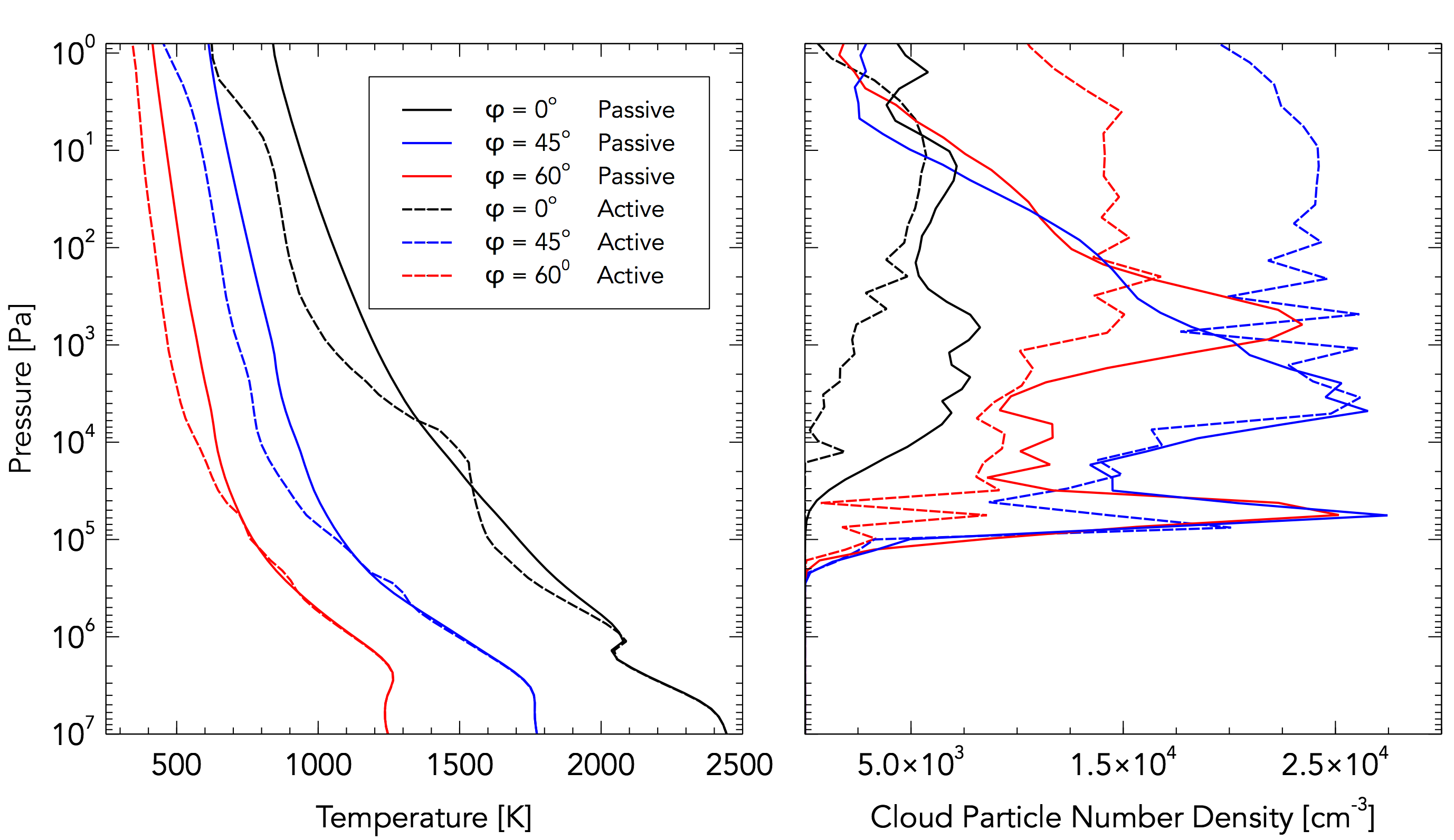}
\caption{Longitudinally averaged gas temperature [K] (left) and particle number density (right) for the hot HD 209458 b model at the end of the transparent cloud stage at t$_{\textrm{cloud}}$ = 50 days (see Table \ref{tab:params} and Section \ref{sec:na}) and the end of the radiatively active stage at t$_{\textrm{cloud}}$ = 100 days. For each time, three profiles are plotted that cover latitudes of $\phi$ = 0$^{\textrm{o}}$ (black), $\phi$ = 45$^{\textrm{o}}$ (blue) and $\phi$ = 60$^{\textrm{o}}$ (red).}
\label{fig:h209_tdiffnddiff}
\end{figure*}



\begin{figure}
\includegraphics[scale=0.6,angle=0]{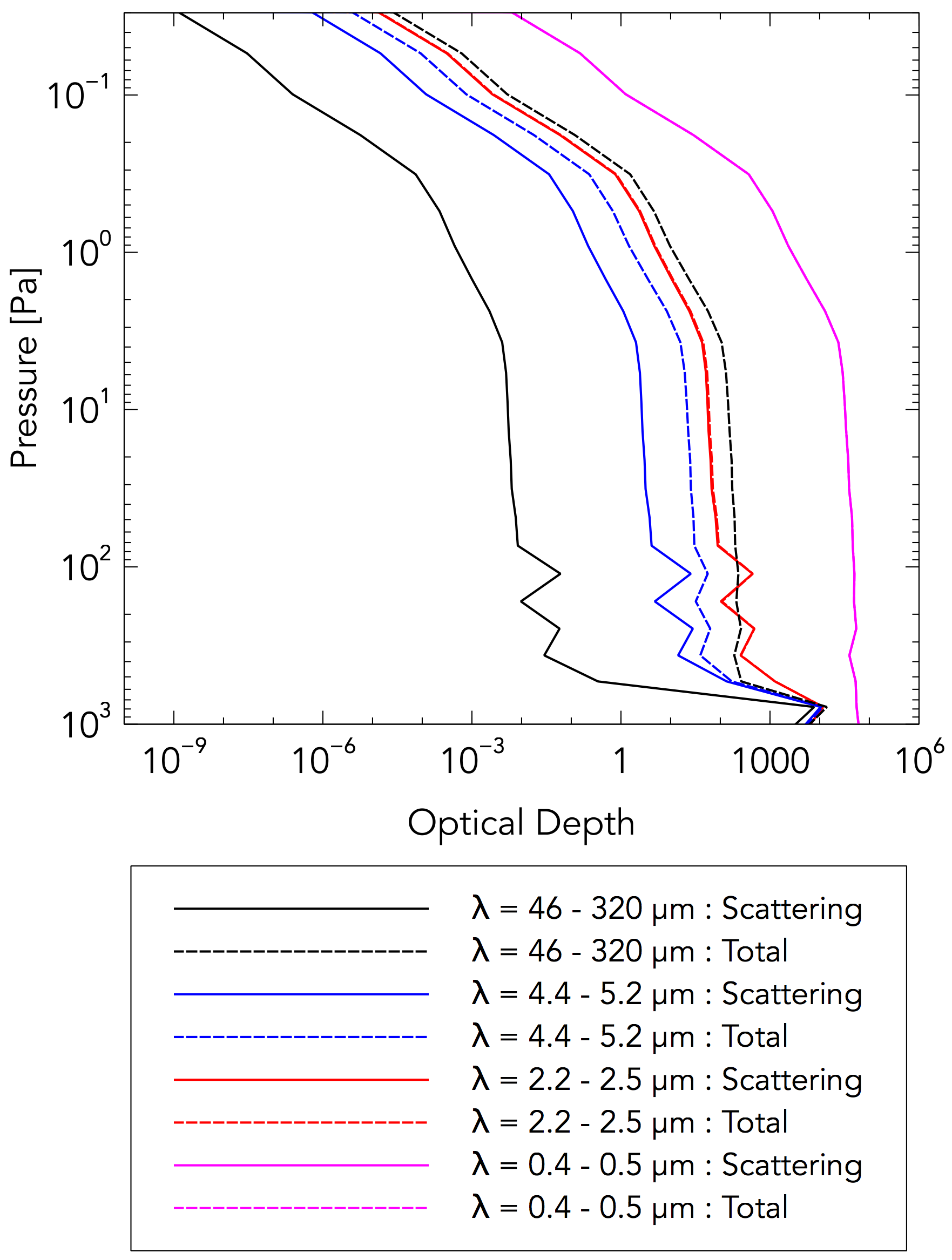}
\caption{Meridional mean at $\lambda$ = 180$^{\textrm{o}}$ of the cloud unitless optical depth. The scattering component of the optical depth are shown by the solid lines and the total optical depth (scattering and absorption) from the cloud is shown by the dashed lines.}
\label{fig:h209_p_sca}
\end{figure}


\subsubsection{HD 189733 b}

Unlike the standard HD 209458 b model, the temperature does not increase, prior to the inversion, to levels that inhibit cloud formation. Hence, the large number densities from the pressure supported high-growth region at pressure higher than P = 10$^6$ Pa are seen at the base of Figure \ref{fig:passive_vert}.

\subsection{Radiatively Active Cloud}

\label{sec:radiative}

After a period of running the simulations with transparent clouds, we couple particles to the radiation scheme. The following results consider the atmosphere at the most evolved point, after the combination of the transient radiative and full radiative cloud stages. 

\subsubsection{Hot HD 209458 b}

From t$_{\textrm{cloud}}$ = 50 days clouds begin to scatter and absorb radiation. 14 days of a transient radiative cloud stage are required to maintain simulation stability as the atmosphere approaches a steady state, defined such that the temperatures are no longer significantly changing between successive time steps. A further 36 days of fully radiatively active clouds are then simulated. The final atmospheric temperature is shown in Figure \ref{fig:h209} (upper, left) and the zonal wind speed in Figure \ref{fig:h209} (upper, right). The difference in the zonal mean temperature shown in Figure \ref{fig:h209_tdiffnddiff} shows that the temperature of the atmosphere decreases from the upper simulation boundary to pressures of approximately P = 10$^6$ Pa and all latitudes see a temperature decrease of a similar magnitude. This cooling effect is due to the large proportion of the incoming stellar flux that is back-scattered instead of being absorbed (as per cloud free simulations). The largest cooling is found at the top of the atmosphere, in the equatorial region, with $\Delta$T reaching up to -250 K. At $\phi$ = 45$^{\textrm{o}}$ and $\phi$ = 60$^{\textrm{o}}$, the zonally averaged profiles show small heating around 2 x 10$^5$ Pa. This limited heating region traces the cloud base where cloud particles are better absorbers due to their larger ($\sim$ 1 $\mu$m) size, and due to their interaction with the planet's long-wave thermal flux. 

At the equator, a large pressure range of more efficient heating ($\Delta$T $\approx$ +100 K) occurs near 10$^4$ Pa, matching the base of the small particles trapped in the jet. Figure \ref{fig:h209_tdiffnddiff} also reveals the change in particle number density from cloud radiation coupling. At the equator, almost every pressure level shows the particle number density decreasing when the cloud becomes radiatively active, but this decrease halts towards the end of the 50 days, potentially as a result muted dynamics (reduced vertical and meridional wind speeds) that occur during this radiative stage. This indicates that despite the cooler temperatures providing enhanced conditions for particle growth, the transfer of particles out of the jet from lower to higher latitudes is still the dominating factor. Additionally, the jet becomes depleted of metal rich gas which inhibits further nucleation and growth of cloud particles. Conversely, for $\phi$ = 45$^{\textrm{o}}$, probing the cloud decks, the cooler temperatures and advection of particles from the jet cause the particle number density to rise for P $\leq$ 2 x 10$^3$ Pa. At $\phi$ = 60$^{\textrm{o}}$, there is an increase in cloud particle number density above P = 10$^2$ Pa, but this value decreases at all other altitudes. The net effect is to increase the contrast between the vertically extensive and particle number dense mid to high latitude cloud decks, and the less cloudy equatorial jet.

\begin{figure}
\includegraphics[scale=0.09,angle=0]{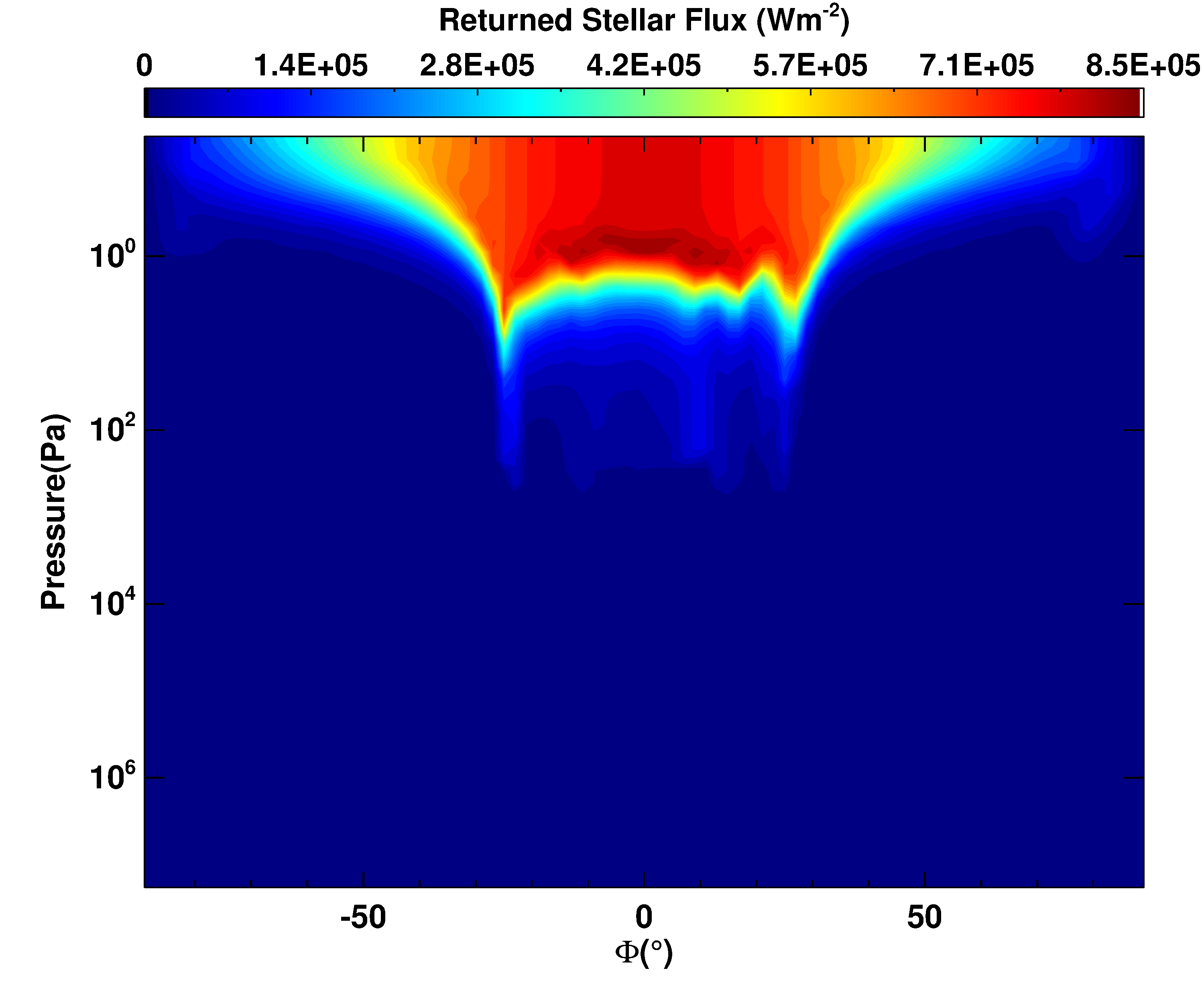}
\caption{Upwards/returned stellar flux (Wm$^{-2}$) for $\lambda$ = 180$^{\textrm{o}}$.}
\label{fig:h209_swout}
\end{figure}

While valuable information regarding the atmosphere's thermal change is obtained from the pressure-temperature profiles, it is useful to understand how the clouds may impact the direct observables. In Figure \ref{fig:h209_p_sca}, the cloud optical depth as a function of atmospheric pressure, for four different wavelength ranges is shown. The efficiency of the Mie scattering from sub-micron particles is reflected in the scattering optical depth increasing with decreasing wavelength, since the Mie scattering strength increases when the particle size matches the interacting radiation. The role of cloud, in the absorption of stellar flux, is limited to the longest wavelengths. For $\lambda$ $<$ 2 $\mu$m, the cloud scatters all incoming stellar irradiation, highlighting the importance of this process in cloudy atmospheres. In Figure \ref{fig:h209_swout}, the reflected stellar flux per layer is shown for varying pressure depth. The difficulty of penetration by the stellar irradiation, due to cloud scattering, is clearly seen with the bulk of the returned stellar flux tracing the cloud top (note that only minimal stellar radiation is absorbed here, as per Figure \ref{fig:h209_p_sca}). Even the most transparent cloud in the equatorial jet allows minimal stellar flux to travel below the millibar pressure level.

\begin{figure}
\includegraphics[scale=0.48,angle=0]{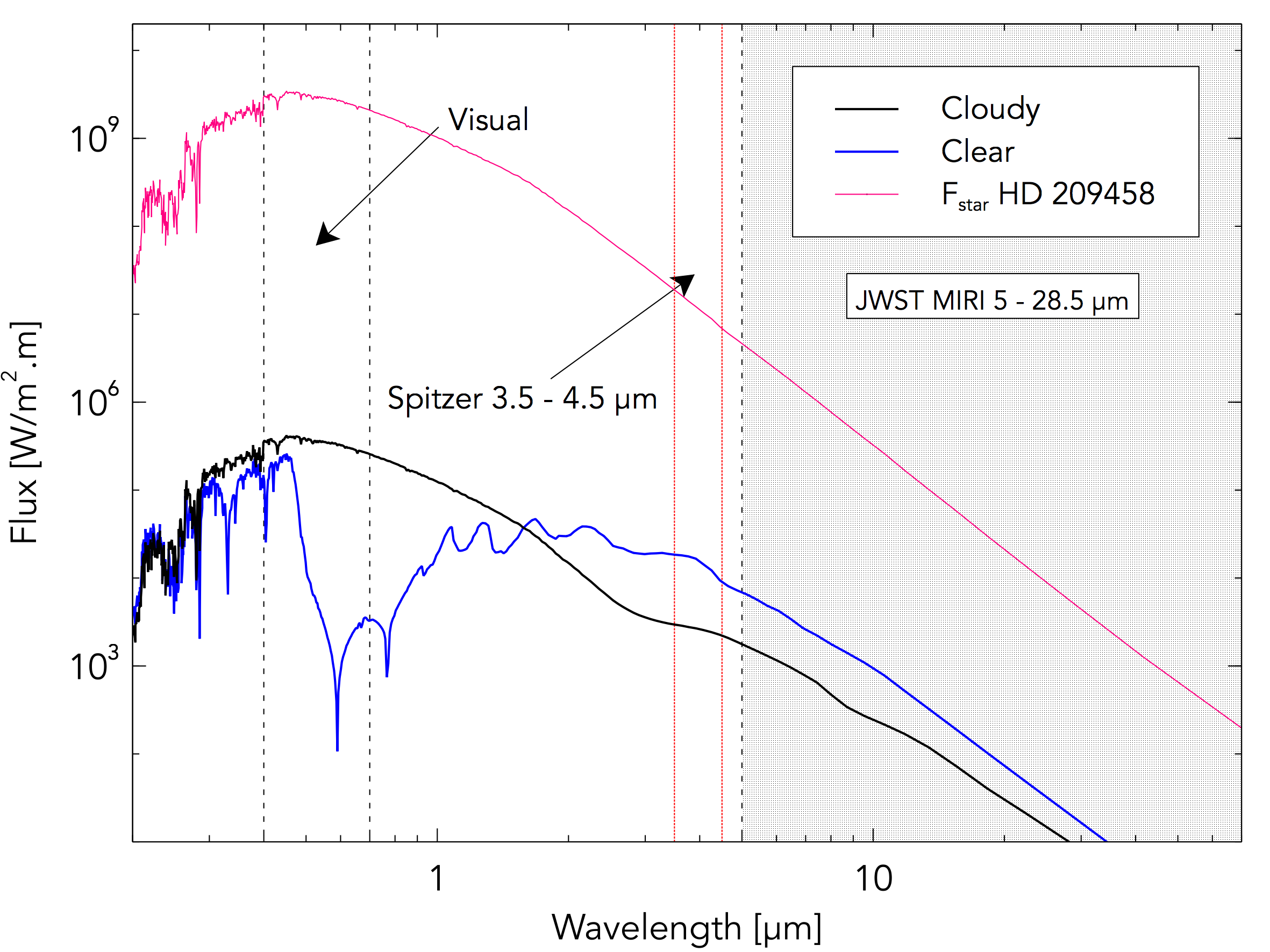}
\caption{Spectral energy distributions of the active, cloudy, hot HD 209458 b (black line) and cloud free (blue line) simulations. The spectrum is taken at secondary eclipse ($\phi$ = 180$^{\textrm{o}}$) and includes the reflected stellar and emitted thermal flux. Cloud free absorption is from Na and K.}
\label{fig:h209_emission}
\end{figure}

Figure \ref{fig:h209_emission}, plots the spectral energy distributions for the cloudy and clear planetary flux, in addition to the received stellar flux, both normalised to Wm$^{-2}$ for an observer at a distance of 1AU. The planetary flux shows the total of the reflected stellar flux and the emitted thermal flux from the planet's atmosphere. We can begin to appreciate the dominance of the scattering processes from the cloud due to the large fractional return of the incoming stellar flux from the emission and reflectance in the SED, from the cloudy atmosphere. In the cloud-free simulation, strong absorption from sodium and potassium is seen in the visual to near-infrared. This absorption is lost in the cloud simulation which sees a proportional return of the incoming stellar flux, and thus highlights the impact of cloud on obscuring chemical species. The reduced stellar heating from the high albedo of the atmosphere leads to the natural consequence of a cooling atmosphere, due to the lowering of the atmosphere's total energy budget. A combination of the cooler atmosphere, and the increased optical depth to thermal radiation due to the cloud, leads to reduced outgoing flux at thermal infra-red wavelengths compared to the cloud free simulations. Indeed, the flux in the Spitzer bandpass between 3.6 $\mu$m and 4.5 $\mu$m is reduced by an order of magnitude when clouds are active.

\begin{figure}
\includegraphics[scale=0.55,angle=0]{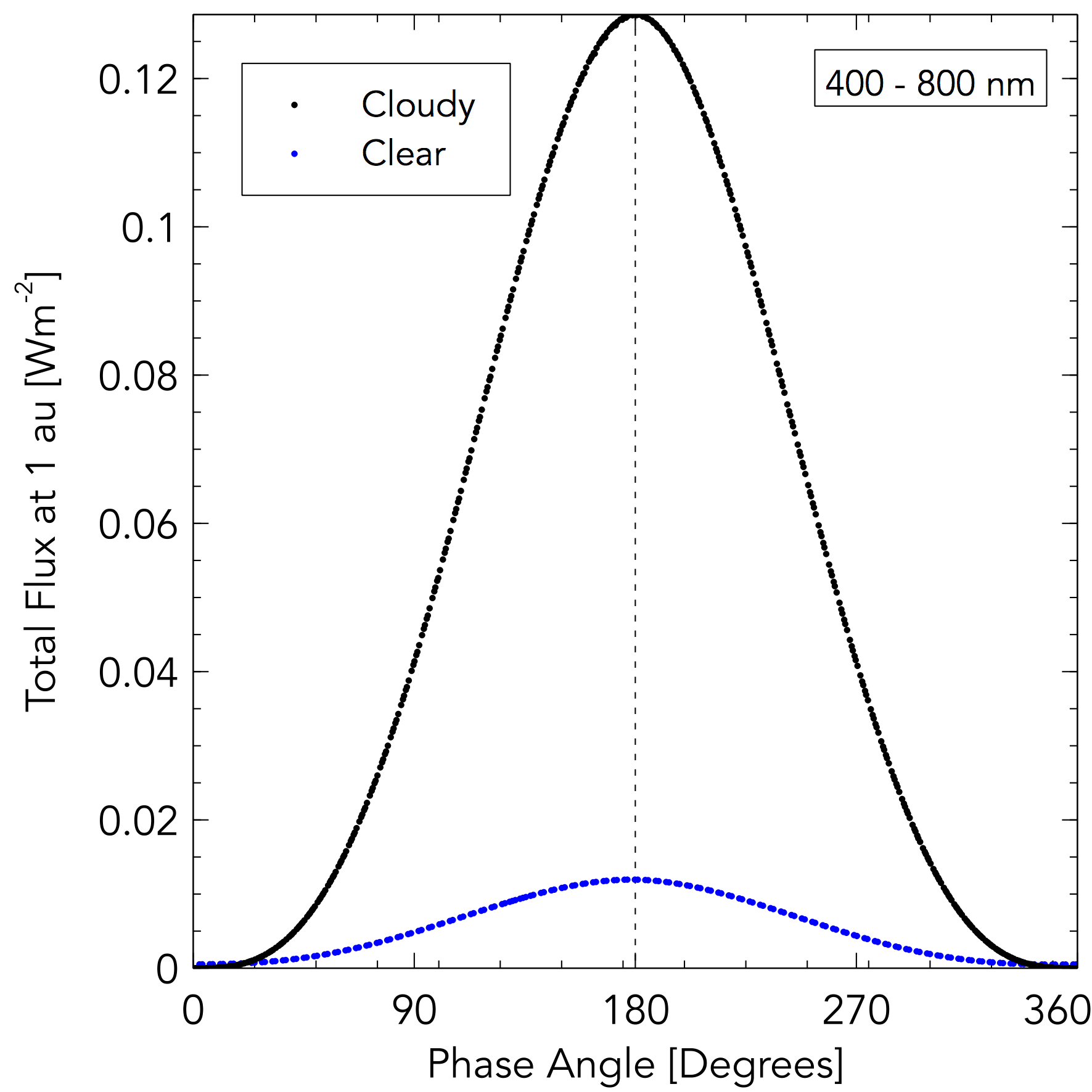}
\caption{Visual phase curve of the active, cloudy, hot HD 209458 b simulation for flux at 1 au received in the Kepler bandpass between 400 - 800 nm. Black points correspond to six complete orbits after the simulation end at t$_{\textrm{cloud}}$ = 100 days (see Table \ref{tab:params} and Section \ref{sec:na}). Blue points follow the flux from a cloud free simulation.}
\label{fig:h209_kphase}
\end{figure}

\begin{figure}
\includegraphics[scale=0.44,angle=0]{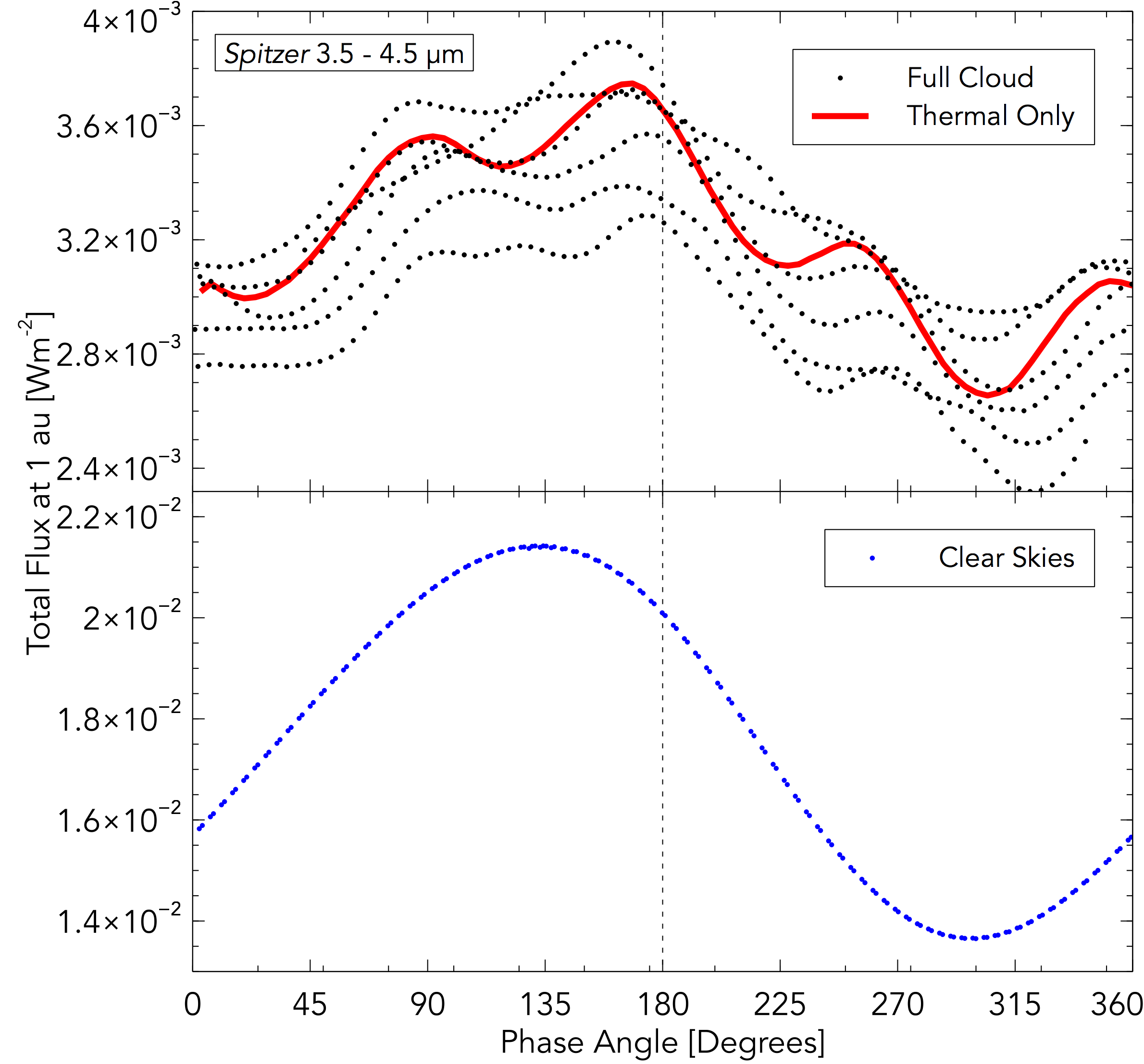}
\caption{Phase curves of the active, cloudy, hot HD 209458 b simulation for flux at 1 au received between 3.5 - 4.5 $\mu$m. Black dots correspond to six complete orbits after the simulation end at t$_{\textrm{cloud}}$ = 100 days. The red line is the flux from the thermal emission of the atmosphere only (and is to be compared with the first orbit (uppermost black dots), and does not included reflected stellar flux.}
\label{fig:h209_phase}
\end{figure}

\begin{figure*}[]

\begin{subfigure}{0.48\textwidth}
\includegraphics[scale = 0.085, angle = 0]{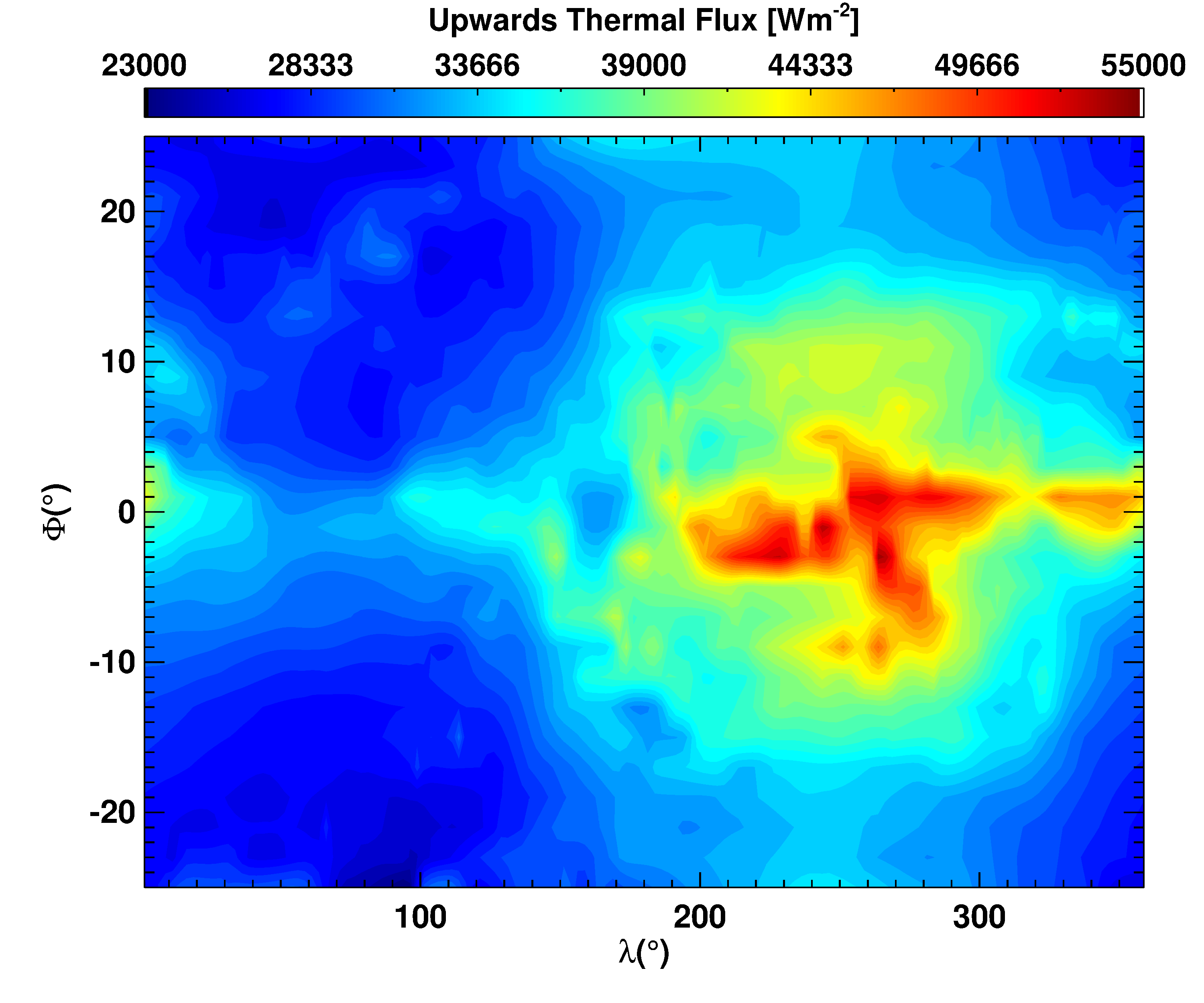}
\end{subfigure}\hspace*{\fill}
\begin{subfigure}{0.48\textwidth}
\includegraphics[scale=0.085,angle=0]{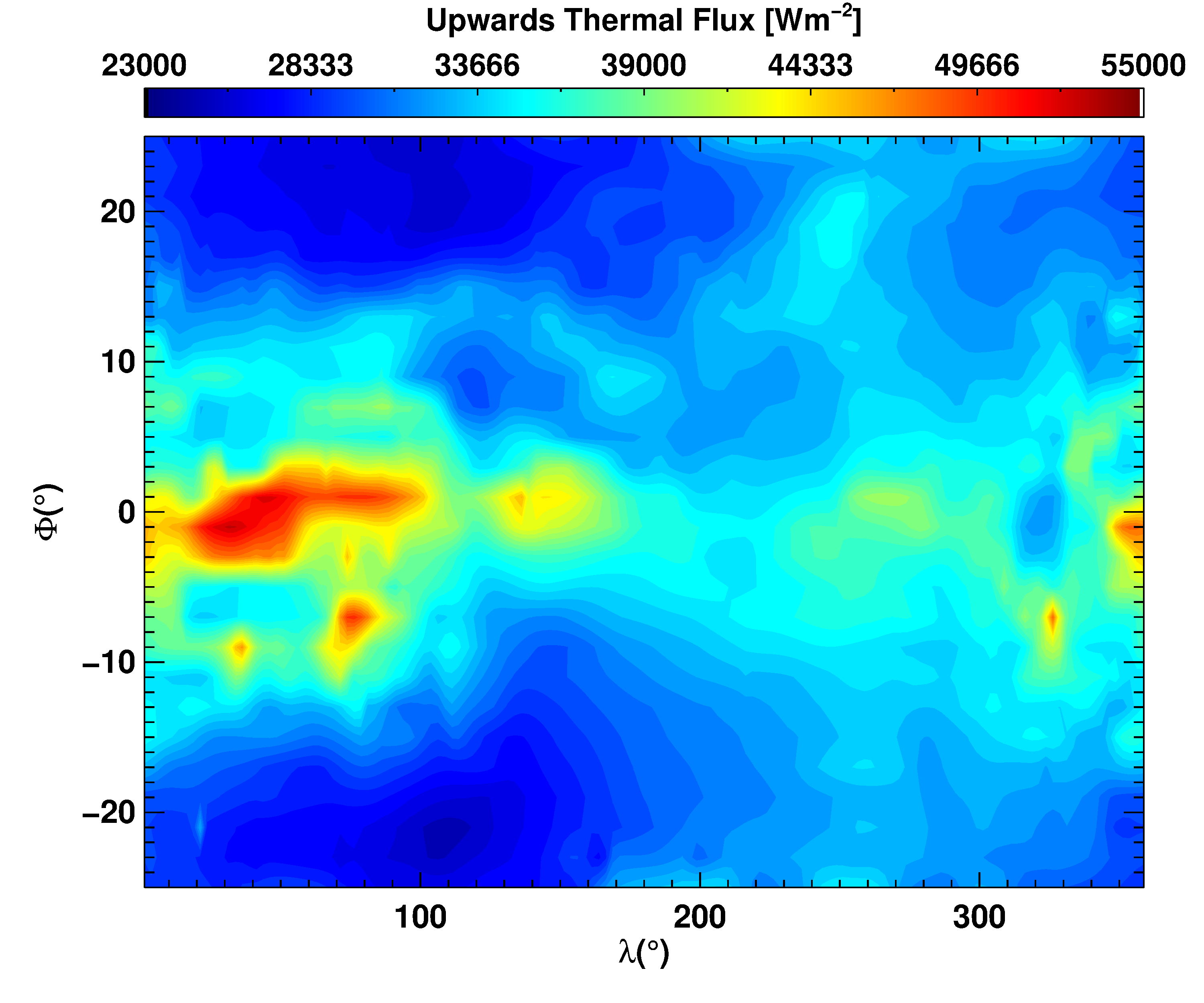}
\end{subfigure}

\begin{subfigure}{0.48\textwidth}
\includegraphics[scale = 0.085, angle = 0]{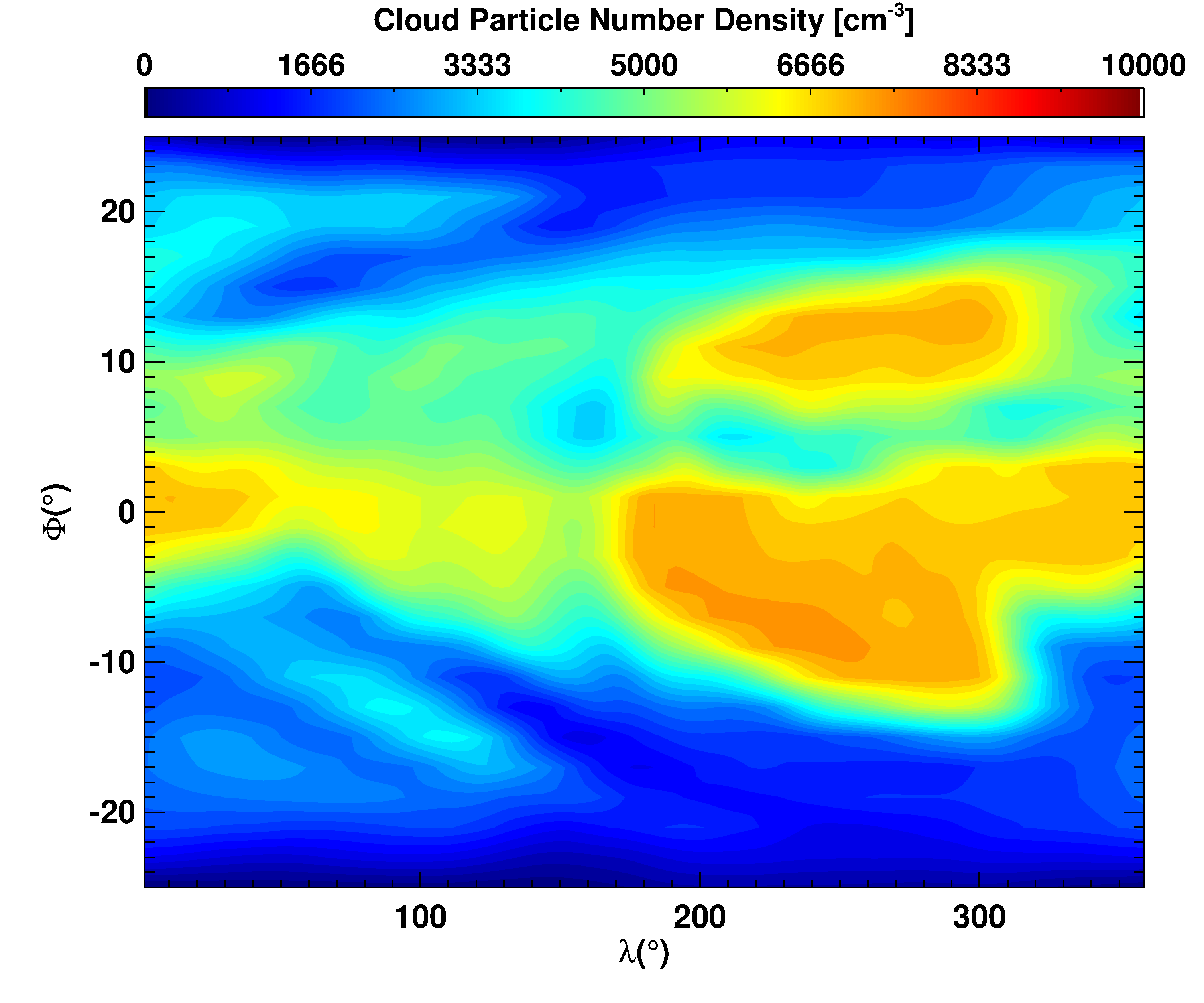}
\end{subfigure}\hspace*{\fill}
\begin{subfigure}{0.48\textwidth}
\includegraphics[scale=0.085,angle=0]{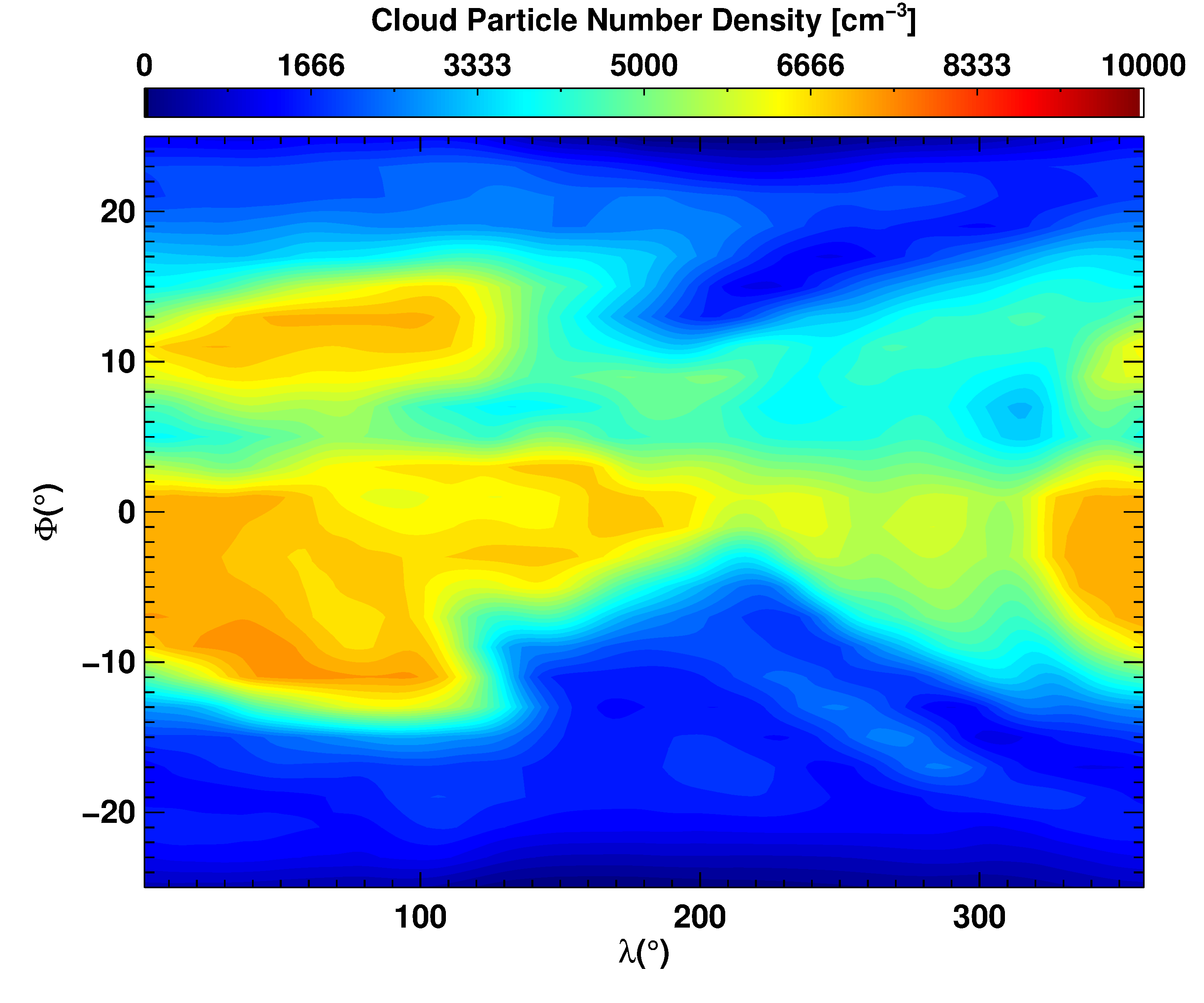}
\end{subfigure}

\caption{Upwards thermal flux (upper) and cloud particle number density (lower) at P = 1 Pa in the hot HD 209458 b atmosphere after t$_{\textrm{cloud}}$ = 106 (left) and 106.5 (right) days. Latitude has been confined to the equatorial jet (-25$^o$ < $\phi$ < 25$^o$)}
\label{fig:h209_variability}

\end{figure*}

With radiatively active clouds, the highly efficient Mie scattering and hence reduced absorption in the visual leads to an increase in flux in the Kepler bandpass, seen in Figure \ref{fig:h209_kphase}. We find no westwards offset in the visual phase curve, but this is expected due to cloud particles persisting across all planetary longitudes; a situation made possible by the high condensation temperature of our silicate particles. While the increased albedo from cloud coverage is detectable in the Kepler bandpass, variability is not. This is not true however for the thermal emission, Figure \ref{fig:h209_phase}, where we not only retrieve the eastwards offset due to the advected hot-spot, but find a variability in the flux. This flux variability is a consequence of the time-dependent nature of the cloud distribution (opacity) in the equatorial jet region. In Figure \ref{fig:h209_variability} we show the evolution of the cloud particle number density and upwards thermal flux for the upper atmosphere pressure level of P = 1 Pa. By analysing these values in the jet at two times, separated by approximately half the jet advective timescale, features in the inhomogeneity of the cloud coverage and emergent flux are offset by roughly $\lambda$ = 180$^o$. Large, non-transient deviations from the mean cloud particle distribution, opacity and hence thermal flux therefore evolve with the advective timescale, driving the variability seen in the thermal phase curves. In addition to the variability, the cooling atmosphere and increased IR opacity results in an order of magnitude reduction, compared to clear-skies, in the Spitzer-band flux. This worsens the agreement with observations of emission, which for the day-side is closely matching the clear-skies predictive flux, suggesting that we are missing some physics from our cloudy model.


\begin{figure}
\includegraphics[scale=0.55,angle=0]{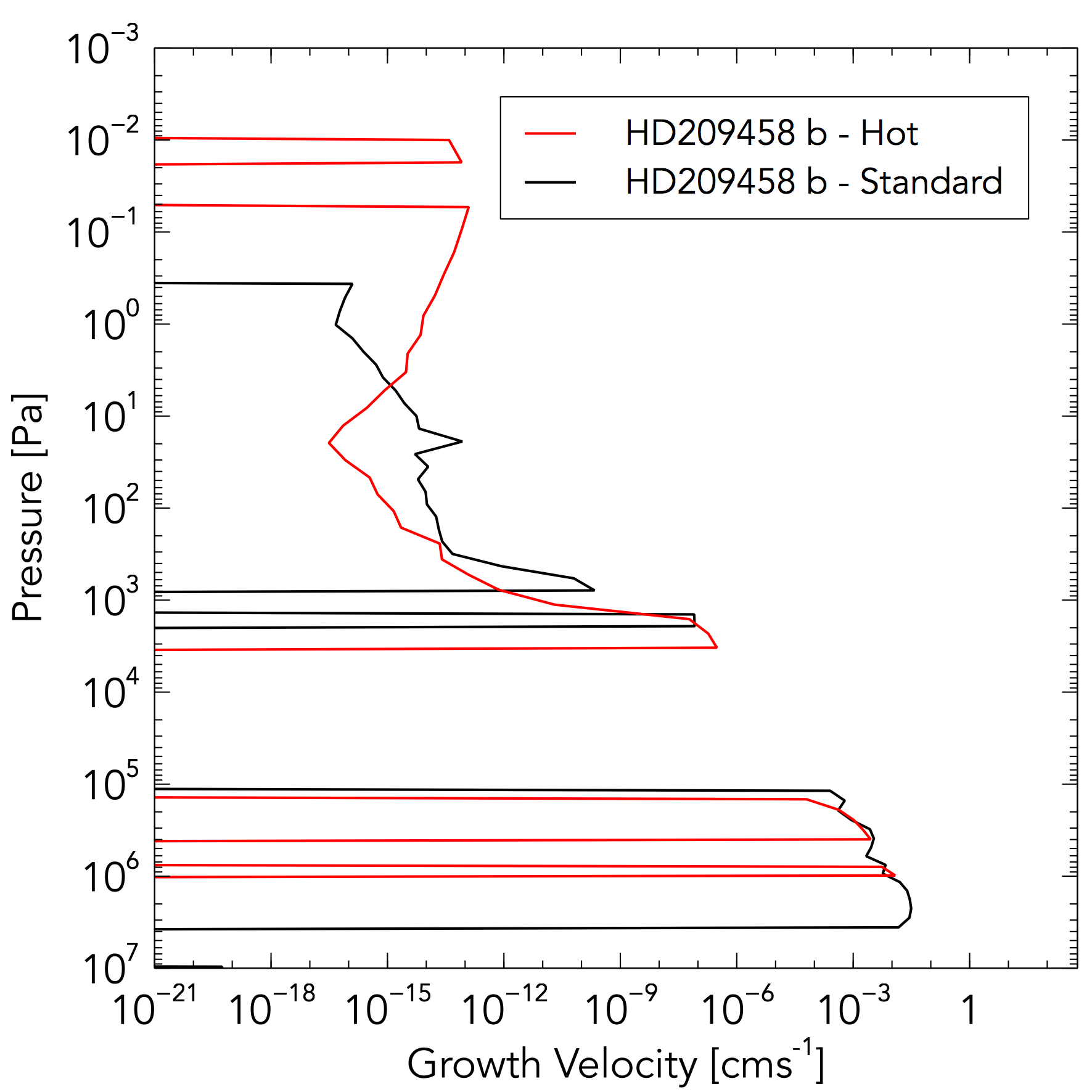}
\caption{Horizontally averaged growth velocity, $\chi^{\textrm{net}}$ for the active hot and standard HD 209458 b atmospheres at t$_{\textrm{cloud}}$ = 100 days (see Table \ref{tab:params} and Section \ref{sec:na}).}
\label{fig:h209_growth}
\end{figure}



\begin{figure*}[]

\begin{subfigure}{0.48\textwidth}
\includegraphics[scale=0.6,angle=0]{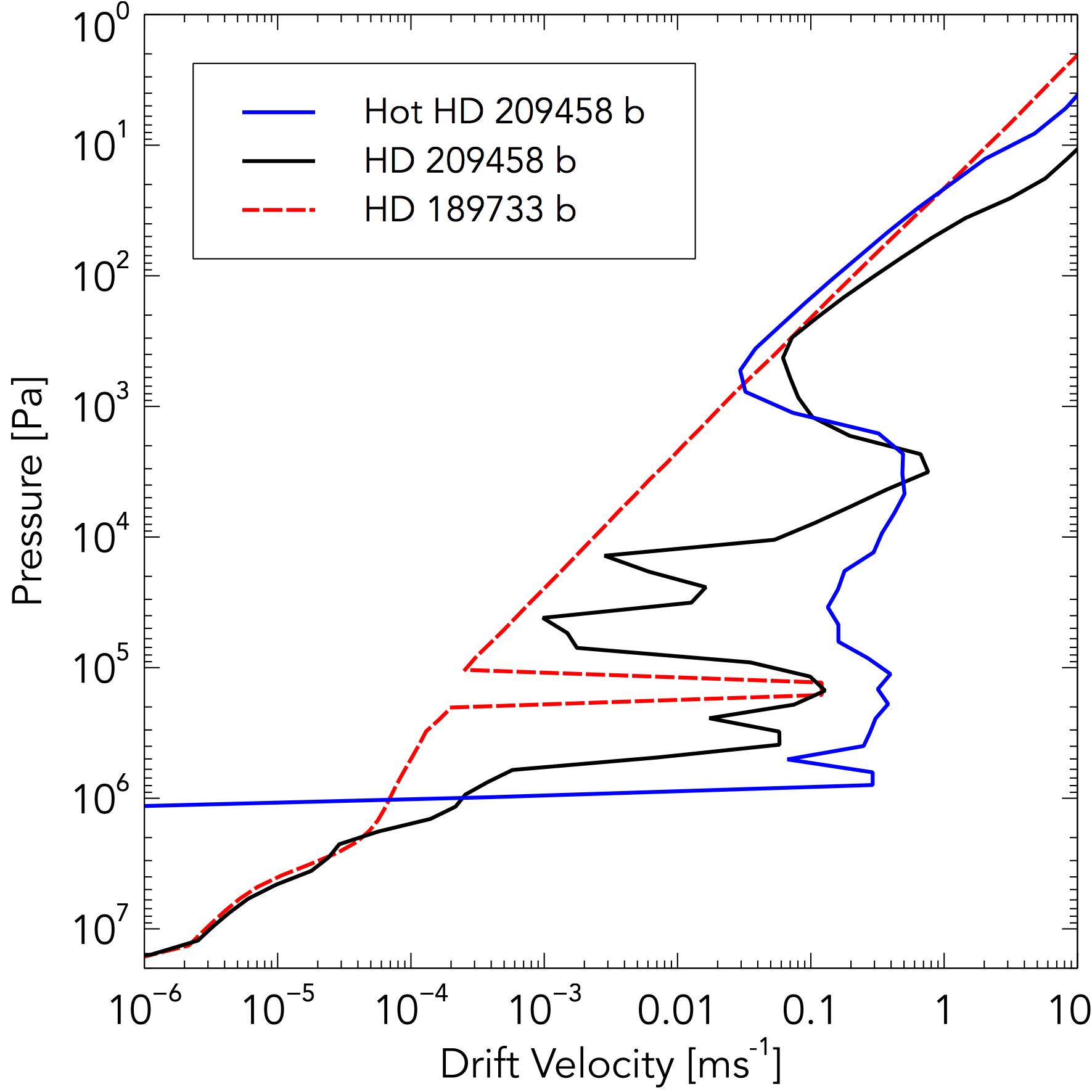}
\caption{Horizontally averaged drift velocities, v$_{\textrm{dr}} $ [ms$^{-1}$], for both the active hot and standard HD 209458 b atmospheres at t$_{\textrm{cloud}}$ = 100 days (see Table \ref{tab:params} and Section \ref{sec:na}).}
\label{fig:h209_drift}
\end{subfigure}\hspace*{\fill}
\begin{subfigure}{0.48\textwidth}
\includegraphics[scale=0.58,angle=0]{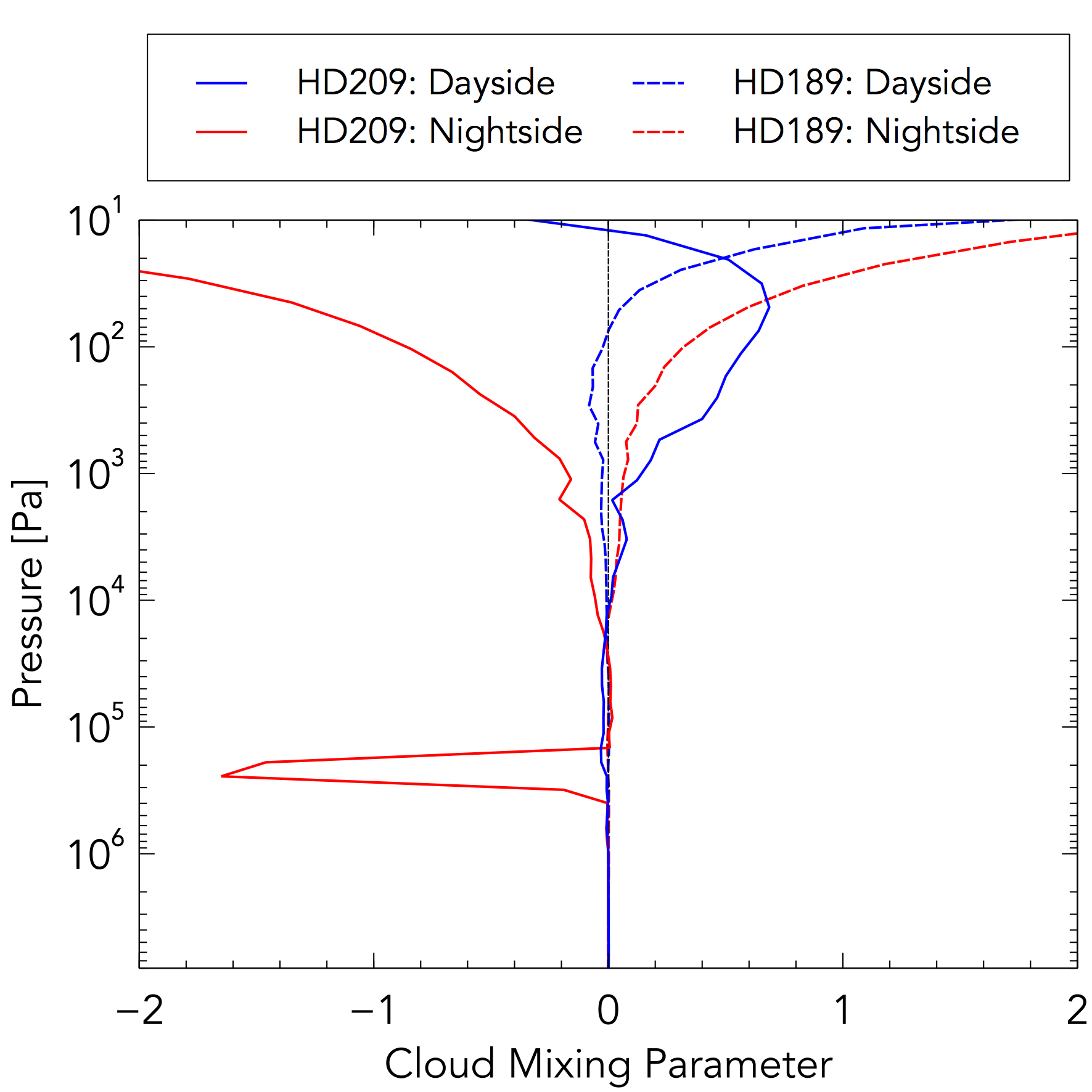}
\caption{Meridional mean of the cloud mixing parameter (ratio of vertical windspeed to drift velocity) as a function of pressure. The red line follows the night-side at $\lambda$ = 0$^{\textrm{o}}$ and the blue line follows the day-side at $\lambda$ = 180$^{\textrm{o}}$ with solid as HD 209458 b and dashed as HD 189733 b.}
\label{fig:h209_mix}
\end{subfigure}

\caption{Vertical transport of cloud, considering the settling velocities (left) and the cloud mixing parameter (right).}

\end{figure*}



\begin{figure*}
\includegraphics[scale=0.75,angle=0]{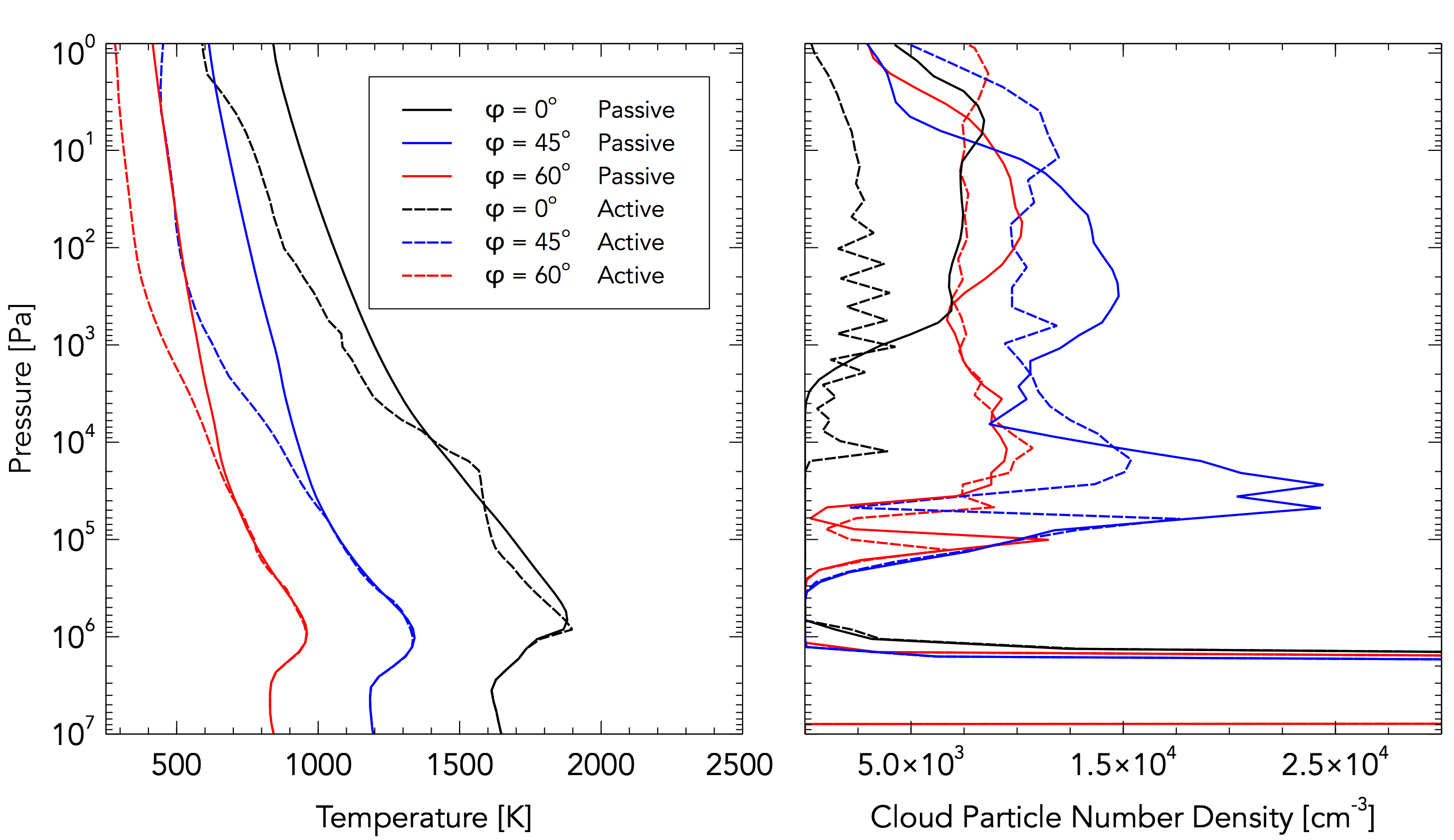}
\caption{Longitudinally averaged gas temperature [K] (left) and particle number density (right) for the standard HD 209458 b model at the end of the transparent cloud stage at t$_{\textrm{cloud}}$ = 50 days (see Table \ref{tab:params} and Section \ref{sec:na}) and the radiatively active stage at t$_{\textrm{cloud}}$ = 100 days. For each time, three profiles are plotted that cover latitudes of $\phi$ = 0$^{\textrm{o}}$ (black), $\phi$ = 45$^{\textrm{o}}$ (blue) and $\phi$ = 90$^{\textrm{o}}$ (red).}
\label{fig:c209_tdiffnddiff}
\end{figure*}


Figure \ref{fig:h209} (upper, right) shows the maximum wind velocity in the equatorial jet has increased by over 100 ms$^{-1}$, but more noticeable is the change of the jet shape, broadening at higher pressures. The out-of-equator westwards, retrograde flow is replaced with eastwards flow such that the majority of the atmosphere is now in prograde rotation (Note: AAM is conserved to better than 0.01$\%$).



\begin{figure*}

\begin{subfigure}{0.48\textwidth}
\includegraphics[scale=0.085,angle=0]{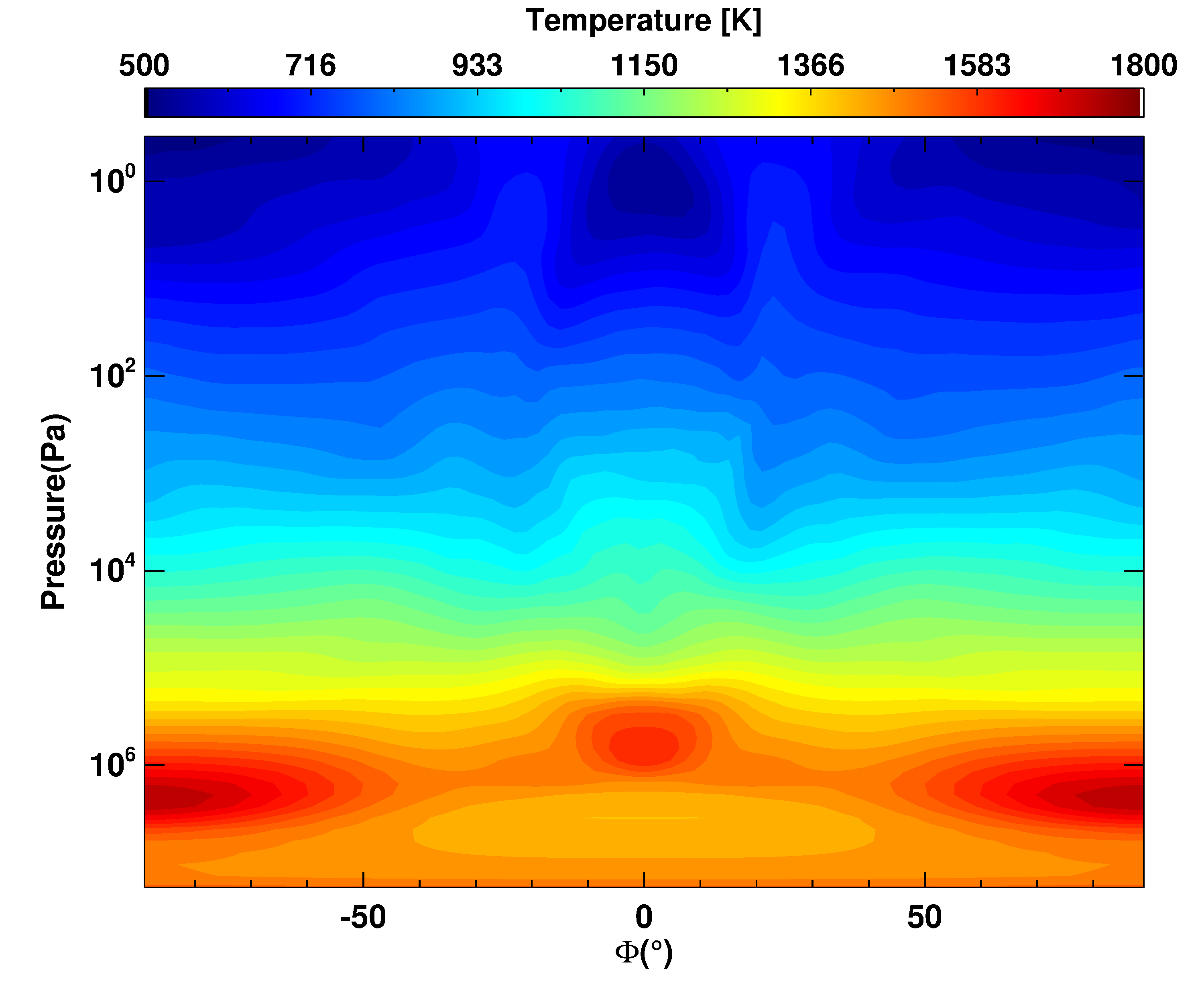}
\end{subfigure}\hspace*{\fill}
\begin{subfigure}{0.48\textwidth}
\includegraphics[scale=0.085,angle=0]{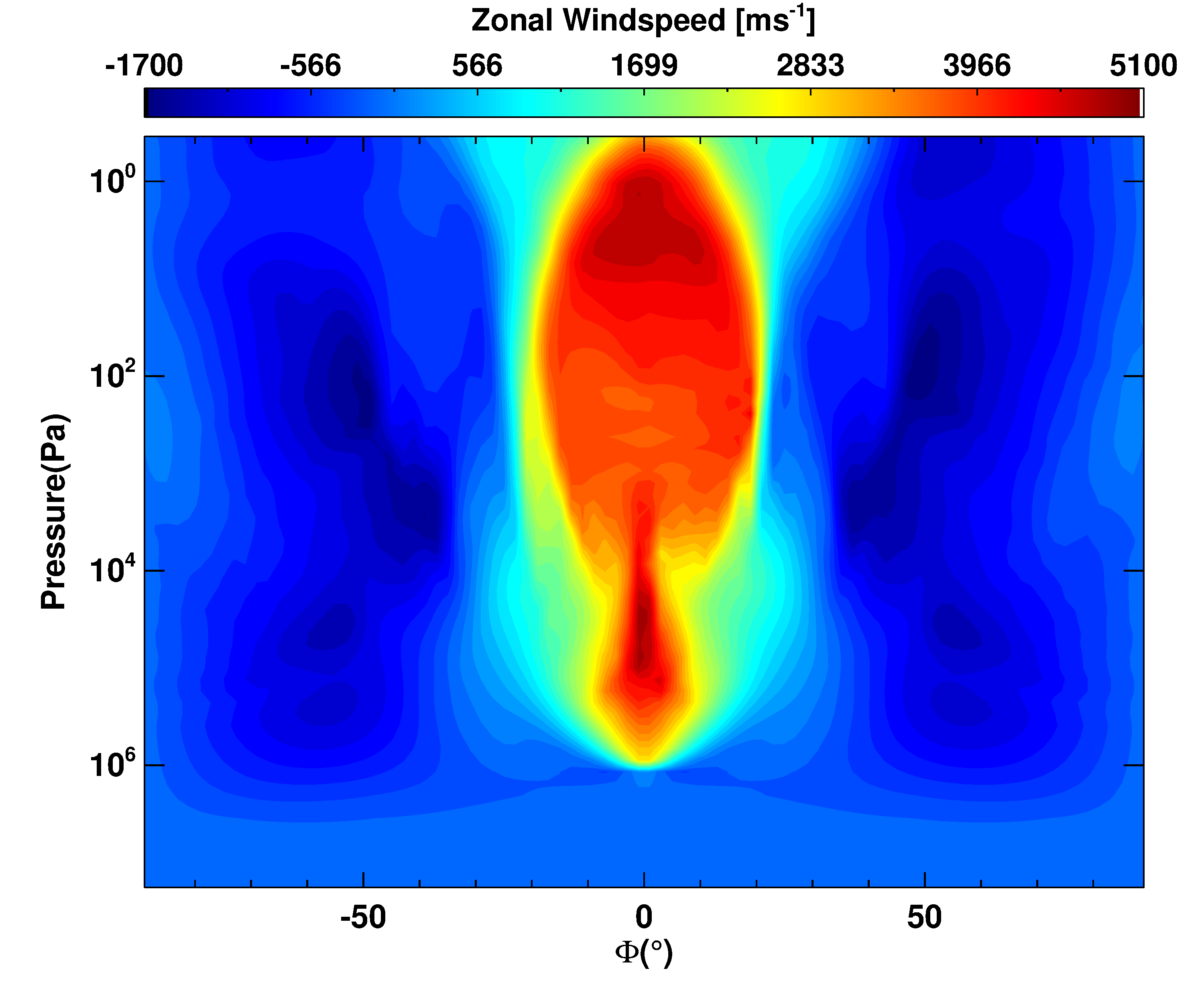}
\end{subfigure}

\begin{subfigure}{0.48\textwidth}
\includegraphics[scale = 0.085, angle = 0]{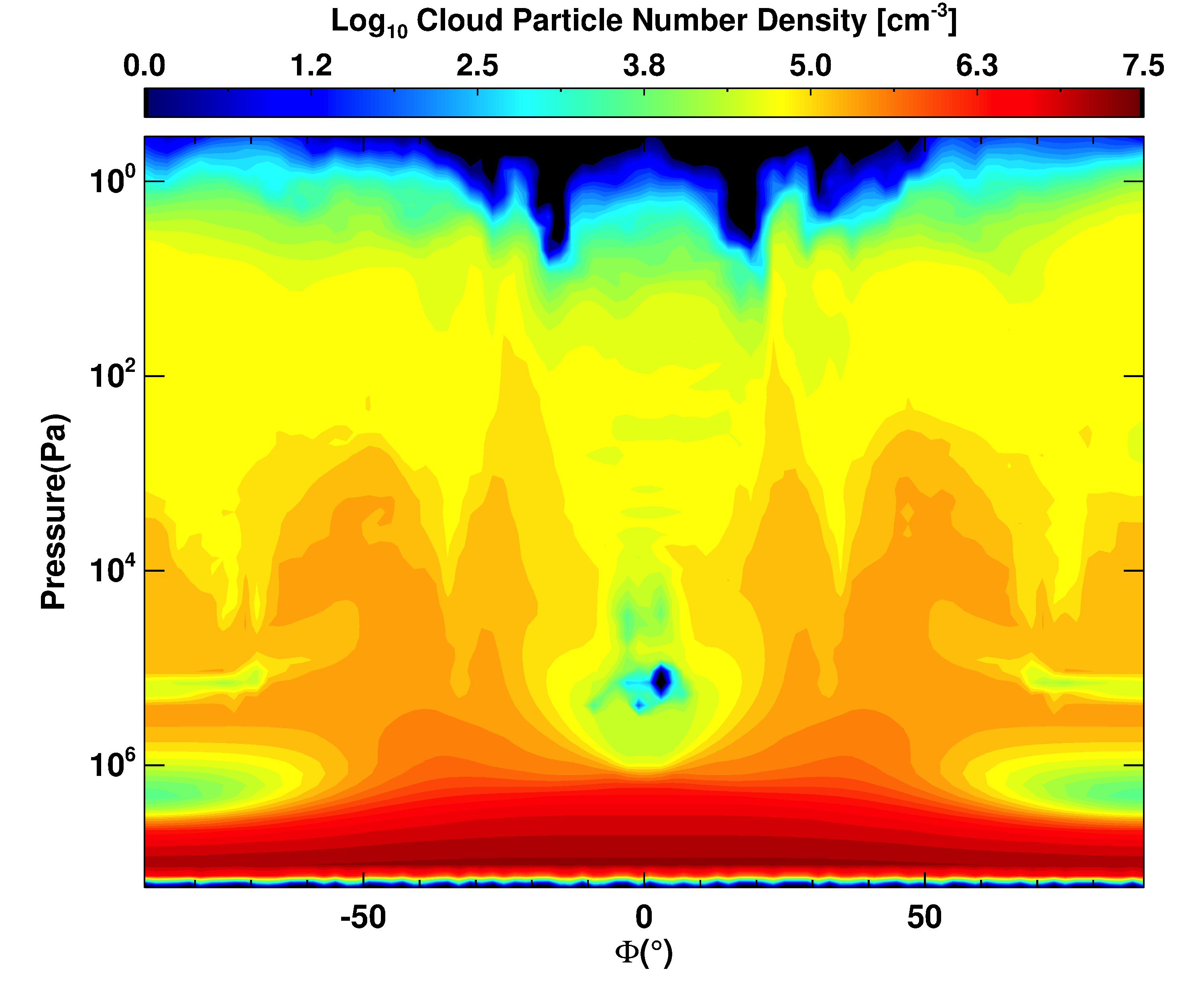}
\end{subfigure}\hspace*{\fill}
\begin{subfigure}{0.48\textwidth}
\includegraphics[scale=0.085,angle=0]{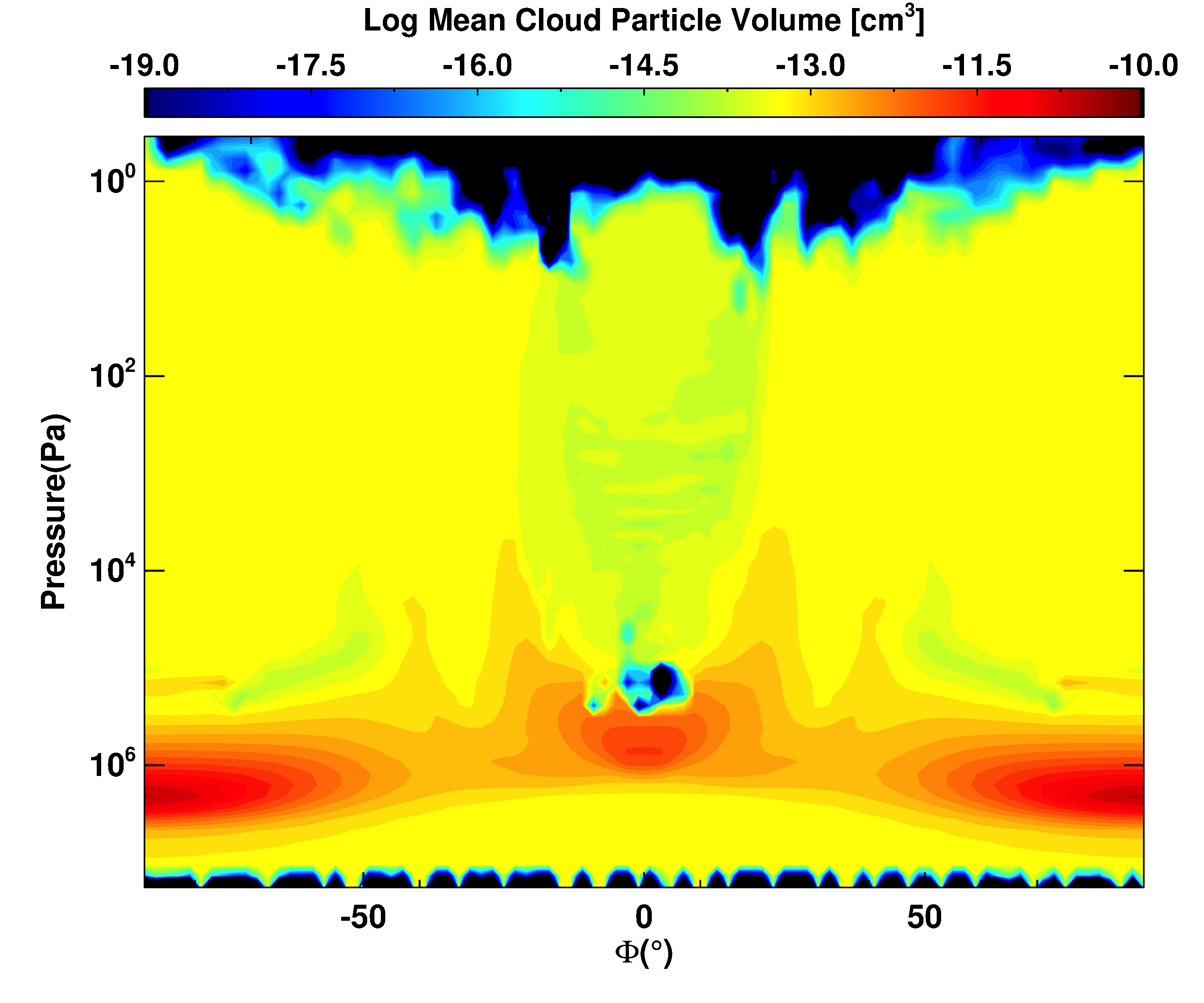}
\end{subfigure}

\begin{subfigure}{0.48\textwidth}
\includegraphics[scale=0.085,angle=0]{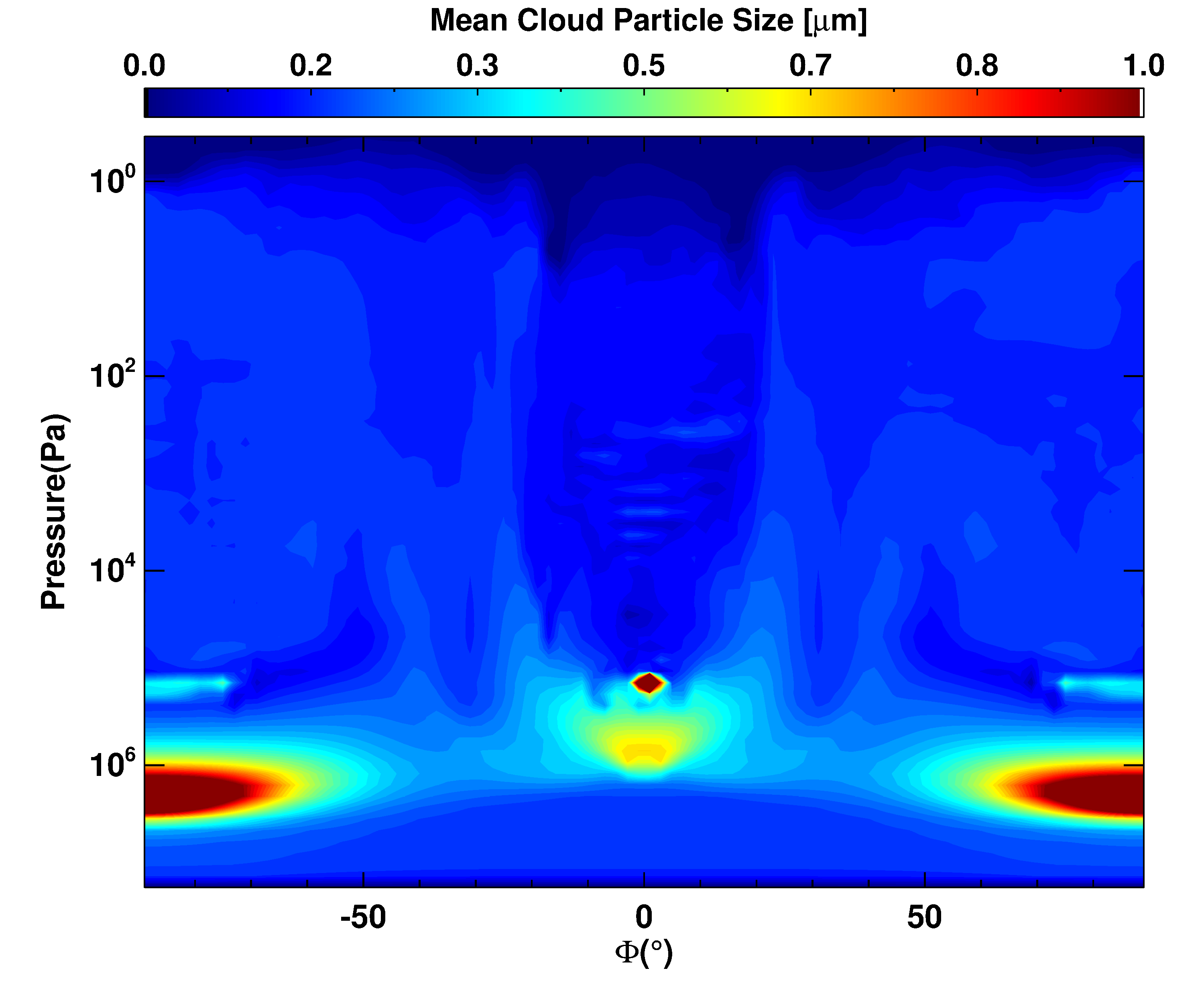}
\end{subfigure}\hspace*{\fill}
\begin{subfigure}{0.48\textwidth}
\includegraphics[scale=0.085,angle=0]{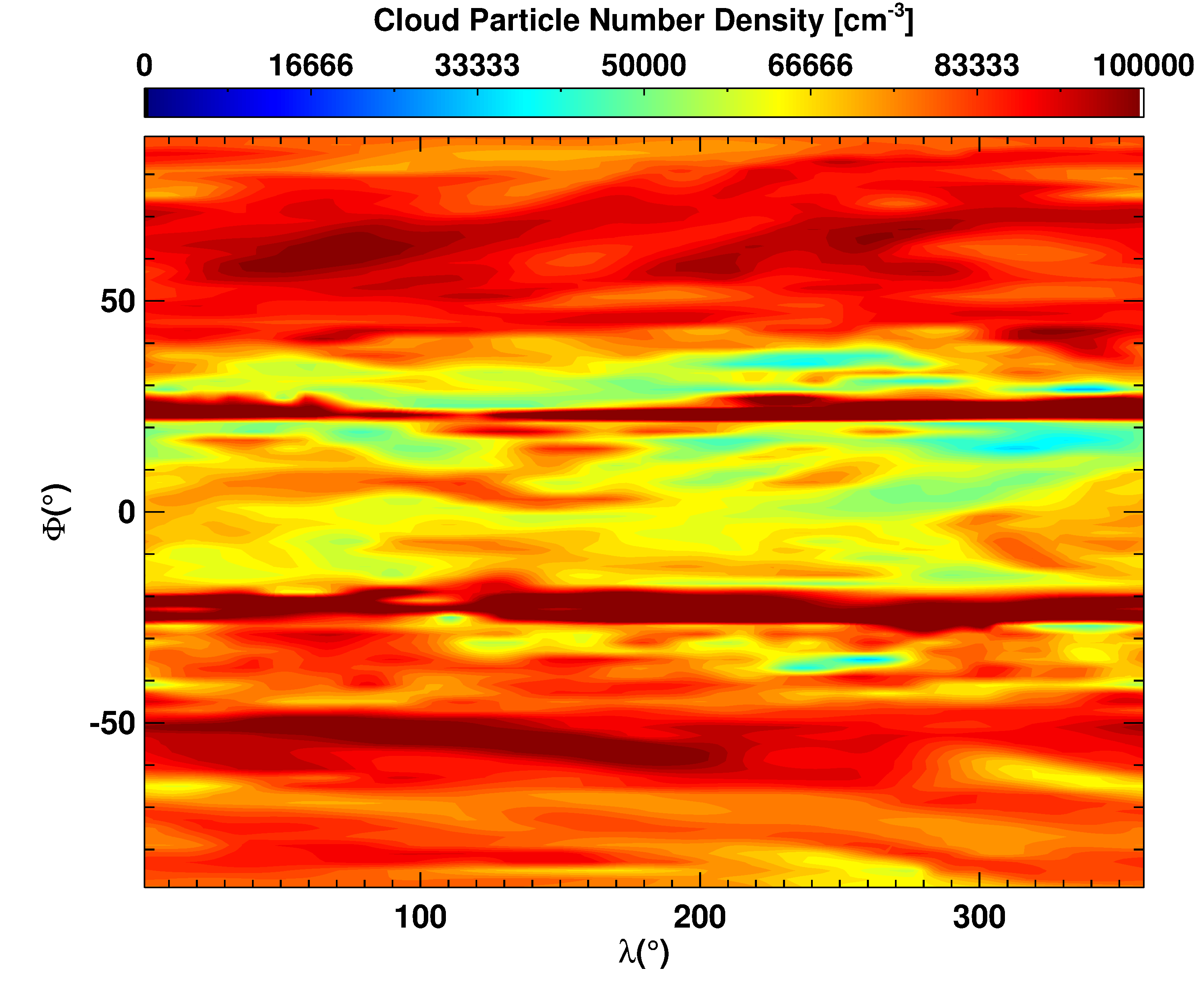}
\end{subfigure}

\caption{Thermal, dynamical and cloud properties of HD 189733 b after t$_{\textrm{cloud}}$ = 100 days (see Table \ref{tab:params} and Section \ref{sec:na}) with 50 days of active clouds. Black regions correspond to areas with values below the colour-bar minimum.}
\label{fig:c189}

\end{figure*}


In Figure \ref{fig:h209} (middle, left) the cloud structure remains in a two deck, out-of-equator configuration. The width of the particle depleted region at the equator increases in response to the widening of the jet itself. The top of atmosphere pressure increases due to the changing thermal profile, but the efficient sedimentation of particles produces a well defined cloud top. The depleted cloud region from P = 10$^4$ Pa still exists, along with the TiO$_2$ layer at the base. The deck loses its layered structure however, with a more stochastic or disorganised nature to the particle number density distribution. The lofted particles in the upper atmosphere in the equatorial region have drifted down to higher pressures, possibly in response to less efficient vertical mixing from the dampened winds. The mean cloud particle volume, shown in Figure \ref{fig:h209} (middle, right) indicates that the large growth velocities at higher pressures continues to result in more voluminous cloud particles near the base, and this is reflected by the increased mean particle size in Figure \ref{fig:h209} (lower, left).


\begin{figure*}
\includegraphics[scale=0.75,angle=0]{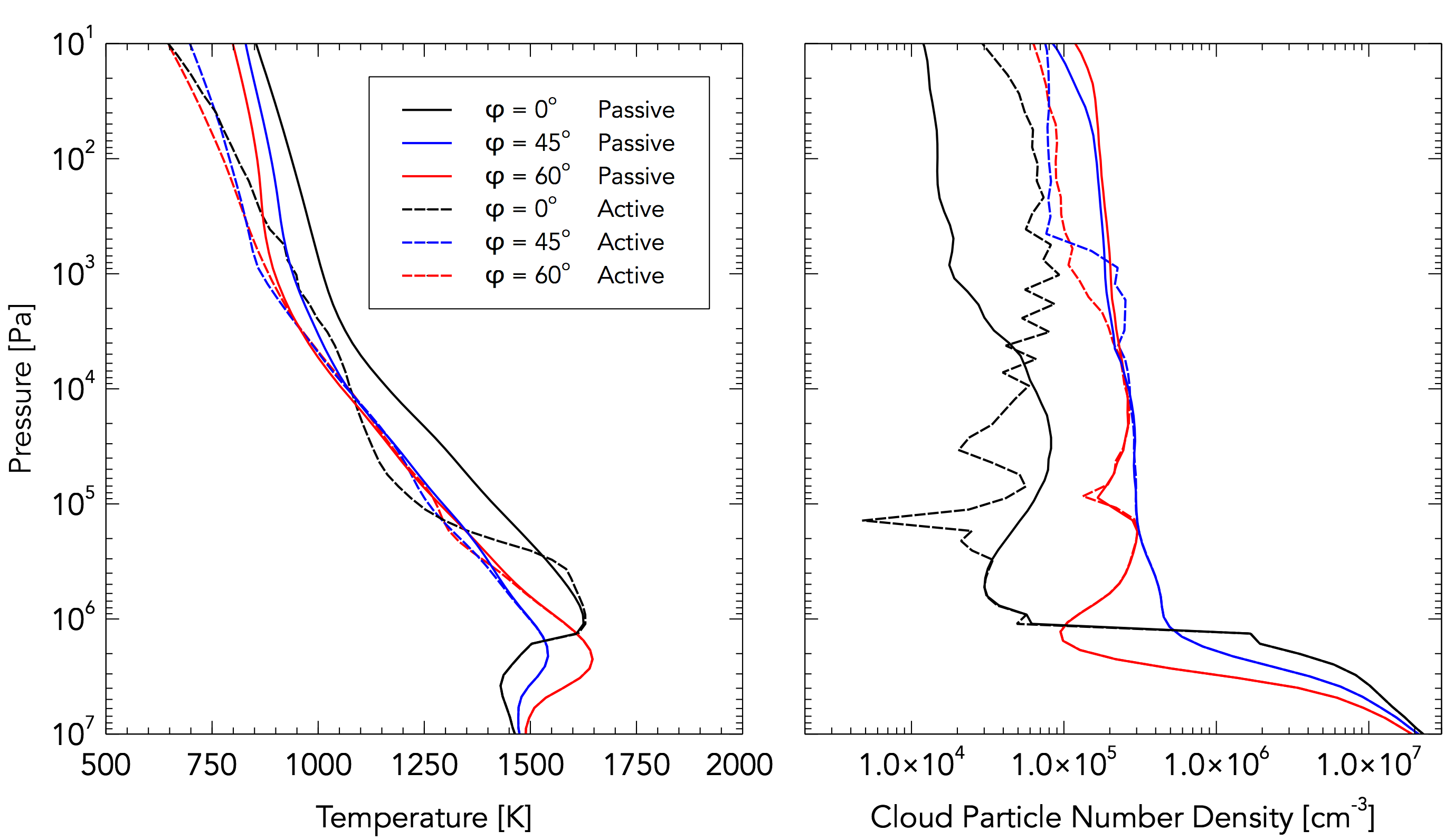}
\caption{Longitudinally averaged gas temperature [K] (left) and particle number density (right) for the standard HD 189733 b model at the end of the transparent cloud stage at t$_{\textrm{cloud}}$ = 50 days (see Table \ref{tab:params} and Section \ref{sec:na}) and the final active stage at t$_{\textrm{cloud}}$ = 100 days. For each time, three profiles are plotted that cover latitudes of $\phi$ = 0$^{\textrm{o}}$ (black), $\phi$ = 45$^{\textrm{o}}$ (blue) and $\phi$ = 90$^{\textrm{o}}$ (red).}
\label{fig:c189_tdiffnddiff}
\end{figure*}


Figure \ref{fig:h209} (lower, right) shows the cloud particle number density for a horizontal slice at P = 10$^2$ Pa. There is little zonal asymmetry in the cloud coverage, clearly indicating the ability for silicate cloud particles to persist across the night and day side of the planet. Since the day-side temperature at this pressure level does not exceed the condensation temperature of the silicate materials used in our simulation, cloud particles can grow and persist at all longitudes. There is a stark contrast with the distribution of cloud particles at the same pressure level for the end of the transparent cloud stage (cf. Figure \ref{fig:passive_h209}a). Between $\phi$ = 30$^{\textrm{o}}$ - 60$^{\textrm{o}}$ there still remains a slight bias towards cloud on the cooler night side, but it is more subtle. For $\phi$ $>$ 60$^{\textrm{o}}$ the reverse is true, with a large contrast between the day and night and regions of sparse cloud coverage on the night side. Interestingly, the low cloud particle number density regions correspond to cold spots that would usually indicate favourable conditions for cloud formation. Particles originally trapped at the equator at t$_{\textrm{cloud}}$ = 50 days still exist but are smoothed out over the full longitude of the atmosphere. On the night side at higher latitudes there are regions of extremely small cloud particle number densities which can be explained by the low night side pressures and hence fast drift velocities that settle the cloud particles more efficiently to higher pressure levels.

The surface growth velocity is displayed in Figure \ref{fig:h209_growth}. Cloud particle growth is still occurring at the end of the simulation, although the velocities are low due to the depletion of elemental abundances. Particles continue to grow between P = 0.01 Pa and P = 5 x 10$^3$ Pa, as well as near the cloud base between P = 10$^5$ Pa and P = 10$^6$. The higher pressures at the base result in more efficient cloud particle growth but this region is also supported by the upwards mixing of elements in the evaporative zone. There is a distinct lack of cloud particle growth between P = 2 x 10$^3$ Pa and P = 10$^5$ that could be attributed to higher, unfavourable temperatures that would inhibit cloud formation without the additional support of high pressures.

Where cloud exists, the drift or settling velocity of the particles (Figure \ref{fig:h209_drift}) never drops below 2 cms$^{-1}$, and is mostly sustained above 10 cms$^{-1}$. The lowest velocities are found at P = 10$^3$ Pa which corresponds to a region of low particle number density. The highest drift velocities (v$_{\textrm{dr}}$ $>$ 1 ms$^{-1}$) are found in the very upper atmosphere due to the low atmospheric densities. To investigate the balance of cloud particle settling and vertical mixing, the difference between the settling velocity and the vertical wind speed is plotted in Figure \ref{fig:h209_mix}, such that the cloud mixing parameter defined as, $\zeta$ = v$_{\textrm{dr}}$ - $w$, (where $w$ is the vertical wind speed) is positive for net upwards transport of cloud particles. On the planet night-side, vertical equilibrium is achieved for P $>$ 10$^4$ Pa, but rarely exceeds this balance. This means that cloud particles around the base are suspended on the night-side, but do not see a net upwards advection. The day-side is far more conducive to vertical mixing due to the faster vertical wind speeds. For 10 Pa $<$ P $<$ 10$^4$ Pa, cloud particles are mixed upwards efficiently, despite precipitation velocities reaching 1 ms$^{-1}$. The transition to net upwards advection correlates with the well defined cloud top at P = 10 Pa. The ability for cloud particles to be lofted into the upper atmosphere, or at least suspended at higher altitudes (note that we do not resolve turbulent mixing in our simulations), is supported by \cite{parmentier13} who perform simple analysis of TiO tracer particles to find that lofting can occur for particles of 5 $\mu$m or less. Such a result appears to be necessary to explain the `hazy' transmission spectra of some planets where condensation curves are not crossed at observable pressures; the dredging of particulates from the deeper atmosphere is required to contaminate the observationally accessible upper atmosphere \citep{sing16}.

\subsubsection{Standard HD 209458 b}

The initial atmospheric temperature profiles, shown in Figure \ref{fig:ptinit}, reveal a similar or in some regions identical thermal structure above P = 10$^6$ Pa between the hot and standard HD 209458 b models. Maybe surprisingly then, the radiatively active and transparent temperatures and cloud particle number densities which are plotted in Figure \ref{fig:c209_tdiffnddiff} show for the standard HD 209458 b, a vastly differing picture to the hotter interior simulation. For the standard simulation, a stronger cooling is found for all latitudes, with the cloud deck at the 1 mbar pressure level leading to a shift in temperature by up to 250 K. For higher latitudes, the thermal properties of the atmosphere do not change below 5 x 10$^4$ Pa, even though the temperature inversion leads to the formation (in the transparent cloud stage) of a expansive cloud deck (n$_{\textrm{d}}$ $\gg$ 10$^6$ cm$^{-3}$). At the equator however, atmospheric cooling takes place for P $\leq$ 10$^6$ Pa. Only the equatorial profile sees a small heating, around P = 2 x 10$^4$ Pa. The largest difference in particle number density occurs for the equator where cloud particles decrease for P $\leq$ 10$^3$. In the higher latitude cloudy regions there is a net tendency to decrease the number of cloud particles, but both latitudes increase in number for the lowest pressures.

\subsubsection{HD 189733 b}

Due to the larger cloud particle number density a 20 day period of transient radiative cloud was necessary to achieve model stability before fully active clouds were enabled for the final 30 days. Changes to the thermal properties of HD 189733 b during the active cloudy stage are shown in Figure \ref{fig:c189} (upper, left) where the zonal mean of the atmospheric temperature is shown. As per both variants of HD 209458 b, highly scattering cloud returns a large proportion of the incoming stellar flux, reducing the net atmospheric energy budget and leading to cooling. This effect is better seen in Figure \ref{fig:c189_tdiffnddiff} where zonal means of both temperature and cloud particle number density are plotted for three latitudes. Unlike HD 209458 b, the temperature does not vary much with latitude. In the transparent cloud stage at the milli-bar pressure level the temperature contrast between the equator and high latitudes is only 200K (compared to 500K for HD 209458 b). Cooling of up to 200K occurs at the atmosphere top (P = 1 Pa) for all latitudes. Temperatures diverge from the transparent cloud profiles for P $\leq$ 10$^4$ Pa for $\phi$ = 45$^{\textrm{o}}$ and only P = 10$^3$ Pa for $\phi$ = 60$^{\textrm{o}}$, but the equator sees a temperature change for P $\leq$ 10$^6$ Pa, showing that the equatorial region is the most effected by the scattering clouds. A maximum atmosphere temperature change of $\Delta$T$_{\textrm{max}}$ = -200K is found in the jet for a wide range of pressures around P = 10$^5$ Pa. Like the standard HD 209458 b simulation, the presence of the thermal inversion means cloud extends all the way down to the lower simulation boundary at 2 x 10$^7$ Pa. The lack of a defined cloud base means no regions of heating, with the exception of a small 50K rise at the equator which coincides with the penetration depth of the jet structure.  

Figure \ref{fig:c189} (upper, right) shows the zonal mean of the longitudinal or zonal velocity, which when compared to the values at the end of the transparent cloud stage in Figure \ref{fig:c189_start} (right) reveals significant changes to the dynamics of the atmosphere. In response to the aforementioned thermal changes, at the lowest pressure the \emph{super-rotating} equatorial jet core appears to narrow from 30$^{\textrm{o}}$ to 20$^{\textrm{o}}$. However, lower velocities outside of the jet core lead to prograde flow that increases to, at its widest point, 40$^{\textrm{o}}$. Similar to what is seen in HD 209458 b, the reverse, westwards flow at higher latitudes decreases in velocity and meridional extent, meaning more of the atmosphere experiences a prograde flow. The jet itself has a lower velocity, likely due to the reduced energy pumping from decreased stellar heating.

The zonal mean of the cloud particle number density and volume are shown in Figures \ref{fig:c189} (middle, left) and \ref{fig:c189} (middle, right), respectively. As per the standard HD 209458 b atmosphere, the temperature inversion at the deepest pressures leads to cloud forming down to the lower simulation boundary. Since the temperatures in the deep atmosphere are still sufficiently low to enable condensation, the high pressures  are supportive of and increase the surface growth (and nucleation) rates leading to a deck with two orders of magnitude more particles than the upper atmosphere, above the inversion. For P $\leq$ 10$^6$ Pa the equator is again less voluminous and numerous in cloud particles than higher latitudes, although there is less of a contrast than for HD 209458 b. The only regions devoid of cloud particles are the lowest pressures at the jet boundary where settling has precipitated the cloud particles faster than it could be mixed upwards. At higher latitudes the cloud extends to the very upper boundary, meaning the cloud top is at P $\approx$ 1 Pa. A horizontal slice of the cloud particle number density at the P = 100 Pa pressure level (Figure \ref{fig:c189} (lower, right)) reveals a more disorganised structure, with a less cloudy jet but high cloud particle number density at the jet boundary and higher latitudes. As expected from the condensation temperature of silicate particles at this pressure, there is little asymmetry in the distribution of cloud particles as a function of longitude. The effect of changing thermal properties of the atmosphere due to active clouds is best viewed via the radiatively transparent and active profiles of the particle number density shown in Figure \ref{fig:c189_tdiffnddiff}. Between the atmosphere top and P = 10$^4$ Pa, the cooler temperatures lead to an increase in particles at the equator but a small decrease at higher latitudes. The moderately higher temperatures at the equator due to the geometry of irradiation means that, prior to active clouds, the nucleation rate is slower than the surrounding higher-latitude and more saturated areas. Hence the decrease in temperature from the scattering clouds means the remaining condensible gases trapped in the jet at the equator can now form seed particles. From P $>$ 5 x 10$^5$ Pa at the equator and for P $>$ 5 x 10$^3$ at higher latitudes, there is no divergence from the pre-active stage seed particle number density. This is because, particularly below the temperature inversion at P $\approx$ 10$^6$ Pa, all condensible gases have formed the maximum number of seed particles. Lower temperatures that could enhance seed nucleation or particle growth are of no use if the elemental reservoir is dry.

The mean particle size is shown in Figure \ref{fig:c189} (lower, left). The largest particles are found around the deep atmosphere polar hot-spots and correspond with a decrease in the number of cloud particles. In these regions the low nucleation rate has produced cells that contain a small number of particles that provided limited sites for further surface growth and as a result exceed 1 $\mu$m in size. However, for the majority of the atmosphere the particles are sub-micron in size, consistent with our HD 209458 b atmospheres. The sensitivity of particle or droplet size with the ratio of nucleation to surface growth, in addition to changing reaction rates across both a vertically or horizontally varying temperature profile reveals one of the significant limitations of post-processing cloud \citep{parmentier16}.

The lower drift velocities for HD 189733 b compared to HD 209458 b, shown in Figure \ref{fig:h209_drift} indicate that gravitational settling is less efficient in this atmosphere. Indeed the mixing parameter in Figure \ref{fig:h209_mix} reveal this to be a partially correct assumption since unlike HD 209458 b, there is a fast net upwards advection of cloud on the planet nightside for P $>$ 1000 Pa. This is not true for HD 209458 b which does not see upwards advection of cloud particles for any pressure on the nightside. However, there is only lofting of cloud particles on the dayside for P $>$ 100 Pa. 

Due to cloud structure similarities (in particular, the lack of zonal asymmetry) between HD 209458 b and HD 189733 b, we do not include in this work, an analysis of the observation properties (emission spectra and phase curves).

\section{Discussion}\label{sec:dis}

All of our simulations show a tendency for clouds to exist in thick decks, at high latitudes, outside of the equatorial jet. It would be easy to assume that the higher temperatures in the jet region contribute to conditions that inhibit the formation of cloud. However, at the millibar pressure level, the maximum temperature for the hot HD 209458 b does not exceed 1100K, well below the T $\sim$ 1500K condensation temperature of our silicate condensates (with the exception of the less thermally stable SiO[s]). At the start of the simulation, both the nucleation and growth velocities around the equator are positive, indicating that cloud formation is active in this region. We find that winds diverging from the equatorial jet lead to the meridional transport of cloud particles to higher latitudes. This advective process also applies to gas which is constantly fed to higher latitudes where seed particle nucleation and surface growth is enhanced from the higher supersaturations due to the lower temperatures. This strong dynamical and thermal interplay does not remove all material from the equator, but leaves a smaller number of cloud particles, particularly for HD 209458 b. The role of meridional (and zonal) winds in this process accentuate the limitation of using cloud that is post-processed from 3D atmospheric profiles; coupled models are necessary to handle the atmosphere dynamics and their influence on cloud particles. 

Since, except for the high temperature deep atmosphere, the atmospheric temperature does not generally exceed the condensation temperature of our silicate materials, cloud particles can form and persist on both the night and day sides of both planets investigated. A possible explanation for the discrepancy between our results and \cite{lee16}, who find that the day-side temperatures evaporate the cloud formed on the cooler night-side leading to a cyclic cloud formation process, is that their day-side temperatures are artificially high due to the non-scattering model where the calculated scattering cross section is added entirely into absorption. Our work indicates that the cloud scattering far outweighs absorption and hence this approximation is unlikely to capture the true radiative balance of the atmosphere. Absorption dominates only at the cloud base where the larger particles are more efficient at absorbing the outgoing thermal radiation, leading to small regions of heating beyond the initial temperature. Our distribution of cloud particles actually shows a slight bias towards higher cloud particle number densities on the day-side, with a depleted region at the southern polar region. Such an effect could be due to the pressure difference of the atmosphere between the day and night side, with cloud particles at a set physical height advecting between higher, cloud formation supporting pressures on the day side to lower pressures on the night side. The effect demonstrates the importance of capturing the horizontal dynamics of the cloud. 

A result consistent with both HD 209458 b and HD 189733 b is that after clouds are coupled to the radiative transfer scheme and thus begin altering the thermal profile of the atmosphere by scattering and absorption, the distribution of cloud particles becomes more disorganised. This is particularly true when looking at the cloud particle number density at P = 100 Pa. Prior to the active stage, the cloud coverage is largely inhomogeneous with latitude due to the jet dynamics and thermal effects, as well as showing slight bias to night-side formation from cooler temperatures. After clouds are allowed to scatter and absorb for 50 days, a much more stochastic distribution of cloud appears, particularly for HD 189733 b. Variability in the cloud opacity causes fluctuating heating rates that lead to changes in atmospheric density. In HD 189733 b, density enhancements running parallel and just outside of the equatorial jet lead to sharp increases in cloud particle numbers, since the number density is a linear function of the gas density.

Due to the symmetric distribution of cloud particle number density and volume between the east and west limbs of the atmosphere, the cloud albedo is close to constant across the stellar-illuminated atmosphere. The consistency in the reflected flux with longitude means that we find no westwards offset in the visual phase curve. The work of \cite{parmentier16} suggests that a planetary equilibrium temperature exceeding 1500K would be necessary to introduce evaporative regions for MgSiO$_3$. For temperatures in our HD 209458 b atmosphere, we would need to include the more volatile metal sulfides as a condensate species to have a similar day-night cloud cycle as found in \cite{lee16}. However, since our particles are mixed composition it is likely the evaporation process would happen on and off the more stable silicate particles we explore here, meaning rather than day-side evaporation of single-specied cloud decks, we would be seeing large bulk changes in the particle volume due to the removal of the volatile components of that particle. Again, the value of capturing the horizontal advection of cloud particles is demonstrated by the advection-driven temporal variability seen in the thermal phase curves, and will be critical in modelling observations of atmospheric variability in planets such as HAT-P-7b \citep{armstrong16}.

Despite the large contrast in cloud particle number density between the equator and higher latitudes, Figure \ref{fig:h209_swout} which shows the outgoing stellar irradiation, reveals the equatorial region is almost as `active' as the more optically thick cloud at higher latitudes, suggesting that only a small covering of tiny particles are necessary to effectively Mie scatter a significant proportion of the irradiation. Our primary result is that our sub-micron sized, silicate-based and hence highly scattering cloud particles result in a large fraction of the stellar irradiation being blocked from penetrating deep into the atmosphere. This stellar irradiation, which would otherwise be absorbed by gas species in our cloud free simulations, no longer contributes to heating. The net energy budget of the atmosphere reduces and the overall effect is for the atmosphere to cool. As a result of the cooler conditions and the increased optical depth due to the cloud, the outgoing thermal radiation is reduced and contributes to a much lower thermal emission flux. The implications for a cooler atmosphere are significant. Firstly, they provide us with what may be the most self-consistent pressure-temperature profiles of HD 209458 b to date. Secondly, the change in conditions invoke us to consider condensates that would not be undergoing active condensation at our pre-active cloud temperatures. It it worth cautioning that our current non-inclusion of condensates with a reduced diaphaneity will likely overestimate the scattering intensity of our cloud. We endeavour to explore a wide range of potential condensates, including iron and corundum, in future work and discuss the implications of this in Section \ref{sec:summary}. A secondary effect to the cooling atmosphere in response to highly scattering, radiatively active clouds is a weakened meridional and vertical flow; the maximum vertical windspeed for HD 209458 b is reduced typically by an order of 5. Interestingly, zonal flows are maintained, with the jet strength stable across the full length of the simulation. The mechanism behind this dynamical change is beyond the scope of this work and will be explored in future studies.

It is a valuable exercise to consider how the results may change when relaxing some of our necessary model limitations. One of these is the simplification that all seed particles are formed via the homo-molecular nucleation of TiO$_2$. In theory, there would be a wide chemical range of additional cloud condensation nuclei, from additional oxides to complex hydrocarbons (aerosols). Despite the low abundance of titanium, seed particles of this composition form freely, in large quantities throughout the atmosphere, in our simulations. The addition of further seed species is therefore unnecessary, for at least providing condensation sites, as there are plentiful cloud condensation nuclei available for surface growth. For this reason, including more nucleation species is unlikely to change the overall horizontal distribution of the cloud. However, should seed particles form in increased abundance due to multiple nucleation pathways, the mean particle size will likely be affected; increased seed particles will result in surface growth occurring over a larger number of particles, diluting the volume per seed. As cloud particles become smaller in size, their vertical dynamics will change; precipitation will be reduced as a result of lower settling velocities, potentially leading to increased cloud coverage in the upper atmosphere. Additionally the size of the particles is a strong function of their radiative properties. The smaller particles will contribute a large scattering (as shown in Figure \ref{fig:h209_p_sca}), potentially altering the thermal balance of the atmosphere. It is also worth considering the effect additional seed species may have on the overall chemical composition of the atmosphere. Even if additional nucleation does not lead to seed that play a large role in subsequent condensation, such seed particles would remove their component elements from the gas phase.

One of the key problems concerning cloud evolution is understanding the interplay between vertical mixing and gravitational settling or precipitation. Our simulations investigate the short term dynamics of the cloud particles, mostly dominated by the fast horizontal advection. On much longer timescales, if clouds are not supported by upwards advection they may settle out of the observable atmosphere and no longer contribute the same optical properties that we explore here. With our simulation times unable to capture the long timescales of the deep atmosphere precipitation, we must consider the balancing of vertical wind and settling velocities as a way of quantifying the net direction of cloud particles. The problem is more complex than this however as once cloud has settled into the evaporative region and returned to its gas phase constituents, there is no longer a particle to settle and the condensible gases are more easily mixed upwards (since they have no associated settling velocity) into cooler regions where they can re-nucleate. Providing there exists even the most minimal updraft near the cloud base, this should be enough to sustain a cloud deck. The deck thickness is then essentially defined by the `convective mixing' and settling timescales much like in the \cite{ackerman01} model. Again, post-processing procedures for diagnostic cloud severely restricts the ability to capture this cyclic process of settling, evaporation and re-nucleation. In our work the vertical mixing effect can partly be seen via the large growth velocities near the cloud base in Figure \ref{fig:h209_growth}. We have also shown via the mixing parameter in Figure \ref{fig:h209_mix}, the balance between settling and updraft velocities, the net direction or mixing efficiency of the cloud. For HD 209458 b in particular, we pay attention to the large net upwards advection experienced by cloud particles on the planet day-side, suspending and in some regions lofting micron to sub-micron particles. This supports the conclusion of \cite{sing16} who find that particulate lofting is necessary for some planets to make the upper, observable atmosphere hazy or cloudy. Additionally, the suspension of cloud particles in our simulations is consistent with the results of \cite{parmentier13} who find that global advection patterns lead to a well-mixed upper atmosphere (P $\gtrsim$ 10$^5$ Pa). For HD 189733 b this particle lofting also extends to the planet nightside, helping to keep cloud particles at high altitudes. A surprising outcome, perhaps, is that despite the diminished vertical updrafts during the radiatively active cloud stage, many regions of the atmosphere can still either suspend (no net vertical motion) or loft (net upwards advection) cloud particles. For the planetary atmospheres we study here at least, this might attenuate the importance of the turbulently transported cloud particles, a process we are unable to model at these numerical resolutions, but is parametrised in many 1D cloud formation models as a vertical eddy diffusion parameter, K$_{\textrm{zz}}$, which includes both bulk and turbulent flow.

In our introduction we discussed the lack of a trend in the level of observable cloudiness with planetary equilibrium temperature. This is particularly true for both HD 209458 b and HD 189733 b where despite very similar atmospheric temperatures, \cite{sing16} shows that while HD 189733 b shows evidence of cloudiness and haziness from the strong Rayleigh scattering slope, HD 209458 b appears to have a less (but still) cloudy atmosphere with more defined alkali metal absorption, reduced Rayleigh scattering and a stronger H$_2$O absorption. Our simulations indicate that both planets would have cloudy atmospheres, although HD 189733 b has much higher cloud particle number densities. The higher surface gravity of HD 189733 b would suggest that precipitation is a much more relevant dynamical component to the evolution of the cloud than for HD 209458 b and therefore it should be expected that cloud particles are more likely to settle out of the observable atmosphere and not contribute towards Rayleigh scattering. However we find that drift velocities in the upper atmosphere of HD 189733 b are comparable to those of HD 209458 b and in fact, coupled with the strong mean vertical atmospheric flows, lead to enhanced lofting of cloud particles in HD 189733 b. Therefore lofting in HD 189733 b, along with the large cloud particle number densities from efficient cloud formation, could lead to the enhanced cloud opacity inferred from the HD 189733 b transmission spectra. It is worth noting that although higher drift velocities may be expected from increased values of the gravitational acceleration, stronger gravity can lead to atmospheric compression, larger local gas densities and hence stifled drift velocities. Therefore it is wise to consider the balance between drift velocities and the vertical wind speeds.

The temperature of the deep atmosphere, below the cloud base, remains constant despite the overall cooling of the atmosphere. This could partly be a result of the long radiative timescales at these pressures which have not yet adjusted to the thermal changes in the upper atmosphere over a short (50 day) period. It is also possible that the cloud base, shown to absorb the intrinsic planetary thermal emission, keeps the temperatures here high by trapping the outgoing convective heat flux. A much longer simulation time, beyond the scope of this work, would be needed to disentangle these effects.

An interesting observation is that more cloud forms in the upper atmosphere of the hot HD 209458 b atmosphere compared to the standard profile. Why should this occur when the thermal structure of these atmospheres is consistent above P = 10$^5$ Pa? The thermochemical conditions should be identical between the two. The answer likely lies in the deep atmosphere where the two differ in their thermal structure. For the standard atmosphere, the temperature inversion allows a thick cloud deck to form, exhausting the elemental abundances. For the hotter atmosphere, precipitating particles evaporate at the condensation temperature transition and enrich the gas phase. These elements may be rapidly advected upwards to lower pressure regions that lead to increased surface growth (and nucleation) velocities that produce either more seed particles or larger grains. This highlights the important link between the deep and observable atmosphere due to cloud mixing and non-equilibrium chemistry; while the deep atmosphere is not observationally probable, understanding the thermodynamical properties of this hidden region is imperative in retrieving the correct distribution of cloud and hence optical properties of the upper observable atmosphere.

\section{Summary $\&$ Further Work}\label{sec:summary}


Clouds form easily in the atmospheres of HD 209458 b and HD 189733 b due to large extents of their atmospheres having temperatures below the condensation temperature of the majority of our silicate-based growth species. The exception is the deepest regions of the hotter HD 209458 b atmosphere, which unlike the standard profile of HD 209458 b, does not see a temperature inversion. Thus, in the deepest atmosphere, temperatures exceed that which would allow for nucleation and particle surface growth. A banded cloud structure forms as a result of increased particle numbers and cloud particle volume at higher latitudes. This effect is a combined result of cloud particle and gas advection from the equator to higher latitudes, the latter leading to increased saturation and hence efficient cloud formation. This meridional variation in cloud coverage can only be seen in 3D (or potentially 2D) models, highlighting the importance of including the full horizontal advection that 1D models are unable to capture.


In cloud-free simulations, the atmosphere maintains its thermal balance by interior thermal emission and the absorption of stellar flux. The presence of cloud has a significant effect on the thermal profile of the planet. Abundant sub-micron sized particles are efficient Mie scatterers that backscatter a large proportion of the incoming stellar short-wave flux. With a greatly reduced stellar absorption, the energy budget of the atmosphere decreases and subsequently temperatures decrease by up to 250 K. The reduced atmospheric temperature and opaque cloud blocking the emerging intrinsic planet flux leads to a significant decrease in the thermal emission. Correspondingly, the increased albedo in the visual, from scattering, boosts the returned stellar flux and makes the planet much brighter in the Kepler bandpass.


Since there is no zonal asymmetry in the distribution of cloud particle number density or volume, with cloud extending in equal abundance across the day and night side, there is no detectable offset in the visual phase curve. The presence of meteorology on the planet however is detectable by both the increased albedo in the visual and also the variable thermal flux originating from the advection of the disorganised cloud structure in the equatorial jet. In addition, the modification of the atmosphere's thermal profile changes the wind speeds on the planet. Particularly, a widening of the equatorial jet is seen leading to prograde flow at higher latitudes.


\cite{sing16} indicates a weak-cloudy atmosphere for HD 209458 b and cloudy and strongly hazy conditions for HD 189733 b. We find that both are cloudy at the observationally important millibar pressure level, with the higher gravity of HD 189733 b not settling cloud particles any more efficiently than in HD 209458 b. The moderately cooler temperatures in HD 189733 b do allow for cloud to form more abundantly, with the particle number density increasing by an order of magnitude over HD 209458 b, making HD 189733 the `cloudier' planet in both particle number and volume. Future studies will include synthetic transmission spectra\footnote{Transmission spectra from the UM is currently in development.} from the UM output which will assist in comparison with the observations of exoplanet atmospheres, such as those from \cite{sing16}. This will, in addition to our already included emission spectra and phase curves, allow for a more complete understanding of cloud's observational signatures. We also endeavour to improve the consistency between chemical phases which will allow for us to identify detectable species (such as the Sodium and Potassium) above the cloud photosphere, in the model spectra.


The success of coupling such a powerful cloud model to the UM naturally invites further investigations into exoplanetary nephology. Our chosen planets occupy a similar temperature range and therefore applying the model to a variety of atmospheric temperatures would be the natural progression; what clouds form and where, when the atmosphere is increasingly irradiated? Further studies would include the simulation of such super-heated atmospheres to explore the day-night condensation-evaporation cycle due to the large horizontal temperature contrasts. The application of the model to Brown Dwarf atmospheres would also be a useful test and may help us to determine why the temperature sequence in hot Jupiters does not correlate with cloud abundance and yet appears to do so for Brown Dwarfs.

This sophisticated coupled model may also be able to tackle further questions raised by observations. For example, WASP-121b has evidence of TiO/VO absorption in transmission \citep{evans16}, and water features that appear in emission on the day-side but absorption on the night-side \citep{evans17} both pointing to the presence of a thermal inversion at low pressures. Such a different thermal profile is likely to have an effect on the presence of cloud. Could a forward 3D model with radiative clouds predict the presence of gas phase TiO/VO at the relevant pressures?


A number of improvements to our model are required in order to tackle, more accurately, these aforementioned questions. These additions are not an added layer of complexity but instead represent the minimum framework for a physically and chemically complete model; e.g. no atmosphere is likely to contain only the limited condensible species in our current work, and some observations may be entirely explained by photochemistry. Such a complete model will be crucial in making the most meaningful comparisons with observations of exoplanetary atmospheres, and we list the most pressing improvements below.

\begin{itemize}  

\item Introducing additional condensates is an essential step in improving the radiative effects from cloud coverage. While we have included a number of silicate species, in addition to titanium dioxide, there are many condensates missing from our model that are predicted to play a significant role in cloud formation.

\item Cloud formation is dependent on, amongst other factors, the availability of gas phase species, which in turn depends on both the prescribed metallicity and the local thermochemical conditions. In future studies, we must explore both a changing metallicity and C/O ratio to determine cloud sensitivities to this expanding parameter space; doing so will be useful to those in the planet formation community who are attempting to constrain their own models from these values.

\item Our cloud model tracks the depletion of elements and solves for the key condensation reactants using a simple equilibrium scheme. These depleted/enriched element abundances are not coupled to our radiative transfer scheme which instead uses a parametrised calculation of the abundances of key gas phase absorbers \citep{burrows99}. We aim to improve chemical consistency in our model by introducing our own coupled flexible Gibbs energy minimisation scheme which will simultaneously provide gas phase abundances for both cloud formation and radiative transfer.

\item Drummond et al., (in prep) also provides us with the option to couple a chemical kinetics scheme to our cloud model. This will provide more accurate determinations of the abundances of condensible gases, but also allow us to simultaneously assess the contributions of both clouds and the cloudless chemical instability theory \citep{tremblin15}. Such a process will be both challenging and time-consuming but a necessary step for full chemical consistency.

\item Highly irradiated planets such as HD 209458 b and HD 189733 b may well have upper atmospheres with active photochemistry. With the importance of the Rayleigh scattering in observed spectra, it will be prudent to investigate the role of photochemically produced hydrocarbons. Adding a photochemical network will help to determine whether the Rayleigh scattering is set primarily by these complex species, tiny suspended condensates in our existing model, or a combination of both.

\item The importance of particle sizes on the cloud's radiative effects also invites us to consider implementing a more physical size distribution, rather than to assume a per-cell uniform particle size. In addition to the more accurate Mie calculations from such an applied size distribution, there will be improved accuracy to the drift velocities. While particle size is a crucial parameter in the cloud's optical properties, the shape of the cloud particles also plays a role. Currently, particles are assumed to condense into perfect spheres. Realistically, dusty particles may have irregular shapes, much like ice crystal in Earth's atmosphere. The assumption of spherical particulates may underestimate the scattering strength and must therefore be removed to understand the tendency of an atmosphere to net absorb or scatter.

\end{itemize}

Our results are an exciting introduction to modelling radiatively active clouds in 3D, but a number of model enhancements are in the pipeline. In addition, a much wider variety of planetary atmospheres should be explored to understand the change in cloud formation processes in super-hot gas giant to cool terrestrial atmospheres. While we present the most sophisticated 3D-coupled model of a cloudy atmosphere, it is complementary to other 1D, 2D and 3D schemes - both coupled and post-processed, which are required to tackle the complex nature of this field.

\begin{acknowledgements}
S.L is funded by and thankful to support from the Leverhulme Trust. The calculations for this paper were performed on the University of Exeter Supercomputer, a DiRAC Facility jointly funded by STFC, the Large Facilities Capital Fund of BIS, and the University of Exeter. Material produced using Met Office Software. BD acknowledges funding from the European Research Council (ERC) under the European Unions Seventh Framework Programme (FP7/2007-2013) / ERC grant agreement no. 336792. D.S.A. acknowledges support from the NASA Astrobiology Program through the Nexus for Exoplanet System Science. GKHL acknowledges support from the Universities of Oxford and Bern through the Bernoulli fellowship program. The authors would also like to thank the anonymous referee for their insightful report.
\end{acknowledgements}


\bibliographystyle{mn_new2}
\bibliography{clouds}

\end{document}